\documentclass[%
  aip,
  jcp,%
  amsmath,%
  amssymb,%
  reprint,%
  citeautoscript,%
  floatfix,%
]{revtex4-2}

\usepackage[version=4]{mhchem}

\usepackage{siunitx}
\DeclareSIUnit{\atomicunit}{a.u.}
\DeclareSIUnit{\atomicmassunit}{u}
\DeclareSIUnit{\angstrom}{\text {Å}}
\DeclareSIUnit{\cal}{cal}

\usepackage{%
  amsthm,%
  amssymb,%
  mathrsfs,%
  mathtools,%
  gensymb,%
  braket,%
  mleftright,%
  nicefrac,%
  interval%
}
\intervalconfig{%
  soft open fences%
}

\mleftright

\usepackage{bm}

\usepackage[usenames, svgnames, x11names, rgb]{xcolor}

\usepackage{graphicx}

\usepackage{subcaption}

\usepackage{%
  booktabs,%
  array,%
  adjustbox,%
  makecell,%
  multirow,%
  bigdelim,%
}
\newcolumntype{M}[1]{>{\centering\arraybackslash}m{#1}}
\newcolumntype{L}[1]{>{\raggedright\arraybackslash}m{#1}}
\newcolumntype{R}[1]{>{\raggedleft\arraybackslash}m{#1}}

\usepackage[inline]{enumitem}

\usepackage[%
hyperfootnotes=false,%
hypertexnames=false,%
hidelinks,%
]{hyperref}
\hypersetup{%
  bookmarksnumbered=true,%
  colorlinks=true,%
  citecolor=teal,%
  linkcolor=Blue4,%
}

\usepackage{mciteplus}

\newcommand*{\D}{\mathrm{d}}
\def\hbarr{{\mskip1.6mu\mathchar'26\mkern-7.6muh}}

\newcommand*{\qsymsq}{\textsc{QSym\textsuperscript{2}}}

\begin{document}

  \author{Bang C. Huynh}
  \email{bang.huynh@chem.ox.ac.uk}
  \affiliation{Physical and Theoretical Chemistry Laboratory, Department of Chemistry, University of Oxford, Oxford OX1 3QZ, United Kingdom}

  \author{Meilani Wibowo-Teale}
  \email{meilani.wibowo@nottingham.ac.uk}
  \affiliation{School of Chemistry, University of Nottingham, Nottingham NG7 2RD, United Kingdom}
  \author{Andrew M. Wibowo-Teale}
  \affiliation{School of Chemistry, University of Nottingham, Nottingham NG7 2RD, United Kingdom}

  \author{Frank De Proft}
  \affiliation{Research Group of General Chemistry (ALGC), Vrije Universiteit Brussel (VUB), Pleinlaan 2, B-1050 Brussels, Belgium}
  \author{Paul Geerlings}
  \email{pgeerlin@vub.be}
  \affiliation{Research Group of General Chemistry (ALGC), Vrije Universiteit Brussel (VUB), Pleinlaan 2, B-1050 Brussels, Belgium}

  \title{Reactivity of Ambident Nucleophiles in Magnetic Fields: a Combined Conceptual DFT and Current-DFT Study}


\begin{abstract}
  The influence of strong external magnetic fields (up to $0.30\,B_0$) on the electronic structure and reactivity of ambident nucleophiles is investigated using the nitrite and thiocyanate anions as prototypical examples.
  To capture magnetic-field-induced changes in reactivity, current-density-functional theory (current-DFT) calculations are interpreted through conceptual density-functional theory (conceptual DFT) descriptors, namely global hardness and local softness via Fukui functions, focusing on the evolution of the electronic structure and associated properties with increasing field strength in different orientations.
  By extending our previous adiabatic treatment of molecules to include higher-spin states, in analogy with earlier work on atoms, and by considering the symmetry properties of relevant quantum-chemical quantities within full magnetic groups using the \qsymsq{} framework, we uncover substantial magnetic-field-induced modulations of both molecular polarity and the shape of the Fukui function.
  Despite these modulations, the ambident nucleophilicity of both \ce{(NO2)-} and \ce{SCN-} is largely preserved, as is the preference for attack by soft electrophiles at the sulfur end of \ce{SCN-}.
  At higher field strengths, however, the Fukui functions become increasingly diffuse, reflecting the growing importance of external magnetic interactions relative to internal electrostatic forces.
  The resulting redistribution of local reactivity not only predicts reaction pathways with unexpected geometries but also raises the possibility that regioselectivity may become progressively less well-defined in the strong-field regime.
\end{abstract}

  \maketitle

\section{Introduction}
\label{sec:intro}

  In recent years, there has been increasing interest in understanding the influence of external electric fields\cite{article:Ciampi2018,article:Shaik2018}, mechanical forces\cite{article:Beyer2005,article:Stauch2016,article:Hickenboth2007}, and very high pressure\cite{article:Grochala2007,article:Rahm2019} on chemical reactivity.
  The ultimate aim is to unveil how these conditions could pave the way for new reactions that would otherwise be inaccessible under normal conditions.
  This shift toward ``new chemistries'' offers experimentalists the opportunity to broaden their portfolio of reaction conditions and expand the boundaries of synthetic design.
  Remarkably, the development of theoretical methods to describe the behavior of atoms and molecules under these extreme or unconventional conditions has often progressed hand-in-hand with the experimental implementation of novel reaction environments.
  Notable examples include studies on oriented external-electric fields, termed ``novel effectors of chemical change'' by Shaik\cite{book:Shaik2021}; the rapidly growing field of mechanochemistry\cite{article:Stauch2016}, marked by the launch of \textit{RSC Mechanochemistry} in 2025\cite{article:Batteas2025}; and recent progress in organic synthesis and materials design under very high pressure\cite{book:Margetic2019,article:Xu2022}.

  It is therefore natural to anticipate similar advances in the study of chemical properties and behaviors of atoms and molecules subject to strong magnetic fields.
  However, this area has long received limited attention from the chemical community\cite{book:Ruder1994,book:Schmelcher2002}, not least because magnetic fields that can be generated and sustained on Earth rarely exceed \SI{50}{\tesla}\cite{article:Hahn2019}.
  The discovery of magnetic fields on the surfaces of white dwarfs\cite{book:Ruder1994,book:Schmelcher2002,article:Angel1974,article:Angel1977} and neutron stars\cite{article:Kong2022}, which are orders of magnitude stronger than those on Earth, has nevertheless motivated physicists to investigate the effects of intense magnetic fields on atoms and small molecules.
  In fact, in the late 1990s and early 2000s, a series of studies examining the energetics, electronic structures, and optical spectra of light atoms and ions were published\cite{article:Ivanov1998,article:Ivanov1999,article:Ivanov2000,article:Ivanov2001,article:Al-Hujaj2004}.
  These works revealed profound changes in both ground and excited electronic states with increasing field strength: due to the electron Zeeman interaction, electron pairs are progressively uncoupled, and high-spin states with many unpaired electrons become stabilized in strong magnetic fields.

  The chemical relevance of these effects was recently highlighted by some of the present authors\cite{article:Francotte2022} in a conceptual density-functional theory (conceptual DFT) analysis\cite{book:Parr1989,article:Parr1995,article:Chermette1999,article:Geerlings2003,article:Ayers2005,article:Geerlings2020,book:Liu2022} of atomic properties in magnetic fields computed by current-density-functional theory (current-DFT)\cite{article:Vignale1987,article:Vignale1988,article:Tellgren2012,article:Tellgren2014,article:Tellgren2018}.
  To enable this study, the conceptual DFT framework was extended to explicitly account for external magnetic fields\cite{article:Geerlings2023,article:Franco-Perez2024}.
  This development builds upon earlier generalizations of conceptual DFT to include finite temperature\cite{article:Franco-Perez2015,article:Gazquez2019}, external electric fields\cite{article:Clarys2021,VanLommel2022}, mechanical forces\cite{article:Bettens2019,article:Bettens2020}, confinement\cite{article:Borgoo2009,bookchapter:Geerlings2021}, and pressure\cite{article:Eeckhoudt2022,article:Eeckhoudt2024}.
  The central idea is to augment the conventional conceptual DFT descriptors --- namely, response functions of the energy with respect to the number of electrons $N_{\mathrm{e}}$ and/or the external potential $v_{\mathrm{ext}}$  (Section~\ref{sec:conceptualdft}) --- with an additional variable characterizing the external magnetic field.
  This \textit{ansatz} is combined with recent electronic-structure methods\cite{article:Vignale1987,article:Vignale1988,article:Tellgren2012,article:Irons2017a,article:Tellgren2008,article:Lange2012,article:Stopkowicz2015,article:Hampe2017} to include arbitrary magnetic field strengths in a non-perturbative manner through the use of London atomic orbitals\cite{article:London1937}.

  To extend our investigation to the electronic structure of molecules in strong magnetic fields\cite{article:Irons2022,article:Wibowo-Teale2024}, we initially focused on \textit{electronegativity}\cite{article:Parr1978}, a first-order global (\textit{i.e.}, position-independent\cite{book:Parr1989,article:Geerlings2003}) descriptor.
  Its central role in characterizing the behavior and ultimate fate of atoms during molecule formation makes it a natural starting point.
  Accordingly, as a first step towards a systematic understanding of molecular chemistry in strong magnetic fields, we revisited the field dependence of atomic electronegativity (and its second-order analogue, \textit{hardness}\cite{article:Parr1983}), building on our earlier work\cite{article:Francotte2022}.
  Using Huheey's equation\cite{book:Huheey2006}, we showed that the magnetic-field-induced trends in electronegativity provide early insight into bond polarity changes in diatomic molecules, in some cases even predicting polarity reversals, as anticipated for the hydrogen halides\cite{article:Irons2022,article:Huynh2024}.
  This behavior follows straightforwardly from the crossing of the electronegativity curves of hydrogen and the halogens at high field strengths.

  These qualitative predictions from conceptual DFT were subsequently confirmed by current-DFT calculations on the hydrogen halides, which quantified the evolution and eventual reversal of the electric dipole moment in strong fields\cite{article:Huynh2024,article:Wibowo-Teale2024}.
  The calculations further revealed a pronounced dependence of the dipole moment variation on field orientation (parallel \textit{versus} perpendicular), which could be meaningfully interpreted in terms of the electron density, the simplest local (\textit{i.e.}, position-dependent) descriptor in conceptual DFT (Section~\ref{sec:conceptualdft}).
  An in-depth symmetry analysis based on the full magnetic symmetry group of the molecule-plus-field system, which takes into account time-reversal symmetry\cite{book:Birss1966,article:Bradley1968,article:Pelloni2011,book:Ceulemans2013,article:Pausch2021}, was also carried out to rationalize the fundamentally different roles that electric and magnetic fields play in molecular electronic structure\cite{article:Wibowo-Teale2024}.
  Extension to simple polyatomics like \ce{H2O} and \ce{NH3} revealed analogous behaviors\cite{article:Irons2022}.

  The current work differs in nature from earlier high-level calculations on diatomics, which primarily focused on geometric and energetic properties --- particularly the evolution of low-lying electronic states with increasing magnetic field strength and the resulting level crossings.
  One such extensive study was due to Lehtola \textit{et al.}\cite{article:Lehtola2019} whereby fully numerical calculations on low-lying electronic states of \ce{H2}, \ce{HeH+}, \ce{LiH}, \ce{BeH+}, \ce{BH}, and \ce{CH+} as a function of magnetic field strength were reported.
  Earlier contributions include the study of \ce{LiH} by Stopkowicz \textit{et al.}\cite{article:Stopkowicz2015}.
  Especially noteworthy is the 2012 work by Lange \textit{et al.}\cite{article:Lange2012} which revealed that a novel ``perpendicular paramagnetic bonding'' mechanism (see also Ref.~\citenum{article:Stopkowicz2017}) is responsible for the unexpected stabilization of the triplet state of \ce{H2} relative to the singlet state at magnetic fields on the order of \SI{e5}{\tesla}.
  This discovery opened the door to fundamentally new bonding concepts, and hence exciting ``new chemistries'', in strong magnetic fields.
  More recently, some of the present authors identified related features of this perpendicular paramagnetic bonding mechanism across a range of geometries in ground and excited states of \ce{H3+} and \ce{H3}\cite{article:Wibowo2023}.

  Returning to conceptual DFT, we recently extended our analysis of $\sigma$-bonded systems\cite{article:Irons2022} to $\pi$-bonded molecules\cite{article:Wibowo-Teale2024}.
  In that work, local conceptual DFT descriptors --- specifically the \textit{Fukui functions}\cite{article:Parr1984} --- were utilized to elucidate how an external magnetic field affects nucleophilic addition to carbonyl groups.
  Key insights emerged from a detailed symmetry analysis of real-valued electron densities and complex-valued molecular orbitals (MOs), automatically assigned with the aid of the symbolic symmetry framework in the recently developed \qsymsq{} program\cite{article:Huynh2024}.
  It is worth emphasizing that, in all studies to date, we have adopted an adiabatic approach: as a first approximation, the field-free electronic configuration is assumed to persist upon application of the magnetic field, with its strength increased gradually and the corresponding electronic state followed by the Maximum Overlap Method\cite{article:Gilbert2008,article:Barca2018}.
  This approach allowed us to understand exactly how the magnetic field influences the spatial distribution of electrons without having to worry about the additional complications arising from the uncoupling of electrons in strong magnetic fields.

  The present study constitutes a two-fold extension of our previous conceptual DFT investigations of molecules in strong magnetic fields.
  Firstly, we continue to analyze chemical reactivity within the conceptual DFT framework, but now focus on reactions in which \textit{chemical hardness} and its local inverse --- the \textit{local chemical softness} --- play a governing role.
  Secondly, we explicitly account for the magnetic-field dependence of the molecular electronic configuration when evaluating relevant molecular descriptors, effectively transitioning to higher spin states in a non-adiabatic manner as the magnetic field increases in strength.

  Chemical hardness is central in discussions of reactivity within the Hard and Soft Acids and Bases (HSAB) principle\cite{article:Pearson1963,article:Pearson1990,book:Pearson1997}.
  At the local level (\textit{i.e.}, when focusing on specific atoms or sites rather than the molecule as a while), the HSAB principle states that if a molecule contains both a hard side A and a soft side B, a soft reagent will preferably interact with site B, whereas a hard reagent will favor site A.
  In conceptual DFT, the local softness is defined as the product of the global softness and the Fukui function\cite{article:Yang1985}, thereby acting as a distributor of softness over the molecule.
  The HSAB principle, at both the global and local levels, has received broad support within the field of conceptual DFT\cite{article:Geerlings2000,article:Ayers2006,article:Torrent-Sucarrat2010}.

  We have previously shown that atomic hardness, like electronegativity, depends strongly on the magnetic field strength\cite{article:Francotte2022}.
  In several cases, pairs of certain atoms even exhibit hardness reversals in strong fields, prompting the question of whether a magnetic field can similarly modulate, or even invert, the hard and sort regions within a molecule.
  This parallels the behaviors observed for electronegativity and molecular electric dipole moment\cite{article:Irons2022}, although the conceptual-DFT descriptors of interest here are different.
  In fact, hardness and local softness now take center stage.
  However, as shall be detailed in Section~\ref{sec:local}, when comparing the local softness between different regions in the same molecule, it suffices to focus only on the Fukui functions since they determine the spatial distribution of local softness.
  This enables us to build directly upon the experience and insights gained from our recent study of magnetic-field effects on the chemical reactivity of $\pi$-systems in carbonyl groups\cite{article:Wibowo-Teale2024}.

  When examining possible changes in molecular electronic configuration, we consider first the electron Zeeman interaction, which constitutes the linear paramagnetic term in the electronic Hamiltonian (Section~\ref{sec:elecstruct-magfields}), and which stabilize unpaired electrons of the same spin.
  As a result, increasing magnetic field strength promotes the successive uncoupling of electron pairs, yielding lower-energy high-spin configurations that often exhibit different spatial symmetries.
  However, at larger fields, this effect may be counterbalanced by the quadratic diamagnetic term in the electronic Hamiltonian (Section~\ref{sec:elecstruct-magfields}).
  In any case, it is evident that not only the energy but all conceptual-DFT descriptors used in the present study (electron density, dipole moment, hardness, softness, and Fukui functions) will be influenced by the state-switching process.

  Although previous studies by Schmelcher \textit{et al.}\cite{article:Al-Hujaj2004,article:Ivanov1999,article:Ivanov1998,article:Ivanov2001,article:Ivanov2000} explored extremely large magnetic fields of up to $10^4\,B_0$ for atoms and ions, we limit the present investigation to fields up to $0.3\,B_0$.
  This range remains chemically relevant, corresponding to fields on the order of \SI{e4}{T}, and is sufficiently strong for electronic state crossings to be expected\cite{article:Lehtola2019}.
  It is also interesting to establish the maximum field strength up to which the adiabatic approach remains meaningful.
  Put differently, we seek to determine the field strength at which higher-spin states (\textit{i.e.}, states with higher $|M_S|$) become sufficiently stabilized to replace the zero-field ground state as the true ground state.

  The systems selected for this proof-of-principle study are two well-known examples of \textit{ambident nucleophiles}.
  Such species possess two distinct nucleophilic sites, one characterized as \textit{hard} and the other as \textit{soft}.
  Both systems have been extensively investigated experimentally and theoretically under zero-magnetic-field conditions.
  The first example is the thiocyanate anion, \ce{SCN-}, which is well-known to form both \textit{thiocyanato} (\ce{M-SCN}) and \textit{isothiocyanato} (\ce{M-NCS}) complexes\cite{article:Burmeister1964,book:Basolo1967}.
  Within the HSAB framework, the sulfur end is generally considered to be the soft site\cite{article:Pearson1963}, preferably interacting with soft electrophiles such as \ce{Ag+} according to the HSAB principle\cite{article:Pearson1963}.
  Conversely, the nitrogen end is normally regarded as the hard site and tends to coordinate with hard electrophiles such as \ce{Co^{3+}}\cite{article:Pearson1963}.
  These experimentally observed preferences have indeed been confirmed by theoretical studies\cite{book:Klopman1974}, often with the aid of conceptual DFT descriptors\cite{article:Lee1988,article:DeProft1996,article:Langenaeker1996,article:DeProft2002}.
  The second system considered in this work is the nitrite anion, \ce{(NO2)-}, whose ambident character is well-known to arise from its ability to coordinate to a metal center either through the nitrogen atom (\textit{nitro} form) or through one of the oxygen atoms (\textit{nitrito} form)\cite{article:Hitchman1982,article:Burmeister1990}.
  Even though the ambident character of \ce{(NO2)-} has been examined within conceptual DFT, its interpretation in terms of the HSAB principle, in order that a clear distinction can be made between the hard and soft sites, is much less straightforward\cite{article:DeProft1996,article:Langenaeker1996,article:DeProft2002}.
  As triatomics containing second- and third-row elements, \ce{SCN-} and \ce{(NO2)-} pose greater challenges for both \textit{ab initio} computations and the interpretation of results in the presence of a magnetic field than the simple diatomics studied previously (see above), with only a few notable exceptions such as \ce{He} and \ce{CH} clusters and molecular benzene\cite{article:Irons2021,article:Pemberton2022}.

  This article is structured as follows.
  In Section~\ref{sec:theory}, we briefly review the key aspects of current-DFT and conceptual DFT relevant to the present work (full details can be found in Refs.~\citenum{article:Francotte2022,article:Irons2022,article:Wibowo-Teale2024}).
  This is followed by a summary of the extended symmetry analysis of local descriptors, \textit{e.g.}, electron density and Fukui functions, in the presence of a magnetic field (see Refs.~\citenum{article:Irons2022,article:Wibowo2023,article:Wibowo-Teale2024} for details).
  Section~\ref{sec:compdetails} outlines the computational strategy for the current-DFT calculations, the conceptual DFT analysis, and the symmetry assignments carried out using \qsymsq{}\cite{article:Huynh2024}, with particular attention given to the selection of electronic configurations required to account for spin-state changes as the external magnetic field increases.
  The obtained results are presented and discussed in Section~\ref{sec:results}, where we analyze the magnetic-field dependence of the state energies, electric dipole moments, and Fukui functions.
  This section concludes with a brief exploration of the influence of an external magnetic field on the electrostatic component of reactivity, examined through the molecular electrostatic potential (MEP).
  Finally, Section~\ref{sec:conclusion} places the conceptual DFT insights obtained for the nucleophilic behavior of the two representative ambident nucleophiles in the broader context of our previous studies on magnetic-field-induced state switching in atoms\cite{article:Francotte2022}, the dipole-moment sensitivity of $\sigma$-bonded systems\cite{article:Irons2022,article:Huynh2024}, and the behavior of the Fukui function in $\pi$-systems under increasing magnetic field strength\cite{article:Wibowo-Teale2024}.

\section{Theory}
\label{sec:theory}

  \subsection{Electronic-Structure Theory in External Magnetic Fields}
  \label{sec:elecstruct-magfields}

    In this Section, we summarize the key aspects of electronic-structure theory that are pertinent to the calculations done in this work.
    Full details have been presented elsewhere\cite{article:Francotte2022,article:Irons2022,article:Wibowo-Teale2024}.

    \subsubsection{The Electronic Hamiltonian}
    \label{sec:elecHamil}

      Consider a molecular system containing $N_{\mathrm{e}}$ electrons and $N_{\mathrm{n}}$ nuclei placed in an external uniform static magnetic field $\mathbf{B}(\mathbf{r}) = \bm{\nabla} \times \mathbf{A}(\mathbf{r})$, where $\mathbf{A}(\mathbf{r})$ denotes a \textit{magnetic vector potential} chosen in such a way that the magnetic field is position-independent, \textit{i.e.}, $\mathbf{B}(\mathbf{r}) = \mathbf{B}$.
      The electronic Hamiltonian describing this molecular system is then given by
      \begin{equation}
        \hat{\mathscr{H}}(\mathbf{B}) =
        \hat{\mathscr{H}}_0 + \hat{\mathscr{H}}_{\mathrm{mag}}(\mathbf{B}).
        \label{eq:H}
      \end{equation}
    
      The first contribution to the electronic Hamiltonian in Equation~\ref{eq:H} is the \textit{zero-field Hamiltonian} $\hat{\mathscr{H}}_0$, which has the form
      \begin{equation*}
        \begin{gathered}
          \hat{\mathscr{H}}_0 =
          \sum_{i=1}^{N_{\mathrm{e}}}
          -\frac{1}{2}\nabla_i^2
          + \sum_{i=1}^{N_{\mathrm{e}}} \sum_{j>i}^{N_{\mathrm{e}}}
          \frac{1}{\lvert \mathbf{r}_i - \mathbf{r}_j \rvert}
          + \sum_{i=1}^{N_{\mathrm{e}}} v_{\mathrm{ext}}(\mathbf{r}_i)
        \end{gathered}
      \end{equation*}
      in atomic units, where $v_{\mathrm{ext}}(\mathbf{r})$ is the multiplicative external potential dictated by the geometric arrangement of the nuclei:
      \begin{equation}
        \begin{gathered}
          v_{\mathrm{ext}}(\mathbf{r}) = \sum_{A=1}^{N_{\mathrm{n}}}
          \frac{-Z_A}{\lvert \mathbf{r} - \mathbf{R}_A \rvert}.
        \end{gathered}
        \label{eq:vext}
      \end{equation}
      In the above expressions, $\mathbf{r}_i$ denotes the position vector of the $i$\textsuperscript{th} electron and $\mathbf{R}_A$ that of the $A$\textsuperscript{th} nucleus.
      We further assume that the center of mass of the nuclear framework,
      \begin{equation}
        \mathbf{O}_{\mathrm{n}} = \frac{%
          \sum_{A=1}^{N_{\mathrm{n}}} M_{A} \mathbf{R}_{A}%
        }{%
          \sum_{A=1}^{N_{\mathrm{n}}} M_{A}%
        },
        \label{eq:com}
      \end{equation}
      where $M_A$ is the mass of the $A$\textsuperscript{th} nucleus, coincides with the origin of the Cartesian coordinate system, \textit{i.e.} $\mathbf{O}_{\mathrm{n}} = \mathbf{0}$, even though this quantity does not appear explicitly anywhere in the Hamiltonian.
      We will see later (Section~\ref{sec:orbitsym}) that this choice helps simplify the analysis of symmetry in the presence of a magnetic field.
    
      The second contribution to the electronic Hamiltonian in Equation~\ref{eq:H} arises from the non-relativistic interaction of the electrons with the external magnetic field:
      \begin{equation}
        \hat{\mathscr{H}}_{\mathrm{mag}}(\mathbf{B}) =
        -i\sum_{i=1}^{N_{\mathrm{e}}} \mathbf{A}(\mathbf{r}_i) \cdot \bm{\nabla}_i
        + \frac{g_{\mathrm{s}}}{2} \sum_{i=1}^{N_{\mathrm{e}}} \mathbf{B} \cdot \hat{\mathbf{s}}_i
        + \frac{1}{2} \sum_{i=1}^{N_{\mathrm{e}}} A^2(\mathbf{r}_i),
        \label{eq:Hmag}
      \end{equation}
      where $\hat{\mathbf{s}}_i$ is the spin angular momentum operator for the $i$\textsuperscript{th} electron and $g_{\mathrm{s}}$ the electron spin $g$-factor.\cite{book:Weil2007, article:Tellgren2018}
      We note that the second contribution to $\hat{\mathscr{H}}_{\mathrm{mag}}$ in Equation~\ref{eq:Hmag} describes the \textit{spin-Zeeman effect} responsible for the switching of spin states that is encountered throughout this work.

    \subsubsection{Non-Perturbative Quantum Chemistry in Strong Magnetic Fields}
    \label{sec:quantumchem-mag}
      The magnetic fields concerned in this work ($\sim B_0$) comprise a region in which the electrostatic forces and direct magnetic effects are of similar orders of magnitude, thereby preventing a perturbative approach\cite{article:Austad2020}.
      A non-perturbative method must thus be relied upon instead.
      To this end, for a uniform magnetic field $\mathbf{B}$, we first \textit{choose} the following magnetic vector potential in the \textit{Coulomb gauge}:
      \begin{equation}
        \mathbf{A}_{\mathbf{O}_{\mathrm{mag}}}(\mathbf{r}) = \frac{1}{2}\mathbf{B} \times (\mathbf{r} - \mathbf{O}_{\mathrm{mag}})
        \label{eq:uniformBfield-vecpot-coulombgauge}
      \end{equation}
      where $\mathbf{O}_{\mathrm{mag}}$ is an arbitrarily chosen gauge origin.
      It can be shown\cite{article:Wibowo-Teale2024} that $\mathbf{A}_{\mathbf{O}_{\mathrm{mag}}}(\mathbf{r})$ is divergence-free and that $\mathbf{B}$ is independent of the gauge origin $\mathbf{O}_{\mathrm{mag}}$.

      To ensure that physical properties such as electron densities and electric dipole moments can be calculated in a gauge-origin-independent manner, we employ field-dependent London atomic-orbital (LAO) basis functions\cite{article:London1937} which have been shown to yield gauge-origin-invariant computational results for physical properties even with minimal numbers of atomic-orbital (AO) functions. \cite{article:London1937,article:Hameka1962,article:Ditchfield1972,article:Helgaker1991a}
      Each LAO $\omega_{\mu}(\mathbf{r}; \mathbf{R}_{\mu})$ centered at position $\mathbf{R}_{\mu}$ is a product of a conventional Gaussian AO $\varphi_{\mu}(\mathbf{r}; \mathbf{R}_{\mu})$ with the London phase factor $\exp[-i \mathbf{A}_{\mathbf{O}_{\mathrm{mag}}}(\mathbf{R}_{\mu}) \cdot \mathbf{r}]$:
      \begin{equation}
        \omega_{\mu}(\mathbf{r}; \mathbf{R}_{\mu})
        = \varphi_{\mu}(\mathbf{r}; \mathbf{R}_{\mu}) \exp[-i \mathbf{A}_{\mathbf{O}_{\mathrm{mag}}}(\mathbf{R}_{\mu}) \cdot \mathbf{r}].
        \label{eq:lao}
      \end{equation}
      The London phase factor takes both the uniform magnetic field $\mathbf{B}$ and the gauge origin $\mathbf{O}_{\mathrm{mag}}$ into account by Equation~\ref{eq:uniformBfield-vecpot-coulombgauge}.
      The gauge-origin invariance guaranteed by LAOs gives us the freedom to choose $\mathbf{O}_{\mathrm{mag}} = \mathbf{0}$ in this work without altering any of the calculated physical observables.
      This also lets us drop the $\mathbf{O}_{\mathrm{mag}}$ subscript in subsequent notations of magnetic vector potentials for the sake of brevity.
      Furthermore, efficient algorithms for evaluating molecular integrals over LAOs have been devised\cite{article:Honda1991,article:Tellgren2008,article:Irons2017a,article:Reynolds2015} to enable a wide range of \textit{ab initio} electronic-structure methods such as Hartree--Fock (HF),\cite{article:Tellgren2008} current-DFT,\cite{article:Vignale1987,article:Vignale1988,article:Tellgren2012,article:Tellgren2018} configuration interaction (CI),\cite{article:Lange2012} and coupled-cluster (CC)\cite{article:Stopkowicz2015} to be used for non-perturbative calculations in strong-magnetic-field regimes where $\lvert \mathbf{B} \rvert \sim B_0 = \hbarr e^{-1}a_0^{-2} \approx \SI{2.3505e5}{\tesla}$.

    \subsubsection{Current-Density-Functional Theory}
    \label{sec:currentdft}
      The interaction between a magnetic field and electrons gives rise to additional electronic effects such as spin polarization\cite{article:Rajagopal1973,article:Gunnarsson1976} and induced currents\cite{article:Rajagopal1973}.
      To account for these, the conventional formalism of density-functional theory (DFT)\cite{article:Hohenberg1964,article:Kohn1965,article:Lieb1983} must be extended to consider both the electron density $\rho(\mathbf{r})$:
      \begin{multline}
        \rho(\mathbf{r}) =
        N_{\mathrm{e}} \int
        \Psi(\mathbf{r}, s, \mathbf{x}_2, \ldots, \mathbf{x}_{N_{\mathrm{e}}})^*
        \Psi(\mathbf{r}, s, \mathbf{x}_2, \ldots, \mathbf{x}_{N_{\mathrm{e}}})
        \\ \D s\ \D\mathbf{x}_2 \ldots \D\mathbf{x}_{N_{\mathrm{e}}},
        \label{eq:rho}
      \end{multline}
      and the magnetization current density $\mathbf{j}_{\mathrm{m}}(\mathbf{r})$\cite{article:Tellgren2018}:
      \begin{equation*}
        \mathbf{j}_{\mathrm{m}}(\mathbf{r}) =
        \mathbf{j}_{\mathrm{p}}(\mathbf{r}) +  g_{\mathrm{s}} \bm{\nabla} \times \mathbf{m}(\mathbf{r}),
      \end{equation*}
      which combines the \textit{paramagnetic current density} describing currents induced by orbital effects:
      \begin{multline*}
        \mathbf{j}_{\mathrm{p}}(\mathbf{r}) =
        N_{\mathrm{e}} \Im \int
        \Psi(\mathbf{r}, s, \ldots, \mathbf{x}_{N_{\mathrm{e}}})^*
        \bm{\nabla}_{\mathbf{r}}
        \Psi(\mathbf{r}, s, \ldots, \mathbf{x}_{N_{\mathrm{e}}})
        \\ \D s\ \D\mathbf{x}_2 \ldots \D\mathbf{x}_{N_{\mathrm{e}}},
      \end{multline*}
      and the \textit{spin-current density} associated with the spin-Zeeman interaction [the second contribution in Equation~\ref{eq:Hmag}] where $\mathbf{m}(\mathbf{r})$ is the \textit{magnetization} of the system:
      \begin{multline*}
        \mathbf{m}(\mathbf{r}) = N_{\mathrm{e}} \int
        \Psi(\mathbf{r}, s, \ldots, \mathbf{x}_{N_{\mathrm{e}}})^*
        \hat{\mathbf{s}}_s
        \Psi(\mathbf{r}, s, \ldots, \mathbf{x}_{N_{\mathrm{e}}})
        \\ \D s\ \D\mathbf{x}_2 \ldots \D\mathbf{x}_{N_{\mathrm{e}}},
      \end{multline*}
      with $\hat{\mathbf{s}}_s$ acting only on the spin coordinate $s$.

      This formulation, first introduced by Vignale and Rasolt\cite{article:Vignale1987,article:Vignale1988} and later refined by Tellgren \textit{et al.}\cite{article:Tellgren2012,article:Tellgren2018}, takes $(\rho, \mathbf{j}_{\mathrm{m}})$ as the basic densities and $(u, \mathbf{A})$ as the basic external potentials, where
      \begin{equation*}
        u = v_{\mathrm{ext}} + \frac{1}{2}A^2
      \end{equation*}
      is the modified scalar potential.
      The ground-state energy then follows from the Legendre--Fenchel conjugation between the concave extrinsic energy functional
      \begin{multline}
        \mathcal{E}[u, \mathbf{A}] =
        \\ \inf_{\rho, \mathbf{j}_{\mathrm{m}}} \left\lbrace
        \mathcal{F}[\rho, \mathbf{j}_{\mathrm{m}}]
        + \int \rho(\mathbf{r}) u(\mathbf{r})\D\mathbf{r}
        + \int \mathbf{j}_{\mathrm{m}}(\mathbf{r}) \cdot \mathbf{A}(\mathbf{r})\D\mathbf{r}
        \right\rbrace
        \label{eq:Efunctional}
      \end{multline}
      and the convex intrinsic energy functional
      \begin{multline}
        \mathcal{F}[\rho, \mathbf{j}_{\mathrm{m}}] =
        \\ \sup_{u, \mathbf{A}} \left\lbrace
        \mathcal{E}[u, \mathbf{A}]
        - \int \rho(\mathbf{r}) u(\mathbf{r})\D\mathbf{r}
        - \int \mathbf{j}_{\mathrm{m}}(\mathbf{r}) \cdot \mathbf{A}(\mathbf{r})\D\mathbf{r}
        \right\rbrace,
      \end{multline}
      in direct analogy with conventional DFT\cite{article:Lieb1983}.

      Suppose we have an $N_{\mathrm{e}}$-electron fully interacting system with electron density $\rho$ and magnetization current density $\mathbf{j}_{\mathrm{m}}$.
      In the practical orbital-based Kohn--Sham formulation of current-DFT\cite{article:Vignale1987,article:Vignale1988}, we consider an auxiliary system containing $N_{\mathrm{e}}$ \textit{non-interacting} electrons with ground-state wavefunction $\Psi_0[\rho, \mathbf{j}_{\mathrm{m}}]$ yielding the same electron density $\rho$ and magnetization current density $\mathbf{j}_{\mathrm{m}}$.
      In this auxiliary system, the intrinsic energy functional $\mathcal{F}[\rho, \mathbf{j}_{\mathrm{m}}]$, whose closed form is unknown, can be decomposed into more manageable contributions:
      \begin{equation}
        \mathcal{F}[\rho, \mathbf{j}_{\mathrm{m}}] \equiv
        T_s[\rho, \mathbf{j}_{\mathrm{m}}]
        + J[\rho]
        + E_{\mathrm{xc}}[\rho, \mathbf{j}_{\mathrm{m}}].
        \label{eq:Fdecomposed}
      \end{equation}
      The first term,
      \begin{equation}
        T_s[\rho, \mathbf{j}_{\mathrm{m}}] \equiv \braket{
          \Psi_0[\rho, \mathbf{j}_{\mathrm{m}}] | \hat{T} | \Psi_0[\rho, \mathbf{j}_{\mathrm{m}}]
        }, \quad
        \hat{T} = \sum_{i=1}^{N_{\mathrm{e}}} -\frac{1}{2}\nabla_i^2,
        \label{eq:Ts}
      \end{equation}
      is the \textit{non-interacting kinetic energy}.
      The second term,
      \begin{equation*}
        J[\rho] =
        \frac{1}{2} \int \frac{%
          \rho(\mathbf{r})\rho(\mathbf{r}')%
        }{%
          \lvert \mathbf{r} - \mathbf{r}' \rvert%
        } \ \D\mathbf{r}\D\mathbf{r}',
      \end{equation*}
      is the usual Coulomb interaction energy.
      And the last term, $E_{\mathrm{xc}}[\rho, \mathbf{j}_{\mathrm{m}}]$, is the unknown \textit{exchange-correlation energy} that must be approximated.

      By definition, the non-interacting ground-state wavefunction $\Psi_0[\rho, \mathbf{j}_{\mathrm{m}}]$ in Equation~\ref{eq:Ts} is a single Slater determinant:
      \begin{equation}
        \Psi_0[\rho, \mathbf{j}_{\mathrm{m}}](\mathbf{x}_1, \ldots, \mathbf{x}_{N_{\mathrm{e}}}) =
        \sqrt{N_{\mathrm{e}}!} \hat{\mathscr{A}} \left[\prod_{i=1}^{N_{\mathrm{e}}} \psi_i(\mathbf{x}_i) \right],
        \label{eq:slaterdet}
      \end{equation}
      where $\hat{\mathscr{A}}$ is the antisymmetrizer acting on the composite spatial--spin coordinates $\mathbf{x}_i$ in terms of which the spin-orbitals $\psi_i$ are written.
      For a particular choice of approximation of $E_{\mathrm{xc}}[\rho, \mathbf{j}_{\mathrm{m}}]$, the optimization of the extrinsic energy functional in Equation~\ref{eq:Efunctional} in the auxiliary non-interacting system (\textit{i.e.}, the wavefunction must assume the single-determinantal form in Equation~\ref{eq:slaterdet}) with respect to variations in the spin-orbitals $\psi_i$ (subject to orthonormality constraints) yields a set of $N_{\mathrm{e}}$ eigenvalue equations to be solved self-consistently for $\psi_i$:
      \begin{equation*}
        \hat{f} \psi_i(\mathbf{x}) = \epsilon_i \psi_i(\mathbf{x}), \qquad
        i = 1, \ldots, N_{\mathrm{e}},
      \end{equation*}
      where
      \begin{multline}
        \hat{f} =
        \frac{1}{2} \left(-i\bm{\nabla} + \mathbf{A}_s \right)^2
        + v_{J} + v_{\mathrm{ext}} + v_{\mathrm{xc}}
        \\ + \frac{1}{2} (A^2 - A_s^2)
        + g_{\mathrm{s}} (\bm{\nabla} \times \mathbf{A}_s) \cdot \hat{\mathbf{s}}
        \label{eq:fockian}
      \end{multline}
      is the one-electron Kohn--Sham-like operator.
      In Equation~\ref{eq:fockian}, $v_J = \int \rho(\mathbf{r}') \lvert \mathbf{r} - \mathbf{r}' \rvert^{-1} \D \mathbf{r}'$ is the well-known Hartree potential, $v_{\mathrm{xc}} = \delta E_{\mathrm{xc}}[\rho, \mathbf{j}_{\mathrm{m}}] / \delta \rho$ the exchange-correlation scalar potential, $\mathbf{A}_s = \mathbf{A} + \mathbf{A}_{\mathrm{xc}}$ the effective vector potential, and $\mathbf{A}_{\mathrm{xc}} = \delta E_{\mathrm{xc}}[\rho, \mathbf{j}_{\mathrm{m}}] / \delta \mathbf{j}_{\mathrm{m}}$ the exchange-correlation vector potential.

    \subsubsection{Current-Density-Functional Approximations}
    \label{sec:currentdfa}
      To ensure accurate and meaningful calculations, the unknown exchange-correlation energy $E_{\mathrm{xc}}[\rho, \mathbf{j}_{\mathrm{m}}]$ above must be approximated in an appropriate manner.
      However, in practice, constructing approximations for $E_{\mathrm{xc}}$ as functionals of the magnetization current density $\mathbf{j}_{\mathrm{m}}$ (and also the electron density $\rho$) is difficult,\cite{article:Capelle1997,article:Tellgren2012} and so the spin-resolved formulation due to Vignale and Rasolt\cite{article:Vignale1988}, using only $\mathbf{j}_{\mathrm{p}}$, shall be used instead.
      In this work, we shall utilize the explicit current dependence at the meta-generalized-gradient-approximation (meta-GGA) level via a modified kinetic energy density\cite{article:Dobson1993,article:Becke1996,article:Bates2012,article:Furness2015},
      \begin{equation}
        \tilde{\tau}(\mathbf{r}) =
        \frac{1}{2}
          \sum_{\sigma} \sum_{i = 1}^{N_{\sigma}}
          \left\lvert
            \nabla \psi^\sigma_{i}(\mathbf{r})
          \right\rvert^2
        - \frac{%
            \lvert\mathbf{j}_{\mathrm{p}}(\mathbf{r})\rvert^2
          }{%
            2\rho(\mathbf{r})
          },
        \label{eq:tautilde}
      \end{equation}
      where $ \psi^\sigma_{i}(\mathbf{r})$ are the Kohn--Sham orbitals with spin $\sigma$ and $N_{\sigma}$ the number of electrons with spin $\sigma$, to ensure that the exchange-correlation energy remains properly gauge-independent in a magnetic field.
      In particular, we employ cTPSS, the current-DFT analog of the Tao--Perdew--Staroverov--Scuseria functional\cite{article:Tao2003,article:Sagvolden2013}.
      We also use r\textsuperscript{2}SCAN0, the global hybrid form which incorporates 25\% HF exchange into the regularized strongly constrained and appropriately normed semi-local density functional\cite{article:Sun2015a,article:Furness2020} and which has been shown to improve the description of molecular properties\cite{article:Bursch2022}.

  \subsection{Conceptual Density-Functional Theory}
  \label{sec:conceptualdft}
    In conceptual DFT, chemical reactivity is described through response functions of the electronic energy $E$ with respect to variations in the number of electrons $N_{\mathrm{e}}$ and the external potential $v_{\mathrm{ext}}(\mathbf{r})$ (Equation~\ref{eq:vext})\cite{book:Parr1989,article:Parr1995,article:Chermette1999,article:Geerlings2003,article:Ayers2005,article:Geerlings2020,book:Liu2022}.
    These quantities arise naturally from the perturbation expansion of the energy functional $E[N_{\mathrm{e}}; v_{\mathrm{ext}}]$, truncated here at second order:
    \begin{subequations}
      \begin{equation}
        \Delta E \approx \Delta E^{(1)} + \Delta E^{(2)}
      \end{equation}
      where
      \begin{multline}
        \Delta E^{(1)} =
          \underbrace{%
            \left(
              \frac{\partial E}{\partial N_{\mathrm{e}}}
            \right)_{v_{\mathrm{ext}}}
          }_{\mu:\ \textrm{chemical potential}}
          \Delta N_{\mathrm{e}}
          \\ + \int \underbrace{%
            \left[
              \frac{\delta E}{\delta v_{\mathrm{ext}}(\mathbf{r})}
            \right]_{N_{\mathrm{e}}}
          }_{\rho(\mathbf{r}):\ \textrm{electron density}}
          \Delta v_{\mathrm{ext}}(\mathbf{r}) \D \mathbf{r}
      \end{multline}
      and
      \begin{multline}
        \Delta E^{(2)} =
            \underbrace{%
              \left(
                \frac{\partial^2 E}{\partial N_{\mathrm{e}}^2}
              \right)_{v_{\mathrm{ext}}}
            }_{\eta:\ \textrm{chemical hardness}}
            \Delta N_{\mathrm{e}}^2 \\[6pt]
          + \int
            \underbrace{%
              \left(
                \frac{\partial}{\partial N_{\mathrm{e}}}
              \right)_{v_{\mathrm{ext}}}
              \left[
                \frac{\delta E}{\delta v_{\mathrm{ext}}(\mathbf{r})}
              \right]_{N_{\mathrm{e}}}
            }_{f(\mathbf{r}):\ \textrm{Fukui function}}
            \Delta v_{\mathrm{ext}}(\mathbf{r})
            \D \mathbf{r}
            \ \Delta N_{\mathrm{e}} \\[6pt]
          + \int
            \underbrace{
              \left[
                \frac{\delta^2 E}{%
                  \delta v_{\mathrm{ext}}(\mathbf{r})
                  \delta v_{\mathrm{ext}}(\mathbf{r}')%
                }
              \right]_{N_{\mathrm{e}}}
            }_{\chi(\mathbf{r}, \mathbf{r}'):\ \textrm{linear response function}}
            \Delta v_{\mathrm{ext}}(\mathbf{r}) \Delta v_{\mathrm{ext}}(\mathbf{r}') \D \mathbf{r} \D \mathbf{r}'.
      \end{multline}%
      \label{eq:deltaEexpansion}%
    \end{subequations}
    Clearly, the expansion coefficients in Equation~\ref{eq:deltaEexpansion} govern the system's response to perturbations and are given by mixed functional and partial derivatives of $E$ with respect to $N_{\mathrm{e}}$ and/or $v_{\mathrm{ext}}$.
    These \textit{response functions} characterize the \textit{intrinsic} reactivity of the system.
    The same perturbation framework has also been extended to take into account external influences such as electric and magnetic fields, mechanical forces, confinement, and pressure\cite{article:Clarys2021,article:Bettens2019,article:Bettens2020,article:Alonso2024,article:Borgoo2009,booksection:Geerlings2021,article:Eeckhoudt2022,article:Geerlings2023}.

    In our first study incorporating external magnetic fields into conceptual DFT for atoms\cite{article:Francotte2022}, we focused on the first- and second-order responses of $E$ with respect to $N_{\mathrm{e}}$ at a constant $v_{\mathrm{ext}}$.
    The first derivative defines the \textit{chemical potential},\cite{article:Parr1978}
    \begin{equation*}
      \mu \equiv \left(
      \frac{\partial E}{\partial N_{\mathrm{e}}}
      \right)_{v_{\mathrm{ext}}},
    \end{equation*}
    which corresponds to the negative of the electronegativity $\chi$ in the Iczkowski--Margrave definition\cite{article:Iczkowski1961}, and reduces, in finite-difference form, to Mulliken's expression:\cite{article:Mulliken1934}
    \begin{equation*}
      \mu \xrightarrow{\textrm{f.d.}} \frac{1}{2}(I + A) = -\chi,
    \end{equation*}
    where $I$ and $A$ denote the first ionization potential and the electron affinity, respectively.
    The second derivative,
    \begin{equation*}
      \eta \equiv \left(
      \frac{\partial^2 E}{\partial N_{\mathrm{e}}^2}
      \right)_{v_{\mathrm{ext}}},
    \end{equation*}
    corresponds to Pearson's \textit{chemical hardness}\cite{article:Parr1983,book:Pearson2005} which in finite-difference form becomes
    \begin{equation*}
      \eta \xrightarrow{\textrm{f.d.}} \frac{1}{2}(I - A).
    \end{equation*}
    Both quantities are \textit{global} descriptors, \textit{i.e.}, independent of position.

    In the follow-up studies where we shifted our attention from atoms to molecules\cite{article:Irons2022,article:Wibowo-Teale2024}, we made use of two \textit{local} descriptors: the \textit{electron density} $\rho(\mathbf{r})$\cite{book:Parr1989,article:Geerlings2003},
    \begin{equation*}
      \rho(\mathbf{r}) = \left[
      \frac{\delta E}{\delta v_{\mathrm{ext}}(\mathbf{r})}
      \right]_{N_{\mathrm{e}}},
    \end{equation*}
    and its $N_{\mathrm{e}}$-derivative at constant external potential, the \textit{Fukui function}\cite{article:Parr1984},
    \begin{subequations}
      \begin{equation}
        f(\mathbf{r})
        = \left(
        \frac{\partial}{\partial N_{\mathrm{e}}}
        \right)_{v_{\mathrm{ext}}}
        \left[
        \frac{\delta E}{\delta v_{\mathrm{ext}}(\mathbf{r})}
        \right]_{N_{\mathrm{e}}}
        = \left[
        \frac{\partial \rho(\mathbf{r})}{\partial N_{\mathrm{e}}}
        \right]_{v_{\mathrm{ext}}},
        \label{eq:fukuitwosided}
      \end{equation}
    which describes the change in the electron density at a given point in space upon a perturbation in the total number of electrons in the system.
    However, the piece-wise linear behavior of the $E$ versus $N_{\mathrm{e}}$ curve\cite{article:Perdew1982} precludes the existence of the $\partial / \partial N_{\mathrm{e}}$ derivative in Equation~\ref{eq:fukuitwosided}.
    Instead, the Fukui functions must be defined as one-sided derivatives:
      \begin{equation}
        f^+(\mathbf{r}) \equiv \left[
        \frac{\partial \rho(\mathbf{r})}{\partial N_{\mathrm{e}}}
        \right]^+_{v_{\mathrm{ext}}},\qquad
        f^-(\mathbf{r}) \equiv \left[
        \frac{\partial \rho(\mathbf{r})}{\partial N_{\mathrm{e}}}
        \right]^-_{v_{\mathrm{ext}}},
        \label{eq:fukuionesided}
      \end{equation}
      \label{eq:fukui}
    \end{subequations}
    where $f^+(\mathbf{r})$ and $f^-(\mathbf{r})$ describe how the density responds upon electron addition or removal, respectively.
    The Fukui functions generalize the frontier MOs which play a vital role in Fukui's reactivity theory.\cite{article:Fukui1952,article:Yang1984,article:Yang2012}
    When multiplied by the \textit{global softness} $S = \eta^{-1}$, they yield the \textit{local softness functions} $s^+(\mathbf{r})$ and $s^-(\mathbf{r})$, which measure the softness with respect to electron addition or removal at a given point, respectively.
    The remaining second-order derivative in Equation~\ref{eq:deltaEexpansion}, $[\delta^2 E / \delta v_{\mathrm{ext}}(\mathbf{r})\delta v_{\mathrm{ext}}(\mathbf{r}')]_{N_{\mathrm{e}}}$, defined as the \textit{linear response function} $\chi(\mathbf{r}, \mathbf{r}')$,\cite{book:Parr1989,article:Geerlings2014} is non-local and therefore not considered in the present study.
    However, it may open up new avenues for future investigations in view of the recent interest in its chemical content.\cite{bookchapter:Geerlings2022,article:Geerlings2023a,article:Wang2023}

    An external uniform magnetic field $\mathbf{B}$ can be incorporated in the evaluation of the Fukui functions straightforwardly via a finite-difference approximation similarly to our previous approach for electronegativity and hardness.\cite{article:Francotte2022}
    In the particular case of $f^-(\mathbf{r})$, which is especially relevant for the study of ambident nucleophiles, subtracting the density for the $(N_{\mathrm{e}} - 1)$-electron system, $\rho_{N_{\mathrm{e}} - 1}(\mathbf{r}; \mathbf{B})$, from the corresponding density of the $N_{\mathrm{e}}$-electron system, $\rho_{N_{\mathrm{e}}}(\mathbf{r}; \mathbf{B})$, yields the following working equation:
    \begin{equation}
      f^-(\mathbf{r}; \mathbf{B}) \xrightarrow{\textrm{f.d.}}
      \rho_{N_{\mathrm{e}}}(\mathbf{r}; \mathbf{B})
      - \rho_{N_{\mathrm{e}} - 1}(\mathbf{r}; \mathbf{B}).
      \label{eq:fm-fd}
    \end{equation}
    The evaluation of the required densities, and hence the Fukui functions, in the presence of an external magnetic field $\mathbf{B}$ for both \ce{(NO2)-} and \ce{SCN-} is performed under the current-DFT framework presented in Section~\ref{sec:currentdft} (see also Section~\ref{sec:compdetails}).

    Finally, to probe the electrostatic component of chemical reactivity\cite{book:Murray1996,article:Geerlings2000,article:Ayers2007} of ambident nucleophiles and its variation with magnetic field, the electron density $\rho_{N_{\mathrm{e}}}(\mathbf{r}; \mathbf{B})$ can be used to compute the \textit{molecular electrostatic potential (MEP)}\cite{article:Bonaccorsi1970} at various field strengths:
    \begin{equation}
      V_{\textrm{MEP}}(\mathbf{r}; \mathbf{B}) =
        \sum_{A=1}^{N_{\mathrm{n}}} \frac{Z_A}{|\mathbf{r} - \mathbf{R}_A|}
        - \int \frac{\rho_{N_{\mathrm{e}}}(\mathbf{r}'; \mathbf{B})}{|\mathbf{r} - \mathbf{r}'|} \D\mathbf{r}',
      \label{eq:mep}
    \end{equation}
    as will be discussed in Section~\ref{sec:charge-orbital-control}.
    The value of the MEP at a position $\mathbf{r}$ gives the interaction energy between the charge distribution of a molecule and a unit positive charge located at $\mathbf{r}$, neglecting all geometrical and electronic rearrangements induced by the interaction.

  \subsection{Symmetry Analysis in External Magnetic Fields}
  \label{sec:symanalysis}

    In this Section, we describe the symmetry concepts that are utilized extensively to classify the electronic states of \ce{(NO2)-} and \ce{SCN-} in an external magnetic field.

    \subsubsection{Unitary and Magnetic Symmetry Groups}
    \label{sec:symgroups}

      The electronic Hamiltonian in Equation~\ref{eq:H} determines the \textit{unitary symmetry group} $\mathcal{G}$ of the system as the group consisting of all \textit{unitary symmetry operations} $\hat{u}$ of the system\cite{article:Irons2022,article:Wibowo-Teale2024} where $\hat{u} \hat{\mathscr{H}} \hat{u}^{-1} = \hat{\mathscr{H}}$.
      Owing to the additive form of the Hamiltonian in Equation~\ref{eq:H}, $\mathcal{G}$ is the intersection of $\mathcal{G}_0$ and $\mathcal{G}_{\mathrm{mag}}$, which are the unitary symmetry groups of $\hat{\mathscr{H}}_0$ and $\hat{\mathscr{H}}_{\mathrm{mag}}$, respectively.
      In this work, these symmetries are further restricted to be \textit{point transformations} acting on the \textit{configuration space} in which physical systems such as atoms, molecules, and fields are described\cite{book:Altmann1986}.

      When magnetic effects are included,\cite{article:Dimmock1962,article:Bradley1968,article:Lazzeretti1984,article:Keith1993,article:Pelloni2011} antiunitary symmetry operations $\hat{a}$ may also be present, where $\hat{a} \hat{\mathscr{H}} \hat{a}^{-1} = \hat{\mathscr{H}}$.
      In this case, the full symmetry of the system is described by the \textit{magnetic symmetry group} $\mathcal{M}$, which extends $\mathcal{G}$ by including all antiunitary symmetry operations that are not present in $\mathcal{G}$.
      In fact, $\mathcal{G}$ is a normal subgroup of index $2$ in $\mathcal{M}$, allowing the decomposition
      \begin{equation}
        \mathcal{M} = \mathcal{G} + \hat{a}_0\mathcal{G},
        \label{eq:maggroup}
      \end{equation}
      where $\hat{a}_0$ can be any of the antiunitary elements in $\mathcal{M}$ but once chosen must be fixed.\cite{article:Bradley1968}
      The left coset $\hat{a}_0\mathcal{G}$ contains all antiunitary elements of $\mathcal{M}$, and $\mathcal{G}$ is called the \textit{unitary halving subgroup} of $\mathcal{M}$.

      One key antiunitary operation in the symmetry characterization of systems in the presence of magnetic fields is time reversal, $\hat{\theta}$.
      time reversal classifies magnetic symmetry groups into \textit{gray groups} which contain $\hat{\theta}$:
      \begin{equation}
        \mathcal{M} = \mathcal{G} + \hat{\theta}\mathcal{G} \equiv \mathcal{G}',
        \label{eq:greygroup}
      \end{equation}
      and \textit{black-and-white groups} which do not:
      \begin{equation}
        \mathcal{M} = \mathcal{G} + \hat{\theta}\hat{u}'\mathcal{G},
        \label{eq:bwgroup}
      \end{equation}
      where $\hat{u}'$ is a unitary operation \textit{not} in $\mathcal{G}$\cite{article:Cracknell1965,article:Cracknell1966,article:Bradley1968}.
      In the absence of an external magnetic field, $\hat{\theta}$ is a symmetry and the system's magnetic symmetry group must be a gray group.
      In contrast, when an external magnetic field is applied, $\hat{\theta}$ ceases to be a symmetry because of the time-odd nature of the magnetic field vector\cite{book:Birss1966,article:Bradley1968,article:Irons2022}.
      Therefore, if the system possesses any antiunitary symmetry operations at all, its magnetic symmetry group must be a black-and-white group; otherwise, it only has a unitary symmetry group.

      For any magnetic group $\mathcal{M}$, one can consider an isomorphic \textit{unitary} group $\mathcal{M}'$.
      If $\mathcal{M}'$ can be identified with a subgroup of the full rotation-inversion group in three dimensions $\mathsf{O}(3)$, it may be labeled using a Sch\"onflies symbol and the magnetic group $\mathcal{M}$ may be written as $\mathcal{M}'(\mathcal{G})$.\cite{article:Cracknell1966,article:Pelloni2011}
      Otherwise, $\mathcal{M}$ is uniquely specified using its coset form with respect to the unitary halving subgroup $\mathcal{G}$ and a representative antiunitary operation $\hat{a}_0$, as done in Equations~\ref{eq:maggroup}--\ref{eq:bwgroup}.

    \subsubsection{Orbit-Based Symmetry Analysis Framework}
      \label{sec:orbitsym}
      To characterize the symmetry transformation under a symmetry group $\mathcal{H}$ of a quantity $\mathbf{w}$ belonging to a complex linear space $V$, we consider the subspace $W \subseteq V$ spanned by the \textit{orbit}\footnote{This is a group-theoretic concept describing a set of symmetry-related objects that must not be confused with \textit{orbitals}, which are one-electron wavefunctions.} \textit{of $\mathbf{w}$ generated by $\mathcal{H}$}:
      \begin{equation}
        \mathcal{H} \cdot \mathbf{w} = \{
        \hat{h}_i \mathbf{w} \mid \hat{h}_i \in \mathcal{H}
        \},
        \label{eq:orbit}
      \end{equation}
      and decompose $W$ as
      \begin{equation}
        W = \bigoplus_i \Gamma_i^{\otimes k_i},
        \label{eq:Wdecomposed}
      \end{equation}
      where $\Gamma_i$ are known irreducible representations or corepresentations\cite{book:Wigner1959,article:Dimmock1963,article:Cracknell1965,article:Bradley1968,article:Newmarch1981} of $\mathcal{H}$, depending on whether $\mathcal{H}$ is a unitary or magnetic symmetry group, respectively, and $k_i$ their multiplicities.
      The mathematical details of this procedure are described in Section~2.4 of Ref.~\citenum{article:Huynh2024}.
      If $\mathcal{H} = \mathcal{G}$, then the decomposition in Equation~\ref{eq:Wdecomposed} is called the \textit{unitary symmetry} of $\mathbf{w}$, and if $\mathcal{H} = \mathcal{M}$, it is instead called the \textit{magnetic symmetry} of $\mathbf{w}$.

      For computational convenience, we place both the center of mass of the nuclear framework (Equation~\ref{eq:com}) and the gauge origin of the magnetic vector potential (Equation~\ref{eq:uniformBfield-vecpot-coulombgauge}) at the origin of the Cartesian coordinate system: $\mathbf{O}_{\mathrm{n}} = \mathbf{O}_{\mathrm{mag}} = \mathbf{0}$.
      This is so that when point symmetry transformations are considered, these origins remain fixed, thereby eliminating the need to account for their transformations.
      The invariance of physical quantities with respect to these origins ensures that this particular choice does not alter the results and conclusions of our work in any way.

    \subsubsection{Magnetic Symmetry}
      \label{sec:magsym}
      \paragraph{Corepresentation theory for real and complex quantities.}
        Given a magnetic group $\mathcal{M}$, Wigner's corepresentation theory\cite{book:Wigner1959,article:Dimmock1963,article:Bradley1968} and its corresponding character theory\cite{article:Newmarch1981,article:Newmarch1983} state that every irreducible corepresentation of $\mathcal{M}$ (Equation~\ref{eq:maggroup}) must be uniquely induced by one or two irreducible representations of its unitary halving subgroup $\mathcal{G}$ in one of three ways:
        \begin{enumerate}[label=(\roman*)]
          \item $D[\Delta]$ is an \textit{irreducible corepresentation of the first kind} of $\mathcal{M}$ that is induced \emph{once} by the irreducible representation $\Delta$ of $\mathcal{G}$, and so $\dim D[\Delta] = \dim \Delta$.
          \item $D[2\Delta]$ is an \textit{irreducible corepresentation of the second kind} of $\mathcal{M}$ that is induced \emph{twice} by the irreducible representation $\Delta$ of $\mathcal{G}$, and so $\dim D[2\Delta] = 2\dim \Delta$.
          \item $D[\Delta_1 \oplus \Delta_2]$ is an \textit{irreducible corepresentation of the third kind} of $\mathcal{M}$ that is induced by two \textit{inequivalent} irreducible representations $\Delta_1$ and $\Delta_2$ of $\mathcal{G}$, and so $\dim D[\Delta_1 \oplus \Delta_2] = \dim \Delta_1 + \dim \Delta_2$.
        \end{enumerate}
        The character table of $\mathcal{M}$ can therefore be constructed entirely from that of $\mathcal{G}$ and includes only its unitary elements, since characters of antiunitary elements are not invariant under general unitary transformations of basis (unless the unitary transformations are also real\cite{article:Bradley1968,article:Newmarch1981}), and thus cannot be tabulated.
        However, the character tables of $\mathcal{M}$ and $\mathcal{G}$ are not necessarily identical due to differences in their conjugacy class structures\cite{article:Newmarch1981,article:Newmarch1983}.

      \begin{figure}
        \centering
        \includegraphics[width=.75\linewidth]{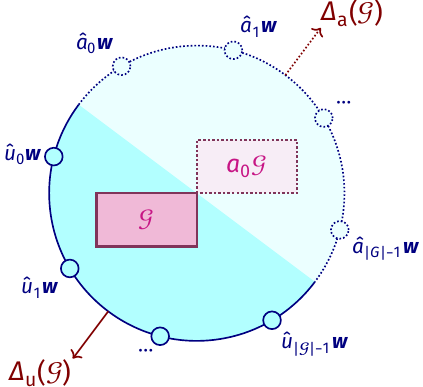}
        \caption{%
          Orbit of $\mathbf{w}$ generated by a magnetic group $\mathcal{M} = \mathcal{G} + a_0\mathcal{G}$.
          This orbit can be split into a unitary half $\mathcal{G} \cdot \mathbf{w}$ (solid arc) and an antiunitary half $a_0\mathcal{G} \cdot \mathbf{w}$ (dotted arc).
          The unitary elements $u_0, u_1, \ldots, u_{|\mathcal{G}|-1}$ belong to $\mathcal{G}$, and the antiunitary elements $a_i$ are given by $a_0 u_i$.
          The unitary half $\mathcal{G} \cdot \mathbf{w}$ spans a representation $\Delta_{\mathrm{u}}$ of $\mathcal{G}$, and the antiunitary half $a_0\mathcal{G} \cdot \mathbf{w}$ spans a representation $\Delta_{\mathrm{a}}$ of $\mathcal{G}$.
          The relationship between $\Delta_{\mathrm{u}}$ and $\Delta_{\mathrm{a}}$ determines the corepresentation of $\mathcal{M}$ spanned by the full orbit $\mathcal{M} \cdot \mathbf{w}$.
        }
        \label{fig:mag-orbit}
      \end{figure}

        In any case, for a real or complex quantity $\mathbf{w}$, the multiplicities $k_i$ in Equation~\ref{eq:Wdecomposed} can be determined using only the characters of $W$ under the operations in $\mathcal{G}$ as per Corollaries 1 and 2 of Theorem 10 in Ref.~\citenum{article:Newmarch1981} --- this is in fact implemented in \qsymsq{}.\cite{article:Huynh2024}.
        An intuition for this method can be gained by a consideration of Figure~\ref{fig:mag-orbit} which represents pictorially the orbit of $\mathbf{w}$ generated by a magnetic group $\mathcal{M}$.
        This orbit comprises a unitary half $\mathcal{G} \cdot \mathbf{w}$ and an antiunitary half $a_0\mathcal{G} \cdot \mathbf{w}$.
        Suppose that the former spans a (reducible or irreducible) representation $\Delta_{\mathrm{u}}$ of the unitary halving subgroup $\mathcal{G}$, and the latter spans a (reducible or irreducible) representation $\Delta_{\mathrm{a}}$ of $\mathcal{G}$.
        Suppose further that both $\Delta_{\mathrm{u}}$ and $\Delta_{\mathrm{a}}$ are irreducible --- the more general cases where $\Delta_{\mathrm{u}}$ and $\Delta_{\mathrm{a}}$ are reducible are straightforward generalization of this.
        The three kinds of irreducible corepresentations of $\mathcal{M}$ then depend on the relationship between $\Delta_{\mathrm{u}}$ and $\Delta_{\mathrm{a}}$\cite{article:Bradley1968,book:Wigner1959}.
        \begin{enumerate}[label=(\roman*)]
          \item An irreducible corepresentation of the first kind $D[\Delta]$ occurs when $\Delta_{\mathrm{u}}$ is equivalent to $\Delta_{\mathrm{a}}$ (so both can be denoted as $\Delta$) such that there exists a unitary matrix $\mathbf{P}$ satisfying $\mathbf{\Delta}_{\mathrm{a}}(u) = \mathbf{P}^{-1} \mathbf{\Delta}_{\mathrm{u}}(u) \mathbf{P}$ and $\mathbf{P}\mathbf{P}^* = +\mathbf{\Delta}_{\mathrm{u}}(a_0^2)$.
          Physically, this means that the antiunitary operations of $\mathcal{M}$ do not cause any doubling of the degeneracy that would be obtained if only $\mathcal{G}$ were considered.

          \item An irreducible corepresentation of the second kind $D[2\Delta]$ occurs when $\Delta_{\mathrm{u}}$ is also equivalent to $\Delta_{\mathrm{a}}$ (so both can be denoted as $\Delta$) such that there exists a unitary matrix $\mathbf{P}$ satisfying $\mathbf{\Delta}_{\mathrm{a}}(u) = \mathbf{P}^{-1} \mathbf{\Delta}_{\mathrm{u}}(u) \mathbf{P}$, but now $\mathbf{P}\mathbf{P}^* = -\mathbf{\Delta}_{\mathrm{u}}(a_0^2)$.
          Physically, this means that even though the unitary half $\mathcal{G} \cdot \mathbf{w}$ and the antiunitary half $a_0\mathcal{G} \cdot \mathbf{w}$ span equivalent irreducible representations of $\mathcal{G}$, they are linearly independent of each other.
          The antiunitary operations of $\mathcal{M}$ thus cause a doubling of the degeneracy that would be obtained if only $\mathcal{G}$ were considered.
          In fact, if $\mathcal{M}$ is a magnetic gray group where $a_0$ can be chosen as $\theta$, this doubling of degeneracy, attributable solely to time-reversal symmetry, is an instance of what is known as \textit{Kramers' degeneracy}\cite{article:Bradley1968}.

          \item An irreducible corepresentation of the third kind $D[\Delta_{\mathrm{u}} \oplus \Delta_{\mathrm{a}}]$ occurs when $\Delta_{\mathrm{u}}$ is inequivalent to $\Delta_{\mathrm{a}}$.
          Physically, this means that the unitary half $\mathcal{G} \cdot \mathbf{w}$ and the antiunitary half $a_0\mathcal{G} \cdot \mathbf{w}$ span different irreducible representations of $\mathcal{G}$.
          The antiunitary operations of $\mathcal{M}$ therefore introduce additional symmetries on top of those that would be obtained if only $\mathcal{G}$ were considered.
        \end{enumerate}

        When a magnetic field is introduced, MOs generally become complex-valued.
        If a magnetic symmetry group $\mathcal{M}$ exists, corepresentation theory is used to fully classify their symmetry under $\mathcal{M}$.
        In the rare cases where the molecule-plus-field system admits only a unitary symmetry group $\mathcal{G}$, conventional representation theory will be used instead to classify their symmetry under $\mathcal{G}$.
        Although the magnetic symmetry groups encountered in this work involve only irreducible corepresentations of the first kind (and thus convey the same information as representations in $\mathcal{G}$) (Section~S2 in the Supplementary Material), we retain the corepresentation framework to make explicit the fact that the antiunitary symmetry elements do not introduce additional degeneracy, and also to accommodate future cases in which such antiunitarity may lead to degeneracy doubling.

      \paragraph{Magnetic symmetry via representations for real quantities.}
        It turns out that corepresentation theory is unnecessary when $\mathbf{w}$ is real-valued\cite{article:Bradley1968,article:Erb2020} and the linear space $V$ containing $\mathbf{w}$ is restricted to be over real numbers only, which shall be denoted as $V_{\mathbb{R}}$.
        In this case, the antiunitary operations of $\mathcal{M}$ act \textit{linearly} on $V_{\mathbb{R}}$, their characters remain invariant under any change of basis, and standard representation theory applies.
        The irreducible representations of $\mathcal{M}$ on $V_{\mathbb{R}}$ are equivalent to those of the isomorphic unitary group $\mathcal{M}'$ on $V$ restricted to $V_{\mathbb{R}}$ (Section~\ref{sec:symgroups}).

        On the other hand, if $\mathbf{w}$ is non-real and $V$ is a complex linear space, using irreducible representations of the unitary group $\mathcal{M}'$ on $V$ in Equation~\ref{eq:Wdecomposed} to characterize the symmetry of $\mathbf{w}$ under $\mathcal{M}$ can yield misleading information, since characters of antiunitary operations are not invariant on complex linear spaces.\cite{article:Bradley1968,article:Newmarch1981}
        Consequently, a non-real $\mathbf{w}$ cannot be classified as even or odd under time reversal, either because it is not an eigenfunction of $\hat{\theta}$ or because its eigenvalue is phase-dependent under complex scaling\cite{article:Uhlmann2016}.

        In the presence of a magnetic field, conceptual-DFT quantities such as electron densities and Fukui functions remain real-valued.
        Consequently, their symmetry is most appropriately characterized using the irreducible representations of the full magnetic group $\mathcal{M}$ (or its isomorphic unitary counterpart $\mathcal{M}'$), which, as will be seen in Section~\ref{sec:dipole-sym}, provide a more complete classification than analysis within the unitary halving subgroup $\mathcal{G}$.

      \paragraph{Actions of time reversal on complex wavefunctions.}
        In Section~4.2.1.3 of Ref.~\citenum{article:Wibowo-Teale2024}, we discussed two distinct actions of time reversal on a complex MO in the one-electron Hilbert space: one that includes spin and one that neglects it.
        In this work, we adopt the latter and treat time reversal as though it were the conventional complex conjugation operation.
        This is justified for spin–collinear MOs, as is indeed the case for all calculations in this work.
        Consequently, we only require \textit{single-valued} irreducible corepresentations of $\mathcal{M}$ for all magnetic symmetry classifications (Ref.~\citenum{article:Wibowo-Teale2024} for a detailed explanation).

\section{Computational Details}
\label{sec:compdetails}

  \subsection{Electronic-Structure Calculations}

\begin{figure}
      \centering
      \includegraphics[width=.75\linewidth]{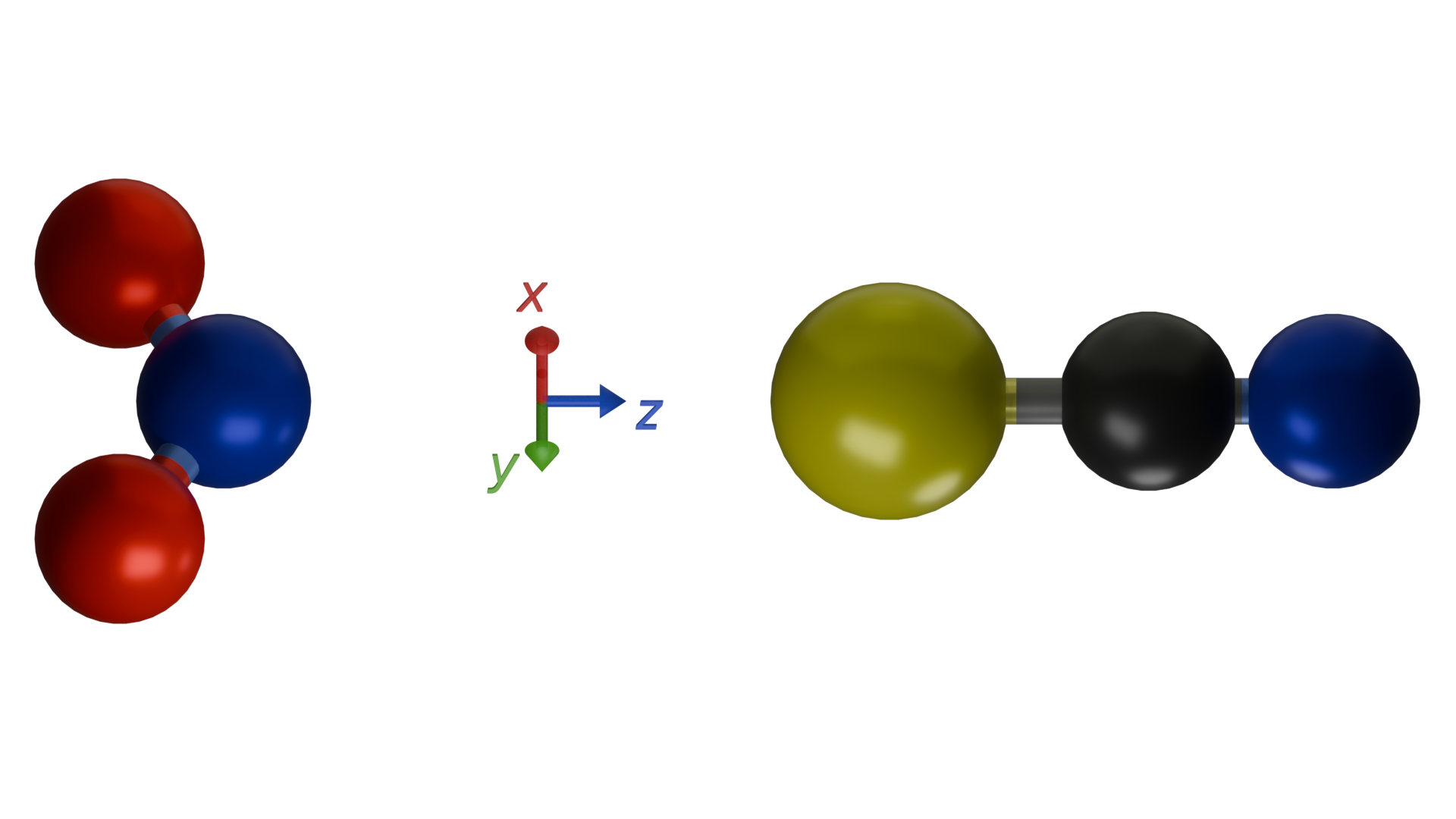}
      \caption{%
        Geometries of \ce{(NO2)-} (left) and \ce{SCN-} (right) used in all calculations.
        In both systems, the principal axis ($C_2$ for \ce{(NO2)-} and $C_{\infty}$ for \ce{SCN-}) is aligned with the $z$-axis, and the center of mass coincides with both the gauge origin of the magnetic vector potential and the Cartesian origin.
        Additionally, \ce{(NO2)-} lies in the $yz$-plane.
        Atom colors: \ce{C} --- black, \ce{N} --- blue, \ce{O} --- red, \ce{S} --- yellow.
      }
      \label{fig:mols}
    \end{figure}

    Geometry optimizations for \ce{(NO2)-} and \ce{SCN-} were performed at zero field in the aug-cc-pVTZ basis set\cite{article:Kendall1992} with the cTPSS\cite{article:Bates2012} and r\textsuperscript{2}SCAN0\cite{article:Furness2020}  exchange--correlation functionals.
    The resulting geometries were nearly identical, and the cTPSS-optimized structure was chosen for all subsequent calculations.
    The molecular orientations are shown in Figure~\ref{fig:mols}.
    For both systems, the center of mass and the gauge origin of the magnetic vector potential (at all field strengths) were placed at the Cartesian origin (Section~\ref{sec:orbitsym}).

    For \ce{(NO2)-}, three magnetic-field orientations were considered: $\mathbf{B} \parallel +\hat{\mathbf{x}}$, $\mathbf{B} \parallel +\hat{\mathbf{y}}$, and $\mathbf{B} \parallel +\hat{\mathbf{z}}$.
    For \ce{SCN-}, only $\mathbf{B} \parallel +\hat{\mathbf{x}}$ and $\mathbf{B} \parallel +\hat{\mathbf{z}}$ were examined due to the cylindrical symmetry of the nuclear framework about the $z$-axis.
    In all cases, only positive field directions were considered, because there exists at least one unitary symmetry operation of the field-free molecular system that reverses the field (\textit{e.g.}, in \ce{(NO2)-}, $\sigma^{yz}$ would map $\mathbf{B} = B\hat{\mathbf{z}}$ to $\mathbf{B} = -B\hat{\mathbf{z}}$ while leaving the molecular framework unchanged).
    This is because if $\hat{\mathscr{H}}(\mathbf{B})$ admits an eigenstate with energy $E$ and is related to $\hat{\mathscr{H}}(-\mathbf{B})$ by a unitary transformation $\hat{u}$, then $\hat{\mathscr{H}}(-\mathbf{B})$ also admits an eigenstate with energy $E$:
    \begin{align}
      \hat{\mathscr{H}}(\mathbf{B}) \Psi = E\Psi
      &\implies \hat{u} \hat{\mathscr{H}}(\mathbf{B}) \hat{u}^{-1} \hat{u} \Psi = \hat{u} (E\Psi) \nonumber \\
      &\implies \hat{\mathscr{H}}(-\mathbf{B}) \hat{u} \Psi = E \hat{u} \Psi.
    \end{align}
    It is therefore sufficient to study only the (approximate) eigenspectrum of $\hat{\mathscr{H}}(\mathbf{B})$, from which the (approximate) eigenspectrum of $\hat{\mathscr{H}}(-\mathbf{B})$ can be deduced.

    For each molecule-plus-field configuration, ground states with $M_S = 0, -1, -2$ were computed and tracked as $|\mathbf{B}|$ increased from $0$ to $0.3\,B_0$ in steps of $0.02\,B_0$.
    All calculations were performed using current-DFT with the aug-cc-pVTZ basis set and the cTPSS\cite{article:Bates2012} and r\textsuperscript{2}SCAN0\cite{article:Furness2020} exchange--correlation functionals.
    The Maximum Overlap Method (MOM)\cite{article:Gilbert2008,article:Barca2018,article:Wibowo2023} was employed to avoid variational collapse and undesirable state switching during self-consistent-field (SCF) iterations.
    In the following, only r\textsuperscript{2}SCAN0 results are reported, as they were found to yield easier SCF convergence than their cTPSS counterparts.

\begin{figure}
      \centering
      \includegraphics[width=.25\textwidth]{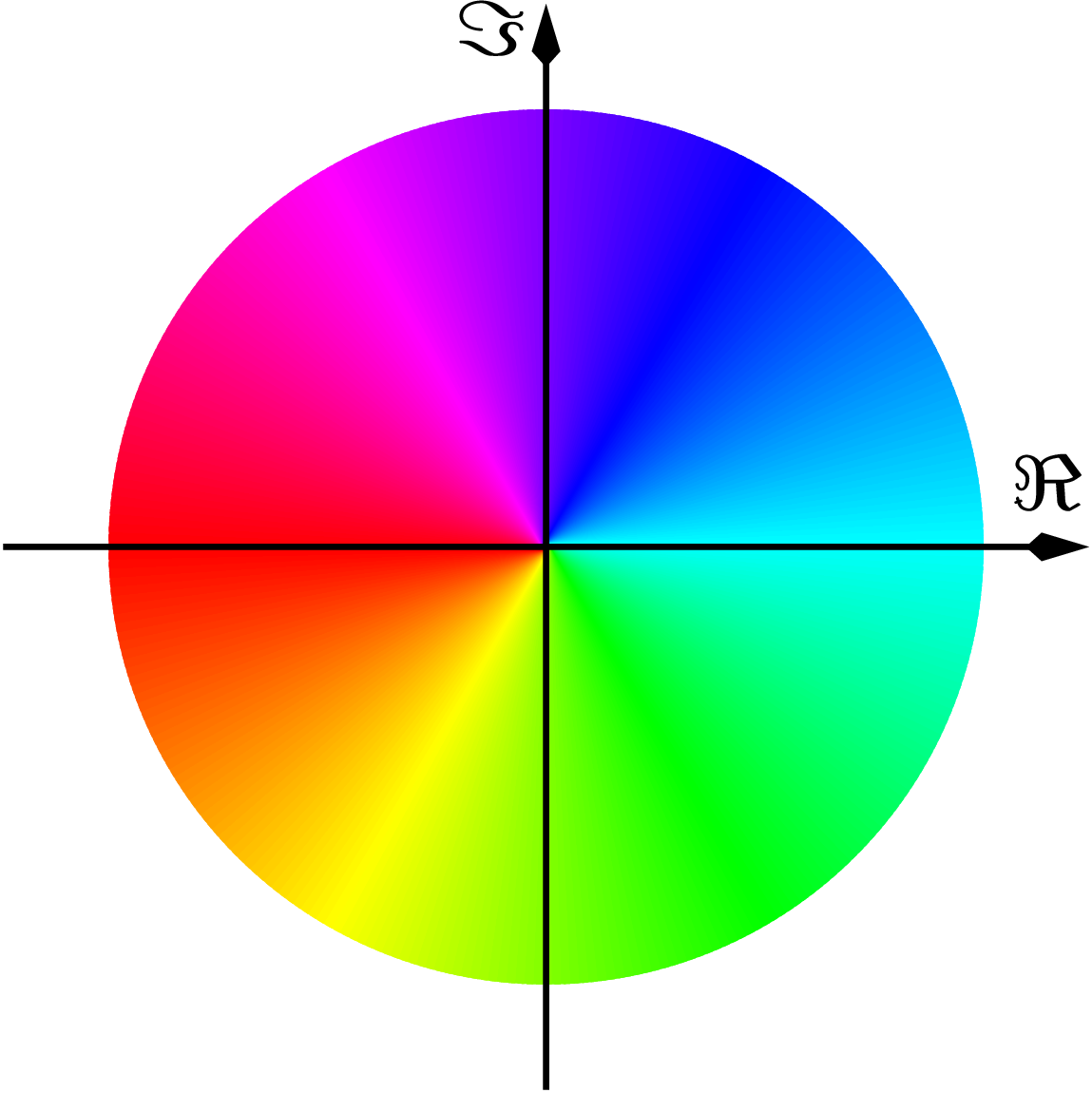}
      \caption{%
        Color wheel used to represent complex phases in isosurfaces.
        In this work, all plotted isosurfaces correspond to real-valued quantities (real MOs at zero magnetic field and real Fukui functions at all magnetic-field strengths); consequently, only the red and light-blue colors are used.
      }
      \label{fig:colorwheel}
    \end{figure}

    All electronic-structure calculations and all evaluations of conceptual DFT quantities were carried out in \textsc{QUEST}\cite{software:Quest2022}.
    All symmetry analysis and symmetry-related calculations were performed through the framework of the \qsymsq{} program (\texttt{v0.12.0})\cite{article:Huynh2024}.
    In this article, all orbital and Fukui function isosurfaces are plotted with colors representing complex phases in $(-\pi, \pi]$ according to the color wheel shown in Figure~\ref{fig:colorwheel}, and all values are reported in atomic units, unless stated otherwise.

  \subsection{Treatment of Wavefunction Degeneracy in the Calculations of Fukui Functions}
    \label{sec:degeneracy}
    The Fukui function calculations for \ce{SCN-} at zero field face a theoretical complication.
    Since the highest occupied MO (HOMO) of zero-field \ce{SCN-} has $\Pi$ symmetry in $\mathcal{C}_{\infty v}$ and is therefore doubly degenerate, the ground-state wavefunction of \ce{SCN^{.}} (\textit{i.e.}, the corresponding $N_{\mathrm{e}} - 1$ system) also has this symmetry and degeneracy.
    Consequently, the corresponding electron density is no longer totally symmetric with respect to $\mathcal{C}_{\infty v}$ and $\mathcal{C}'_{\infty v}$ (see Appendix~\ref{app:denprojection} and Table~S4 in the Supplementary Material).
    A na\"{\i}ve application of the finite-difference approximation in Equation~\ref{eq:fm-fd} would therefore yield a Fukui function with an incorrect symmetry.

    One simple way to circumvent this issue is to consider the $N_{\mathrm{e}} - 2$ system and use $\Delta N_{\mathrm{e}} = 2$ instead in Equation~\ref{eq:fm-fd}.
    The ground-state wavefunction of zero-field \ce{SCN+} is indeed non-degenerate and the corresponding density is totally symmetric, thus ensuring that the obtained Fukui function is also.
    This method has been used in the past when evaluating Fukui functions for degenerate systems\cite{article:Langenaeker1996}.
    However, in this work, we made use of a more elegant approach based on the symmetry projection of electron densities, thereby avoiding large $N_{\mathrm{e}}$ perturbations.
    The full details of this approach are presented in Appendix~\ref{app:denprojection}, but suppose we have a set of $d$ degenerate pure-state densities $\{\rho_i(\mathbf{r}) \mid i = 1, \ldots, d\}$ corresponding to $d$ degenerate wavefunctions spanning a $d$-dimensional irreducible (co)representation of a (unitary or magnetic) symmetry group $\mathcal{H}$, we can construct a unique ensemble density $\bar{\rho}(\mathbf{r})$ that is totally symmetric with respect to $\mathcal{H}$ as
    \begin{multline}
      \bar{\rho}(\mathbf{r})
        = \frac{1}{|\mathcal{H}|}
        \sum_{h \in \mathcal{H}} \hat{h} \rho_i(\mathbf{r})
        = \frac{1}{d} \sum_{j=1}^{d} \rho_j(\mathbf{r})
      \\ \textrm{for all $i = 1, \ldots, d$}.
      \label{eq:rhobar}
    \end{multline}
    We note that since $\bar{\rho}(\mathbf{r})$ is real-valued, its symmetry under $\mathcal{H}$ can be considered using standard representation theory even if $\mathcal{H}$ is a magnetic group (Section~\ref{sec:magsym} and Appendix~\ref{app:denprojection}).

\section{Results and Discussion}
\label{sec:results}

  \subsection{Electric Dipole Moment and Its Symmetry as a Guide to Density and Fukui Function Behaviors}
  \label{sec:dipole-sym}

    We begin by analyzing the effect of the orientation of an external magnetic field on the symmetry of the charge distribution in \ce{(NO2)-} and \ce{SCN-}, as reflected in their electric dipole moment.
    This provides a foundation for the subsequent discussion of density-based conceptual DFT quantities --- particularly the Fukui function --- as well as the symmetry and electronic properties of the relevant configurations and states.

    Table~\ref{tab:dipsym} summarizes the unitary and magnetic symmetry groups of \ce{(NO2)-} and \ce{SCN-} for various magnetic-field orientations, together with the symmetry-allowed components of the electric dipole component $\bm{\mu}$.
    For \ce{(NO2)-}, $\mathbf{B} \parallel \hat{\mathbf{x}}$ is \textit{perpendicular} to the molecular plane, $\mathbf{B} \parallel \hat{\mathbf{z}}$ is \textit{parallel} to the principal axis of the molecule, and $\mathbf{B} \parallel \hat{\mathbf{y}}$ lies within the molecular plane but perpendicular to the principal axis (\textit{in-plane}).
    For \ce{SCN-}, $\mathbf{B} \parallel \hat{\mathbf{x}}$ and $\mathbf{B} \parallel \hat{\mathbf{y}}$ are both \textit{perpendicular} to the principal axis of the molecule, whereas $\mathbf{B} \parallel \hat{\mathbf{z}}$ is \textit{parallel} to it.
    In all cases, the symmetry-allowed components of the electric dipole component $\bm{\mu}$ are those that transform as the totally symmetric irreducible representation of the corresponding symmetry group.
    Since dipole moments are real-valued quantities, their symmetry with respect to magnetic groups can be meaningfully classified using conventional representation theory (Section~\ref{sec:magsym}).

    \begin{table}[h!]
      \centering
      \caption{%
        Symmetry groups and allowed electric dipole components $\bm{\mu}$ of \ce{(NO2)-} and \ce{SCN-}.
        The geometries of these two systems are given in Figure~\ref{fig:mols}.
        $\mathcal{G}$ gives the unitary symmetry group of the molecule-plus-field system and $\mathcal{M}$ the magnetic symmetry group (Section~\ref{sec:symgroups}).
        The allowed electric dipole components are those transforming as the totally symmetric irreducible representation of the corresponding symmetry group.
        Highlighted in magenta are the components constrained to vanish by $\mathcal{M}$ but not by $\mathcal{G}$.
        Character tables for all magnetic groups are given in Section~S2 in the Supplementary Material.
      }
      \label{tab:dipsym}
      \renewcommand\arraystretch{1.25}
      \begin{tabular}{%
        l |%
        l l |%
        l l
      }
        \toprule
        %
        \multicolumn{5}{c}{\ce{(NO2)-}}
        \\
        \midrule
        %
        Field
        %
        & $\mathcal{G}$ 
        & $\bm{\mu}$ 
        %
        & $\mathcal{M}$ 
        & $\bm{\mu}$ 
        \\
        \midrule
        $\mathbf{0}$ 
        %
        & $\mathcal{C}_{2v}$ 
        & $\mu_z$ 
        %
        & $\mathcal{C}'_{2v}$ 
        & $\mu_z$ 
        \\[8pt]
        $\mathbf{B} \parallel \hat{\mathbf{x}}$ 
        %
        & $\mathcal{C}_{s}(yz)$ 
        & $\textcolor{magenta}{\mu_y}, \mu_z$ 
        %
        & $\mathcal{C}_{2v}(\mathcal{C}_s)$ 
        & $\mu_z$ 
        \\[-7pt]
        {\scriptsize perpendicular} &&&&
        \\[8pt]
        $\mathbf{B} \parallel \hat{\mathbf{y}}$ 
        %
        & $\mathcal{C}_{s}(xz)$ 
        & $\textcolor{magenta}{\mu_x}, \mu_z$ 
        %
        & $\mathcal{C}_{2v}(\mathcal{C}_s)$ 
        & $\mu_z$ 
        \\[-7pt]
        {\scriptsize in-plane} &&&&
        \\[8pt]
        $\mathbf{B} \parallel \hat{\mathbf{z}}$ 
        %
        & $\mathcal{C}_{2}$ 
        & $\mu_z$ 
        %
        & $\mathcal{C}_{2v}(\mathcal{C}_2)$ 
        & $\mu_z$ 
        \\[-7pt]
        {\scriptsize parallel} &&&&
        \\
        \bottomrule
        \multicolumn{5}{c}{}\\
        \toprule
        %
        \multicolumn{5}{c}{\ce{SCN-}}
        \\
        \midrule
        %
        Field
        %
        & $\mathcal{G}$ 
        & $\bm{\mu}$ 
        %
        & $\mathcal{M}$ 
        & $\bm{\mu}$ 
        \\
        \midrule
        $\mathbf{0}$ 
        %
        & $\mathcal{C}_{\infty v}$ 
        & $\mu_z$ 
        %
        & $\mathcal{C}'_{\infty v}$ 
        & $\mu_z$ 
        \\[8pt]
        $\mathbf{B} \parallel \hat{\mathbf{x}}$ 
        %
        & $\mathcal{C}_{s}(yz)$ 
        & $\textcolor{magenta}{\mu_y}, \mu_z$ 
        %
        & $\mathcal{C}_{2v}(\mathcal{C}_s)$ 
        & $\mu_z$ 
        \\[-7pt]
        {\scriptsize perpendicular} &&&&
        \\[8pt]
        $\mathbf{B} \parallel \hat{\mathbf{y}}$ 
        %
        %
        & $\mathcal{C}_{s}(xz)$ 
        & $\textcolor{magenta}{\mu_x}, \mu_z$ 
        %
        & $\mathcal{C}_{2v}(\mathcal{C}_s)$ 
        & $\mu_z$ 
        \\[-7pt]
        {\scriptsize perpendicular} &&&&
        \\[8pt]
        $\mathbf{B} \parallel \hat{\mathbf{z}}$ 
        %
        & $\mathcal{C}_{\infty}$ 
        & $\mu_z$ 
        %
        & $\mathcal{C}_{\infty v}(\mathcal{C}_{\infty})$ 
        & $\mu_z$ 
        \\[-7pt]
        {\scriptsize parallel} &&&&
        \\
        \bottomrule
      \end{tabular}
    \end{table}

    The dipole moment selection rules for \ce{(NO2)-} are identical to those previously discussed for \ce{H2CO} and \ce{H2O}\cite{article:Irons2022,article:Wibowo-Teale2024}, due to the fact that these systems share the same symmetry groups.
    At the level of unitary symmetry alone, the $\mathcal{C}_{2v}$ symmetry of \ce{(NO2)-} descends to $\mathcal{C}_s(yz)$, $\mathcal{C}_s(xz)$, and $\mathcal{C}_2$ for the perpendicular, in-plane, and parallel field orientations, respectively.
    Accordingly, the perpendicular and in-plane fields allow two non-zero dipole moment components: one along the principal axis and one perpendicular to it.
    When antiunitary operations are included, however, the full magnetic symmetry groups are isomorphic to $\mathcal{C}_{2v}$ for all three field orientations, thus forcing any dipole moment components perpendicular to the principal axis to vanish.

    In a similar manner, the dipole moment selection rules for \ce{SCN-} are identical to those presented for the halogen halides\cite{article:Irons2022}, reflecting their analogous symmetry.
    As in the case of \ce{(NO2)-}, inclusion of antiunitary symmetry is essential to account for the vanishing of dipole moment components perpendicular to the molecular axis for the perpendicular field orientations.
    Consequently, in all three cases, only the dipole moment component along the principal axis survives.
    The dipole moment results will be discussed in Section~\ref{sec:dipole-moment-variations} for the different electronic configurations and states described in Section~\ref{sec:configs-states}.

  \subsection{Electronic Configurations and States}
  \label{sec:configs-states}

    \subsubsection{Spin States in \ce{(NO2)-}}

\begin{figure*}[h!]
        \centering
        \includegraphics[scale=.72]{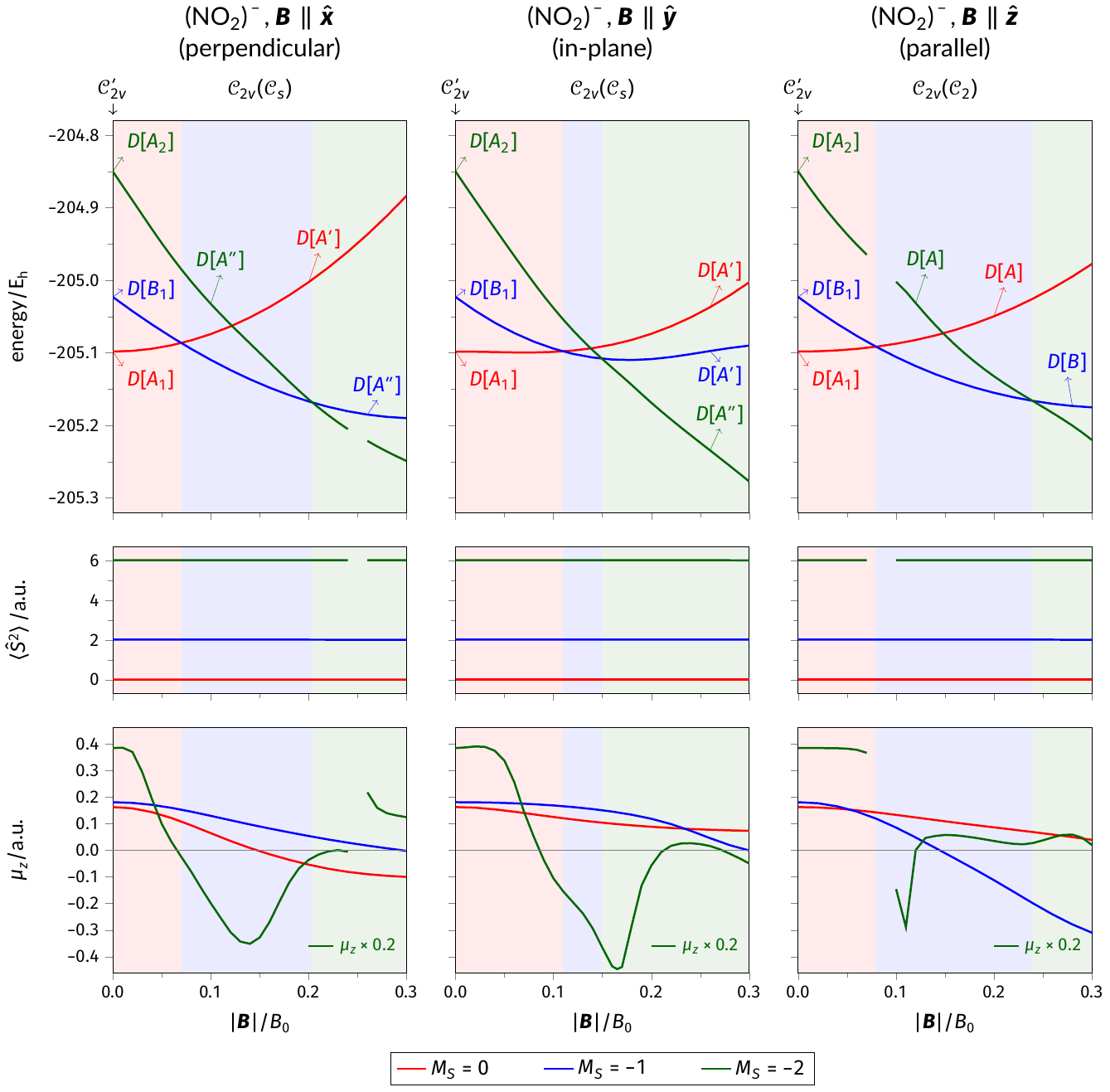}
        \caption{%
          Energy, $\langle \hat{S}^2 \rangle$, and the $z$-component of the electric dipole moment for three $M_S$ states of \ce{(NO2)-} as functions of magnetic field strength, shown for perpendicular ($\mathbf{B} \parallel \hat{\mathbf{x}}$), in-plane ($\mathbf{B} \parallel \hat{\mathbf{y}}$), and parallel ($\mathbf{B} \parallel \hat{\mathbf{z}}$) field orientations.
          Each state is assigned a corepresentation symmetry in the corresponding magnetic group.
          The shaded backgrounds indicate the ground state in each field-strength region.
          The $\mu_z$ values for the $M_S = -2$ state have been scaled by a factor of $0.2$ to fit on the same scale as those of the other two states.
        }
        \label{fig:no2m-e-sz-mu}
      \end{figure*}

\begin{figure*}[h!]
        \centering
        \includegraphics[scale=.7]{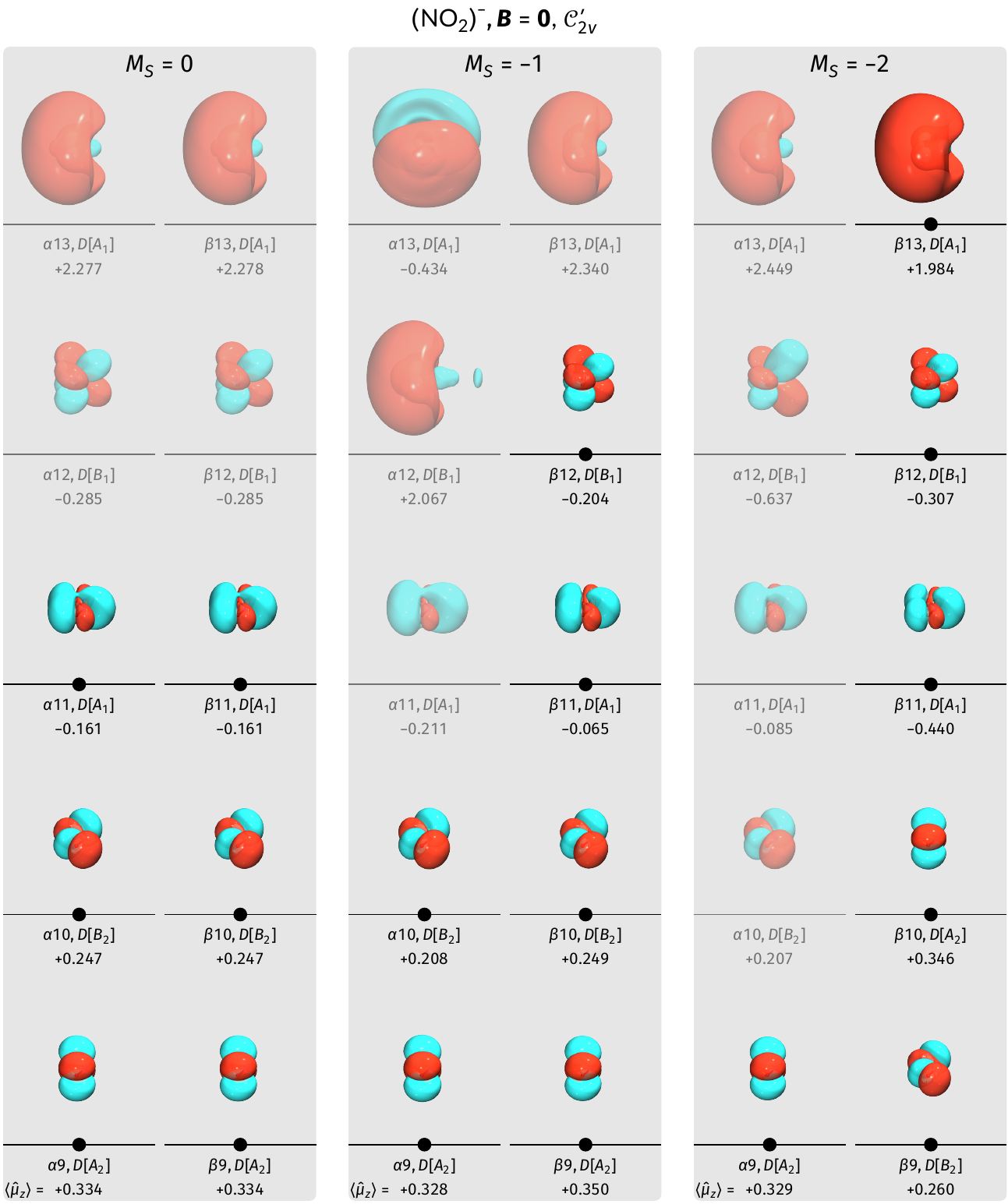}
        \caption{%
          Frontier MOs for three $M_S$ states of \ce{(NO2)-} at zero field.
          MOs are ordered by energy, with occupied MOs indicated by black dots.
          Isosurfaces are plotted at isovalue $|\psi(\mathbf{r})| = 0.015 a_0^{-3/2}$, and the color at each point $\mathbf{r}$ represents the phase angle $\arg \psi(\mathbf{r}) \in (-\pi, \pi]$ at that point according to the color wheel shown in Figure~\ref{fig:colorwheel}.
          Each MO is labeled by its corepresentation symmetry in the magnetic gray group $\mathcal{C}'_{2v}$, and its contribution to the electric dipole moment along the $z$-axis, $\langle \hat{\mu}_z \rangle = \braket{\psi | \hat{\mu}_z |\psi}$, is also reported.
        }
        \label{fig:no2m-mos-zerofield}
      \end{figure*}

      Figure~\ref{fig:no2m-e-sz-mu} shows the magnetic-field dependence of the energy, $\langle \hat{S}^2 \rangle$, and the $z$-component of the electric dipole moment for three \ce{(NO2)-} states with $M_S = 0, -1, -2$ over the range $0 \le |\mathbf{B}| \le 0.3\,B_0$.
      The closed-shell ground state at zero field ($M_S = 0$, $D[A_1]$ symmetry in $\mathcal{C}'_{2v}$) was tracked for various field orientations using the MOM to maintain continuity.
      The open-shell excited states with $M_S = -1$ and $M_S = -2$, having $D[B_1]$ and $D[A_2]$ symmetries in $\mathcal{C}'_{2v}$ respectively, were similarly analyzed.
      Figure~\ref{fig:no2m-mos-zerofield} shows the frontier MOs for these three states at zero field.

      Figure~\ref{fig:no2m-e-sz-mu} shows that, for all three field orientations, increasing the magnetic field strength destabilizes the the closed-shell $M_S = 0$ while stabilizing the open-shell $M_S = -1$ and $M_S = -2$ states.
      The destabilization of the closed-shell state arises from the quadratic diamagnetic term in the Hamiltonian (Equation~\ref{eq:Hmag}), whereas the spin-Zeeman paramagnetic term drives the progressive stabilization of the open-shell high-spin states.
      Although higher-spin states are less stable at zero field, they experience stronger stabilization in a magnetic field due to the greater number of unpaired electrons coupling to the field.

      Figure~\ref{fig:no2m-e-sz-mu} also shows the variation of $\langle \hat{S}^2 \rangle$ for the three states.
      However, these values should be interpreted with utmost caution, as they correspond to Kohn--Sham single-determinantal wavefunctions associated with the auxiliary non-interacting system, rather than the actual fully interacting system.
      Moreover, within the unrestricted Kohn--Sham framework, such states are susceptible to spin contamination\cite{article:Wang1995,article:Cohen2007,article:Perdew2009,article:Jacob2012}.
      Nevertheless, $\langle \hat{S}^2 \rangle$ remains essentially constant at $0.0$, $2.0$, and $6.0$ for the $M_S = 0$, $-1$, and $-2$ states, respectively, as the magnetic field increases.
      This suggests minimal spin contamination and indicates that these states correspond to almost pure spin states belonging to the singlet, triplet, and quintuplet manifolds, respectively.
      In any case, the crossings observed between these three states in Figure~\ref{fig:no2m-e-sz-mu} are symmetry-allowed, as they involve states of different spin symmetry.

      The evolution in the relative stability among the three states mirrors the behavior of atomic ground-state configurations under increasing magnetic fields, as detailed in Ref.~\citenum{article:Francotte2022} and the early work by Schmelcher \textit{et al.}\cite{article:Ivanov1998,article:Ivanov1999,article:Ivanov2000,article:Ivanov2001,article:Al-Hujaj2004}.
      In these systems, states with more negative $M_S$ values are progressively stabilized by the spin-Zeeman effect.
      This behavior was not addressed in our previous studies of molecules in strong magnetic fields\cite{article:Irons2022,article:Wibowo-Teale2024}, which adopted an adiabatic approach that tracked only the zero-field ground state across the entire field range.
      For the \ce{(NO2)-} anion, the adiabatic treatment remains valid up to $0.07\,B_0 \approx \SI{16500}{\tesla}$ for perpendicular fields ($\mathbf{B} \parallel \hat{\mathbf{x}}$), $0.11\,B_0 \approx \SI{25900}{\tesla}$ for in-plane fields ($\mathbf{B} \parallel \hat{\mathbf{y}}$), and $0.08\,B_0 \approx \SI{18800}{\tesla}$ for parallel fields ($\mathbf{B} \parallel \hat{\mathbf{z}}$).
      Given that these thresholds represent extreme field strengths from an experimental standpoint, the adiabatic results starting from the zero-field ground state retain significant chemical relevance.

      Beyond these thresholds, the $M_S = 0$ state ceases to be the ground state.
      Instead, the $M_S = -1$ state becomes lowest in energy up to $0.20\,B_0$, $0.15\,B_0$, and $0.24\,B_0$ for perpendicular, in-plane, and parallel fields, respectively, beyond which the $M_S = -2$ state takes over.
      Since the three states differ in spin symmetry by virtue of the difference in their $M_S$ and $\langle \hat{S}^2 \rangle$ values, their crossings are symmetry-allowed even when they have the same spatial magnetic symmetry (\textit{e.g.} the $D[A'']$ states in $\mathcal{C}_{2v}(\mathcal{C}_s)$ with $M_S = -1$ and $M_S = -2$ in perpendicular fields).

      Regardless of field orientation, the general behavior of the three states is consistent with those reported in the literature.
      For example, Lehtola \textit{at al.}\cite{article:Lehtola2019} observed the same pattern in simple diatomics, where singlet, triplet, and quintuplet states of \ce{LiH}, \ce{BeH+}, \ce{BH}, and \ce{CH+} exhibit analogous evolution signatures with increasing magnetic field strength.
      In light of these observations, conceptual-DFT analyses for chemical reactivity based on an adiabatic approach should be restricted to relatively low field strengths, before spin-Zeeman stabilization of high-spin states and diamagnetic destabilization of closed-shell states significantly alter the character of the ground state.
      For \ce{(NO2)-}, this threshold was found to be $0.07\,B_0$ in our calculations.

    \subsubsection{Spin States in \ce{SCN-}}

\begin{figure*}[h!]
        \centering
        \includegraphics[scale=.72]{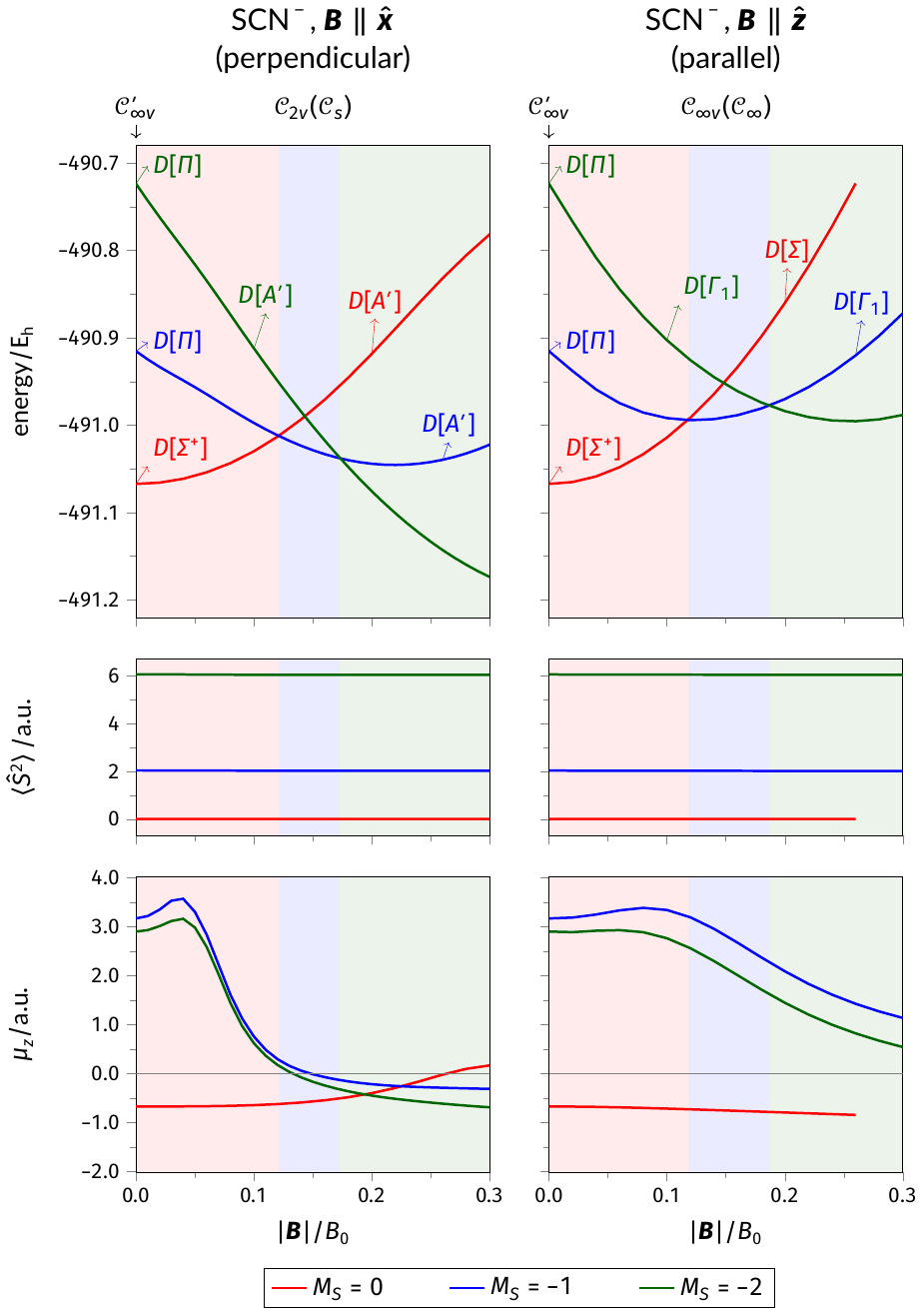}
        \caption{%
          Energy, $\langle \hat{S}^2 \rangle$, and the $z$-component of the electric dipole moment for three $M_S$ states of \ce{SCN-} as functions of magnetic field strength, shown for perpendicular ($\mathbf{B} \parallel \hat{\mathbf{x}}$) and parallel ($\mathbf{B} \parallel \hat{\mathbf{z}}$) field orientations.
          Each state is assigned a corepresentation symmetry in the corresponding magnetic group.
          The shaded backgrounds indicate the ground state in each field-strength region.
          In $\mathcal{C}_{\infty}$, $\Gamma_1$ denotes one of the two complex-conjugate one-dimensional irreducible representations subduced from $\Pi$ in $\mathcal{C}_{\infty v}$, with character $\exp(i \phi)$ for the rotation $C(\phi)$.
        }
        \label{fig:scnm-e-sz-mu}
      \end{figure*}

\begin{figure*}[h!]
        \centering
        \includegraphics[scale=.7]{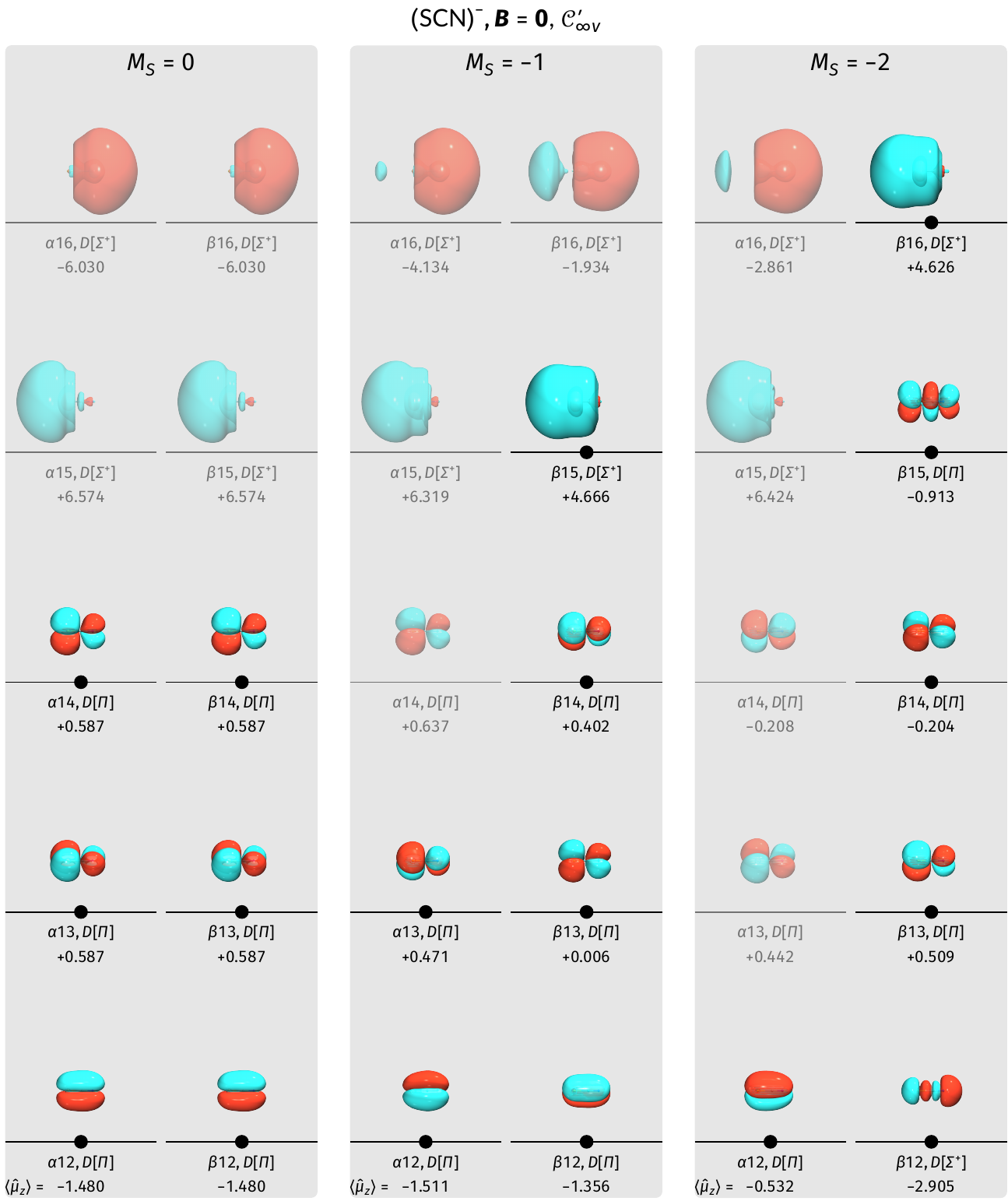}
        \caption{%
          Frontier MOs for three $M_S$ states of \ce{SCN-} at zero field.
          MOs are ordered by energy, with occupied MOs indicated by black dots.
          Isosurfaces are plotted at isovalue $|\psi(\mathbf{r})| = 0.010 a_0^{-3/2}$, and the color at each point $\mathbf{r}$ represents the phase angle $\arg \psi(\mathbf{r}) \in (-\pi, \pi]$ at that point according to the color wheel shown in Figure~\ref{fig:colorwheel}.
          Each MO is labeled by its corepresentation symmetry in the magnetic gray group $\mathcal{C}'_{2v}$, and its contribution to the electric dipole moment along the $z$-axis, $\langle \hat{\mu}_z \rangle = \braket{\psi | \hat{\mu}_z |\psi}$, is also reported.
        }
        \label{fig:scnm-mos-zerofield}
      \end{figure*}

      In an analogous manner to \ce{(NO2)-}, the $M_S = 0$, $M_S = -1$, and $M_S = -2$ states of \ce{SCN-} were tracked from $\mathbf{B} = \mathbf{0}$ to $|\mathbf{B}| = 0.30\,B_0$ for both perpendicular and parallel field orientations (Figure~\ref{fig:scnm-e-sz-mu}); the corresponding frontier MOs at zero field are displayed in Figure~\ref{fig:scnm-mos-zerofield}.
      At zero field, the $M_S = -1$, and $M_S = -2$ states are doubly degenerate with spatial magnetic symmetry $D[\Pi]$ in $\mathcal{C}'_{\infty v}$.
      Examination of the MOs in Figure~\ref{fig:scnm-mos-zerofield} reveals that this degeneracy originates from the uneven occupation of the $d[\pi]$ orbitals.
      Application of a magnetic field lifts this spatial degeneracy and splits each $D[\Pi]$ state into two non-degenerate components: $D[A'] \oplus D[A'']$ in $\mathcal{C}_{2v}(\mathcal{C}_s)$ for perpendicular fields ($\mathbf{B} \parallel \hat{\mathbf{x}}$), and $D[\Gamma_1] \oplus D[\bar{\Gamma}_1]$ in $\mathcal{C}_{\infty v}(\mathcal{C}_{\infty})$ for parallel fields ($\mathbf{B} \parallel \hat{\mathbf{z}}$).
      In $\mathcal{C}_{\infty}$, $\Gamma_1$ denotes one of the two complex-conjugate one-dimensional irreducible representations subduced from $\Pi$ in $\mathcal{C}_{\infty v}$, with character $\exp(i \phi)$ for the rotation $C(\phi)$, and $\bar{\Gamma}_1$ is its complex conjugate.
      In what follows, we focus only on the lower-energy component in each case, effectively considering only the ground state within each $M_S$ sector.

      Figure~\ref{fig:scnm-e-sz-mu} shows that the three states vary in an analogous fashion to those in \ce{(NO2)-}, where the ground state transitions progressively from $M_S = 0$ to $-1$ and then to $-2$ as the field strength increases.
      The corresponding $\langle \hat{S}^2 \rangle$ values also remain almost constant at $0.0$, $2.0$, and $6.0$, indicating minimal spin contamination (subject to the caveat discussed earlier) across most of the field range.
      The $M_S = 0$ state, however, ceases to be trackable as $|\mathbf{B}|$ approaches $0.30\,B_0$ in the parallel orientation; this is fortunately inconsequential since this state is no longer the ground state in this regime.

      The magnetic-field threshold up to which the $M_S = 0$ is still the ground state and an adiabatic approach based on this state remains valid was found to be approximately $0.14\,B_0$ for \ce{SCN-} in our calculations, which is higher than that for \ce{(NO2)-}.
      This system dependence is not unexpected in view of the different zero-field electronic structure of the two systems (Figures~\ref{fig:no2m-mos-zerofield} and \ref{fig:scnm-mos-zerofield}).
    
  \subsection{Electric Dipole Moment Variation}
  \label{sec:dipole-moment-variations}

    Having established the stability of the electronic states under increasing field strength, we now turn to the electric dipole moment, which provides an overall signature of the electron density.
    In all cases, only the $z$-component survives, in accordance with the symmetry restrictions discussed in Section~\ref{sec:dipole-sym} and summarized in Table~\ref{tab:dipsym}.
    Figures~\ref{fig:no2m-e-sz-mu} and \ref{fig:scnm-e-sz-mu} therefore only display the variation of $\mu_z$ with field strength.
    Our previous studies\cite{article:Irons2022,article:Huynh2024,article:Wibowo-Teale2024} have shown that the electric dipole moment, and hence molecular polarity, can be heavily modulated, and even reversed, by strong magnetic fields.
    In what follows, we extend this analysis to a non-adiabatic framework.

    \subsubsection{Dipole Moments in \ce{(NO2)-}}

      Given the orientation of \ce{(NO2)-} as shown in Figure~\ref{fig:mols}, it is expected that, at zero field, the $z$-component of the electric dipole moment for all three states points from the \ce{O} side towards the \ce{N} side.
      This is consistent with the greater electronegativity of \ce{O} relative to \ce{N} in the absence of a magnetic field\cite{article:Francotte2022}.
      Introducing a magnetic field along any of the three principal directions causes $\mu_z$ to decrease and become less positive for all three states, as shown in Figure~\ref{fig:no2m-e-sz-mu}.

      For the $M_S = 0$ and $M_S = -1$ states, this smooth and gradual reduction in $\mu_z$ can be traced back to the field-induced changes in atomic electronegativity\cite{article:Francotte2022}.
      Specifically, the initial response $\partial \chi / \partial |\mathbf{B}|$ is approximately $0.00$ for \ce{N} and $-0.26$ for \ce{O}.
      Therefore, to first order, the magnetic field leaves the electronegativity of \ce{N} largely unchanged while lowering that of \ce{O}, thereby reducing, and eventually reversing, the electronegativity difference between the two elements.

      On the other hand, the irregular behavior of $\mu_z$ for the $M_S = -2$ state arises partly from the occupation of the much more diffuse $\beta 13$ MO (the HOMO, see Figure~\ref{fig:no2m-mos-zerofield}).
      This orbital makes a strong positive contribution to $\mu_z$ ($+1.984$ at zero field), leading to the markedly larger dipole moment observed for the $M_S = -2$ state compared to the lower-spin states.
      Its diffuse character also makes it much more sensitive to the external magnetic field (Section~\ref{sec:local}), which accounts for the pronounced variation of $\mu_z$ with field strength.

      The dipole moments for the three spin states exhibit rather different behaviors across the three field orientations.
      In the perpendicular case, the $M_S = 0$ curve shows a steep decrease that would eventually lead to dipole reversal within an adiabatic picture.
      However, at around $0.07\,B_0$, the $M_S = -1$ state becomes more stable, causing the dipole moment to ``jump'' to a more positive value before decreasing more slowly without reversal.
      At approximately $0.21\,B_0$, the $M_S = -2$ becomes the ground state and the dipole moment switches abruptly once again, but this time to a slightly negative value.
      The dipole moment then exhibits a discontinuity at around $0.25\,B_0$ due to the non-convergence of the current-DFT calculation at this field strength for $M_S = -2$.
      It is unclear whether the SCF procedure has located a different state because the energy profile appears smooth and no change in symmetry nor $\langle \hat{S}^2 \rangle$ is observed across this point.

      In contrast, for the in-plane orientation, the dipole moments of the $M_S = 0$ and $M_S = -1$ states exhibit relatively weak sensitivity towards the applied field within their respective ground-state regimes.
      Beyond $0.15\,B_0$, however, the $M_S = -2$ ground state significantly inverts the dipole moment, again accompanied by irregular behaviors which are also less straightforward to interpret, not least because the transition to the $M_S = -2$ state occurs at a lower field strength than in the other two field orientations.

      In the parallel orientation, the dipole moment of the $M_S = 0$ state is again relatively insensitive to the applied field within its ground-state regime.
      In contrast, the dipole moment of the $M_S = -1$ state shows a much stronger response: it decreases significantly and eventually inverting at around $0.14\,B_0$.
      When the $M_S = -2$ state becomes the ground state at \textit{ca.} $0.24\,B_0$, the dipole moment undergoes a further abrupt change to a small but positive value, thereby inverting once more.

      The differences in dipole behavior among the three states can be understood by considering how their MOs vary, in both energy and symmetry, as the field is introduced in different orientations.
      This is analogous to how the orientation-dependent dipole reversal in hydrogen fluoride was rationalized in our previous work\cite{article:Huynh2024}, although a detailed analysis of these effects for \ce{(NO2)-} lies beyond the scope of the present study.

    \subsubsection{Dipole Moments in \ce{SCN-}}

      For \ce{SCN-} aligned along the $z$-axis as shown in Figure~\ref{fig:mols}, the only non-zero dipole moment component, $\mu_z$, is negative at zero field for the ground $M_S = 0$ state (Figure~\ref{fig:scnm-e-sz-mu}), \textit{i.e.}, it points from \ce{N} to \ce{S}, in accordance with the higher electronegativity of \ce{N} compared to \ce{C} and \ce{S}.
      On the other hand, the $M_S = -1$ and $M_S = -2$ states possess positive and much larger dipole moments, on account of the occupation of diffuse HOMOs that are skewed towards the \ce{S} end of the molecule.
      However, the variation of $\mu_z$ for these states is much less drastic than that observed for the $M_S = -2$ state of \ce{(NO2)-} (Figure~\ref{fig:no2m-e-sz-mu}).
      A possible reason for this difference is the less diffuse character of the HOMOs in the $M_S = -1$ and $M_S = -2$ states of \ce{SCN-} (Figure~\ref{fig:scnm-mos-zerofield}) compared to the highly diffuse HOMO in the $M_S = -2$ state of \ce{(NO2)-} (Figure~\ref{fig:no2m-mos-zerofield}).

      In both field orientations, the dipole moment of the $M_S = 0$ state varies only weakly with field strength within its ground-state regime.
      Beyond this regime, the adiabatic picture predicts that a perpendicular field would induce a dipole reversal at approximately $0.26\,B_0$, whereas a parallel field would not.
      This behavior mirrors that observed for the hydrogen halides\cite{article:Irons2022} and can be rationalized in terms of the field variation in the electronegativities of \ce{S} and \ce{N}\cite{article:Francotte2022}.
      A complementary symmetry-based explanation, akin to that proposed for hydrogen fluoride\cite{article:Huynh2024}, can also be invoked.

      When transitions to the $M_S = -1$ and $M_S = -2$ states are taken into account in a non-adiabatic picture, $\mu_z$ experiences discontinuous changes that lead to complex and nuanced overall behaviors, as also observed for \ce{(NO2)-}.
      In the perpendicular-field case, the transition from the $M_S = 0$ to the $M_S = -1$ regime at \textit{ca.} $0.12\,B_0$ causes the initially negative $\mu_z$ to become slightly positive, after which it gradually decreases and reverses for the second time.
      As the $M_S = -2$ regime is approached, $\mu_z$ exhibits another abrupt change to a slightly more negative value before decreasing further, reaching a value at $0.30\,B_0$ close to that at zero field.
      In the parallel-field case, by contrast, entry into the $M_S = -1$ regime produces an abrupt switch from a negative to a much larger positive  $\mu_z$, which then decreases gradually.
      A further transition into the $M_S = -2$ regime results in a smaller positive value that continues to decrease but does not yet invert by $0.30\,B_0$.

  \subsection{Ambident Nucleophilicity via Conceptual DFT}
  \label{sec:ambidentcharacter}

    \subsubsection{Global Considerations}
    \label{sec:global}
      We remarked previously\cite{article:Francotte2022} that atomic hardness can be strongly influenced by an external magnetic field.
      In Table~\ref{tab:hardness}, we summarize the atomic hardness $\eta$ and its derivative at zero field for \ce{N}, \ce{O}, and \ce{S}, the terminal atoms in \ce{(NO2)-} and \ce{SCN-}.

      \begin{table}[h!]
        \centering
        \caption{%
          Global hardness of isolated atoms as a function of magnetic field strength and its derivative at zero field computed with current-DFT using the cTPSS exchange--correlation functional in the d-aug-cc-pV5Z basis set (data reproduced from Ref.~\citenum{article:Francotte2022}).
        }
        \label{tab:hardness}
        \renewcommand\arraystretch{1.25}
        \begin{tabular}{r | r r r | r r r}
          \toprule
          \multirow{2}{*}{$|\mathbf{B}|$ / $B_0$} & \multicolumn{3}{c |}{$\eta$ / $\si{\hartree}$} & \multicolumn{3}{c}{$\partial \eta / \partial |\mathbf{B}|$ / $\si{\hartree} B_0^{-1}$} \\
          \cmidrule{2-7}
          & \ce{N} & \ce{O} & \ce{S} & \ce{N} & \ce{O} & \ce{S} \\
          \midrule
          $0.00$ & $0.27$ & $0.22$ & $0.15$ & $+0.00$ & $+0.26$ & $+0.25$ \\
          $0.50$ & $0.21$ & $0.21$ & $0.16$ \\
          \bottomrule
        \end{tabular}
      \end{table}

      Consider first \ce{SCN-}.
      At zero field, \ce{N} is a much harder atom than \ce{S}, as indicated by the positive hardness difference $\Delta\eta_{\ce{N}-\ce{S}} = \eta_{\ce{N}} - \eta_{\ce{S}} = \SI[retain-explicit-plus]{+0.12}{\hartree}$.
      As the field strength increases to $0.50\,B_0$, this difference decreases to \SI[retain-explicit-plus]{+0.05}{\hartree}, suggesting that an external magnetic field has the capability to alter the regioselectivity in \ce{SCN-}.
      In fact, this effect is even more pronounced when considering the derivative $\partial \Delta\eta_{\ce{N}-\ce{S}} / \partial |\mathbf{B}| = \SI{-0.25}{\hartree} B_0^{-1}$ at zero field, which highlights the strong sensitivity of the hardness difference to the applied field.

      For \ce{(NO2)-}, the positive zero-field hardness difference between \ce{N} and \ce{O}, $\Delta\eta_{\ce{N}-\ce{O}} = \SI[retain-explicit-plus]{+0.05}{\hartree}$, can be attributed to the exchange stabilization of the $2p^3$ configuration in \ce{N}.
      This difference vanishes at $|\mathbf{B}| = 0.50\,B_0$ when \ce{N} is introduced into a molecular environment, effectively eliminating the already weak regioselectivity of \ce{(NO2)-} compared to \ce{SCN-}.
      The zero-field derivative $\partial \Delta\eta_{\ce{N}-\ce{O}} / \partial |\mathbf{B}| = \SI{-0.26}{\hartree} B_0^{-1}$ likewise indicates that $\Delta\eta_{\ce{N}-\ce{O}}$ decreases with increasing field strength.

      Although these considerations are global and do not take into account local effects arising from symmetry or specific bonding environments, they nonetheless suggest that magnetic fields may modulate regioselectivity, and may even induce hardness/softness reversal, in ambident nucleophiles such as \ce{(NO2)-} and \ce{SCN-}.
      To probe these effects in greater detail, it is necessary to move to a local description and examine the corresponding Fukui functions, which are equivalent to the local softness functions in the present study where reactivity comparisons are made within a single molecule (\textit{cf.} the discussion following Equation~\ref{eq:fukui}).

    \subsubsection{Local Considerations}
    \label{sec:local}

      \paragraph{Symmetry of Fukui functions in \ce{(NO2)-}.}
        Figure~\ref{fig:no2m-fukui} shows how the Fukui function $f^-(\mathbf{r})$ for the three $M_S$ states of \ce{(NO2)-} varies with magnetic-field strength and orientation.
        As Fukui functions are one-sided $N_{\mathrm{e}}$-derivatives of the electron density (Equation~\ref{eq:fukui}), they necessarily share the symmetry of the electron density at all field strengths.
        Since the electron density is real-valued, its symmetry can be classified using representations of the full magnetic group $\mathcal{M}$ (Section~\ref{sec:magsym}).

        For all three field orientations considered, the unitary halving subgroup $\mathcal{G}$ of the molecule-plus-field system is Abelian (Table~\ref{tab:dipsym}).
        Furthermore, in each case, the corresponding magnetic group $\mathcal{M}$ only contains irreducible corepresentations of the first kind (Section~S2 in the Supplementary Material), which are necessarily non-degenerate.
        In other words, antiunitary operations in these groups cannot induce additional wavefunction degeneracies, implying that symmetry-conserved wavefunctions must be non-degenerate in $\mathcal{M}$.
        The associated pure-state electron densities must therefore transform as the totally symmetric irreducible \textit{representation} of $\mathcal{M}$.
        Since this holds true for both the $N_{\mathrm{e}}$- and the $(N_{\mathrm{e}} - 1)$-systems entering the finite-difference expression for $f^-(\mathbf{r})$ for \ce{(NO2)-} (Equation~\ref{eq:fm-fd}), the resulting approximate $f^-(\mathbf{r})$ must likewise be totally symmetric with respect to $\mathcal{M}$.
        This can indeed be seen in all isosurface plots of $f^-(\mathbf{r})$ in Figure~\ref{fig:no2m-fukui}, noting that $f^-(\mathbf{r})$ being real-valued and spin-free is invariant under time reversal.

      \paragraph{Zero-field ambident character of \ce{(NO2)-}.}
        The zero-field picture clearly demonstrates the ambident nucleophilicity of \ce{(NO2)-}: the \ce{N} and \ce{O} centers are comparably nucleophilic due to the \ce{N} $sp^2$ lone pair and the in-plane \ce{O} lone pairs, as evident from the contour and isosurface plots of $f^-(\mathbf{r})$ for the ground $M_S = 0$ state (red-shaded isosurface, top row, left-most column, Figure~\ref{fig:no2m-fukui}).
        This can be compared to results from earlier studies\cite{article:DeProft1996,article:Langenaeker1996,article:DeProft2002}, which reported per-site condensed Fukui functions $f_A^{-}(\mathbf{r})$ of $0.281$ ($A = \ce{N}$) and $0.359$ ($A = \ce{O}$) at the B3LYP level\cite{article:DeProft1996}, and, at a higher level of theory (CISD), $0.310$ and $0.345$, respectively\cite{article:DeProft2002}.
        At first glance, these results indicate that \ce{N} is the harder center, although the difference in softness between the two sites at the CISD level is rather small.
        In Ref.~\citenum{article:Langenaeker1996}, it was however found that the uncondensed Fukui function $f^{-}(\mathbf{r})$ in the molecular plane is more spatially extended around the \ce{N} center, indicating that the \ce{N} site is slightly softer in the plane of nucleophilic attack.
        The contour plot $f^-(\mathbf{r})$ for the ground $M_S = 0$ state at zero field in Figure~\ref{fig:no2m-fukui} indeed confirms this, as measured by the distance between the \ce{N} and \ce{O} nuclei and the outermost contour.
  
        These results reinforce Klopman's observation\cite{book:Klopman1974} that the \ce{N} $sp^2$ lone pair is readily accessible to a soft electrophile, whereas the highest concentration of negative charge resides on the oxygen atoms, as reflected in the dominant resonance structures.
        He therefore concluded that charge-controlled reactions preferably occur at the oxygen sites, whereas orbital-controlled reactions favor the nitrogen site.
        In terms of the HSAB principles, these earlier studies, together with the present results, confirm the pronounced ambident nucleophilicity of  \ce{(NO2)-}, although the distinction between hard and soft preferences remains rather subtle.
        In our view, this subtlety arises because the high-spin $2p^3$ configuration of an isolated \ce{N} atom, which is responsible for its atomic hardness (Section~\ref{sec:global}), is substantially modified upon bond formation via the $sp^2$ hybridization process in \ce{(NO2)-}.
        A complementary discussion of charge-controlled reactivity based on MEPs that further strengthens this interpretation will be presented in Section~\ref{sec:charge-orbital-control}.

      \paragraph{Perpendicular-field ($\mathbf{B} \parallel \hat{\mathbf{x}}$) variation of ambident character of \ce{(NO2)-}.}

        Let us now examine the variation of the Fukui function $f^-(\mathbf{r})$ with magnetic field strength, starting with the perpendicular-field ($\mathbf{B} \parallel \hat{\mathbf{x}}$) case (Figure~\ref{fig:no2m-fukui-bx}).
        In the adiabatic approximation where one considers only the $M_S = 0$ state, the ambident character with a slight preference for a softer \ce{N} site persists with field strength, as is evident by the lack of any significant qualitative changes in the shape of $f^-(\mathbf{r})$ (top row, Figure~\ref{fig:no2m-fukui-bx}).
  
        However, when the $M_S = -1$ state is considered for field strengths between $0.07\,B_0$ and $0.20\,B_0$ where this state becomes the ground state, $f^-(\mathbf{r})$ adopts a drastically different shape (blue-shaded isosurfaces, second row, Figure~\ref{fig:no2m-fukui-bx}), despite remaining totally symmetric within $\mathcal{C}_{2v}(\mathcal{C}_s)$ as required.
        In particular, $f^-(\mathbf{r})$ now exhibits positive lobes that extend above and below the molecular plane, which suggests the possibility of non-planar electrophilic attack towards both the \ce{N} and \ce{O} sites.
  
        The origin of this behavior can be understood from the zero-field frontier MOs shown in Figure~\ref{fig:no2m-mos-zerofield}.
        Relative to the $M_S = 0$ state, the $M_S = -1$ state corresponds to a one-electron excitation from the $\alpha 11$ MO of $D[A_1]$ symmetry to the $\beta 12$ MO of $D[B_1]$ symmetry.
        As the latter orbital is now the HOMO, it directly influences the shape of $f^-(\mathbf{r})$.
        In fact, the in-plane node of this orbital at zero field is responsible for a qualitatively different in-plane contour plot of $f^-(\mathbf{r})$ with substantially reduced values in the molecular plane.
        Although the overall reactivity pattern remains ambident, the slight preference for in-plane attack at the \ce{N} site by a soft electrophile is largely diminished.
  
        When the $M_S = -2$ state becomes the ground state beyond $0.20\,B_0$, the Fukui function $f^-(\mathbf{r})$ expands dramatically (green-shaded isosurface, third row, Figure~\ref{fig:no2m-fukui-bx}): it becomes markedly more diffuse, as evidenced by both the reduced isovalue and the increased spatial extent of the isosurface, and adopts an overall toroidal shape that is likewise visible in the corresponding in-plane contour plot.
        This behavior suggests the presence of induced electron currents circulating perpendicular to the applied magnetic field (\textit{i.e.}, in the $yz$-plane)\cite{article:Sundholm2021,article:Pelloni2011,article:Irons2021a}.
        The pronounced diffuseness of $f^-(\mathbf{r})$ implies that regioselectivity and ambident character are no longer meaningful descriptors, since the nucleophilic region of the system is now spread over the entire molecule rather than localized at specific atomic sites.
        Accordingly, the notion of a preferred direction of attack for an incoming electrophile within the molecular plane also ceases to be relevant.

        The diffuseness of $f^-(\mathbf{r})$ in the $M_S = -2$ state can be traced to the nature of the HOMO, which is now the $\beta 13$ MO (Figure~\ref{fig:no2m-mos-zerofield}).
        Its isosurface reveals a significantly more diffuse character than those of the lower-lying MOs.
        This is consistent with the marked increase in the magnitude of its contribution to the electric dipole moment, $|\braket{\psi | \hat{\mu}_z | \psi}|$, which also explains the anomalous dipole variation of the $M_S = -2$ state shown in Figure~\ref{fig:no2m-e-sz-mu}.
        In fact, the highly diffuse character of the HOMO suggests that the electron occupying this orbital is less tightly bound to the nuclei in the system and therefore considerably more susceptible to magnetic-field-induced current circulation.

        In general, electron--nucleus and electron--electron interactions dominate at low field strengths up to \textit{ca.} $0.05\,B_0 \approx \SI{11800}{\tesla}$.
        As the field approaches $0.1\,B_0 \approx \SI{23500}{\tesla}$, however, the pronounced change in the shape of the Fukui function for \ce{(NO2)-} reflects a subtle shift in the balance between magnetic and electrostatic interactions.
        This regime marks the onset at which the external field becomes sufficiently strong to begin competing with the molecule’s internal electrostatic forces, which are eventually dominated at very high field strengths by virtue of the quadratic term in the interaction between the electrons and the external magnetic field (Equation~\ref{eq:Hmag}).
        A direct consequence of this is the step-wise unpairing of electrons, leading to abrupt changes in conceptual-DFT quantities such as electron densities and Fukui functions, and hence to discontinuous chemical behaviors that cannot be captured within the adiabatic approximation alone.

      \paragraph{In-plane-field ($\mathbf{B} \parallel \hat{\mathbf{y}}$) and parallel-field ($\mathbf{B} \parallel \hat{\mathbf{z}}$) variations of ambident character of \ce{(NO2)-}.}
        Let us now turn our attention to the in-plane-field case with $\mathbf{B} \parallel \hat{\mathbf{y}}$  (Figure~\ref{fig:no2m-fukui-by}).
        Within the adiabatic approximation, the $M_S = 0$ state remains the ground state up to $0.11\,B_0$.
        Over this range, \ce{(NO2)-} retains its in-plane nucleophilic character but exhibits a sharp reduction in reactivity around the \ce{N} region, thereby enhancing selectivity towards the \ce{O} ends.
        This evolution is evident from the red-shaded isosurface plots and the corresponding in-plane contours in the top row of Figure~\ref{fig:no2m-fukui-by}.

        \begin{figure*}[p]
          \centering
          \begin{subfigure}{\textwidth}
            \centering
            \includegraphics[width=\textwidth]{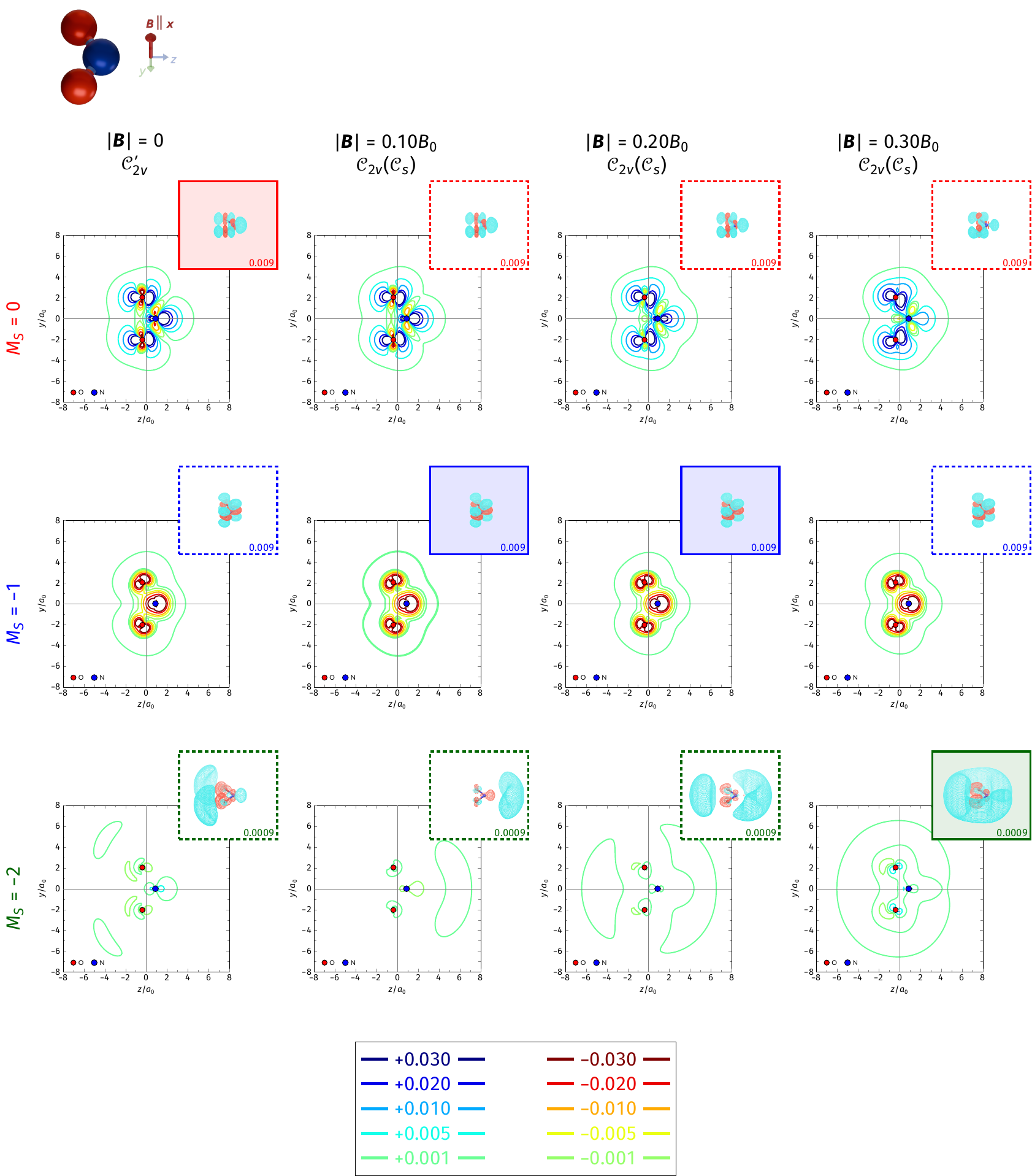}
            \caption{\ce{(NO2)-}, perpendicular field ($\mathbf{B} \parallel \hat{\mathbf{x}}$).}
            \label{fig:no2m-fukui-bx}
          \end{subfigure}
        \end{figure*}
        \begin{figure*}[p]
          \ContinuedFloat
          \begin{subfigure}{\textwidth}
            \centering
            \includegraphics[width=\textwidth]{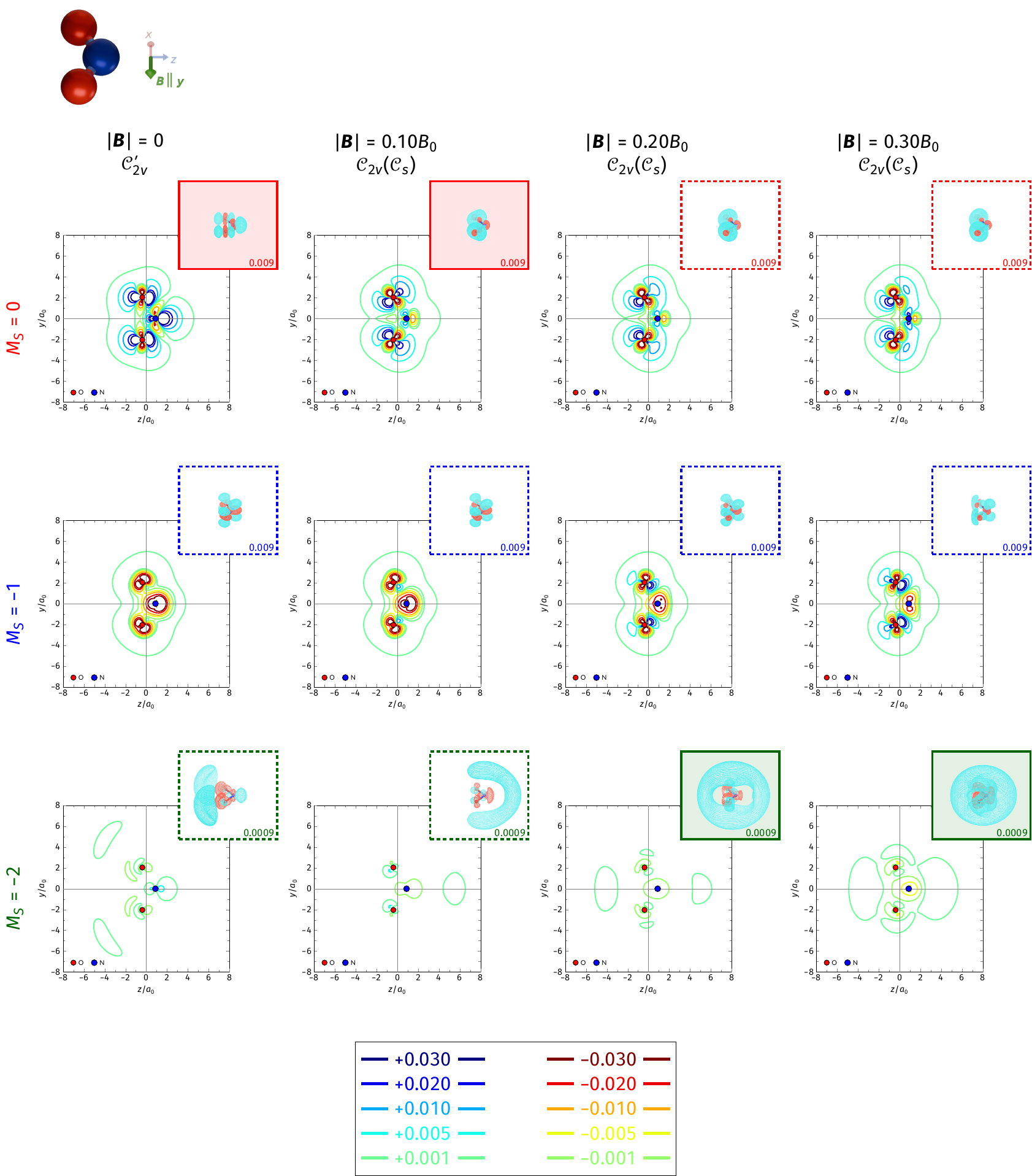}
            \caption{\ce{(NO2)-}, in-plane field ($\mathbf{B} \parallel \hat{\mathbf{y}}$).}
            \label{fig:no2m-fukui-by}
          \end{subfigure}
        \end{figure*}
        \begin{figure*}
          \ContinuedFloat
          \begin{subfigure}{\textwidth}
            \centering
            \includegraphics[width=\textwidth]{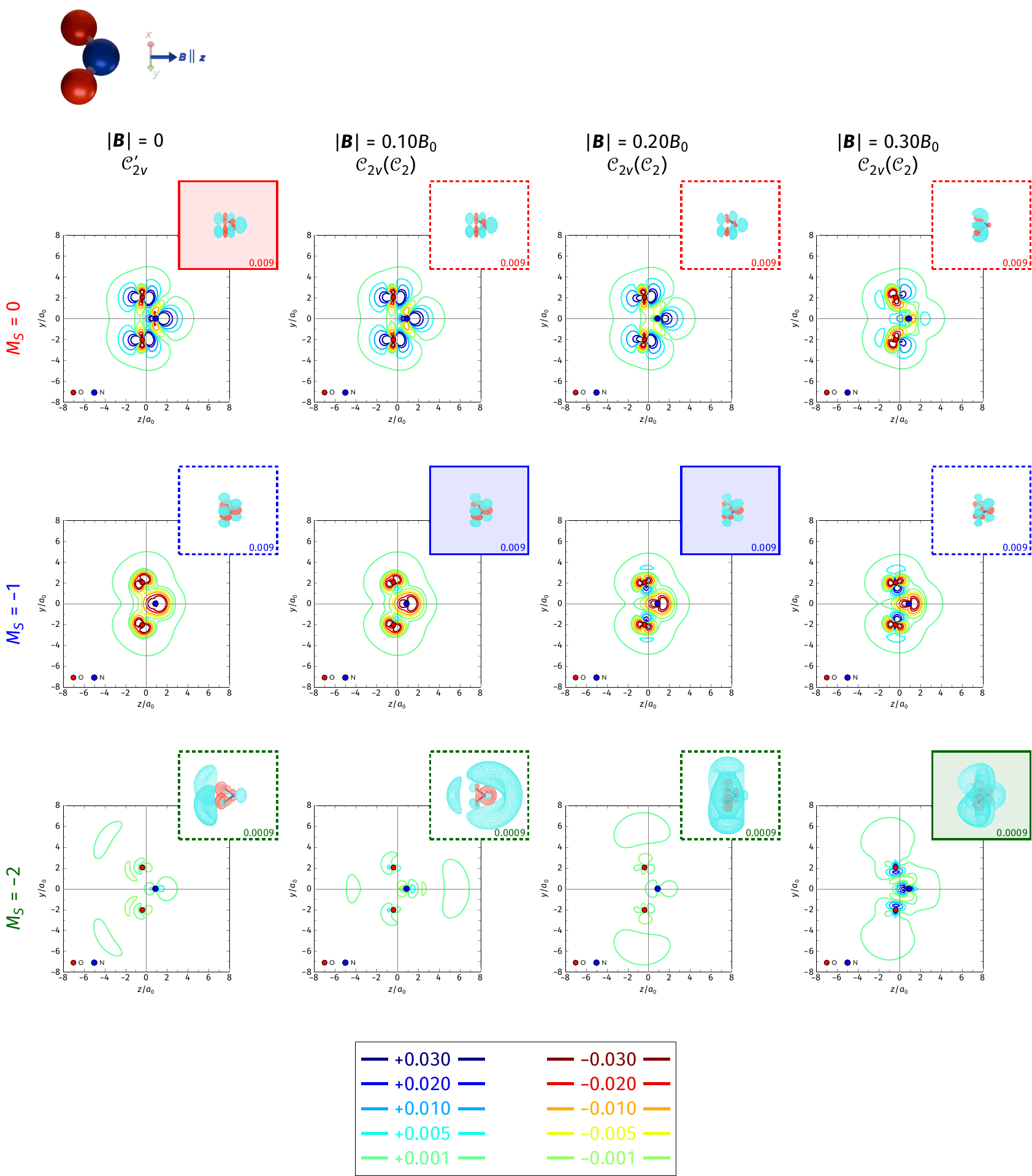}
            \caption{\ce{(NO2)-}, parallel field ($\mathbf{B} \parallel \hat{\mathbf{z}}$).}
            \label{fig:no2m-fukui-bz}
          \end{subfigure}
          \caption{%
            Isosurfaces and contour plots of the Fukui function $f^-(\mathbf{r})$ for \ce{(NO2)-}.
            The isovalue $|f^-(\mathbf{r})|$ used for each isosurface is indicated in the bottom-right corner.
            At each field strength, the isosurface of the ground $M_S$ state is highlighted by a shaded box with solid borders.
          }
          \label{fig:no2m-fukui}
        \end{figure*}
        \clearpage

        At higher field strengths, the $M_S = -1$ state becomes the ground state up to $0.15\,B_0$.
        In this regime, the Fukui function exhibits diminished in-plane reactivity together with enhanced nucleophilicity above and below the molecular plane, while also maintaining ambident behavior between the \ce{N} and \ce{O} centers.
        Beyond $0.15\,B_0$, the $M_S = -2$ state takes over as the ground state, and the Fukui function again becomes highly diffuse and indicative of the presence of induced electron currents circulating in the $xz$-plane perpendicular to the direction of the field (green-shaded isosurfaces, third row, Figure~\ref{fig:no2m-fukui-by}).

        The parallel-field case ($\mathbf{B} \parallel \hat{\mathbf{z}}$) (Figure~\ref{fig:no2m-fukui-bz}) exhibits behaviors similar to the previous two orientations.
        At low field strengths, the adiabatic $M_S = 0$ ground state displays in-plane ambident nucleophilicity with a slight preference for the \ce{N} site (red-shaded isosurface, top row, Figure~\ref{fig:no2m-fukui-bz}).
        At intermediate field strengths, where the $M_S = -1$ state becomes the ground state, the in-plane nucleophilicity weakens and is replaced by predominantly out-of-plane ambident nucleophilicity (blue-shaded isosurfaces, second row, Figure~\ref{fig:no2m-fukui-bz}).
        Finally, at high field strengths, the ground state is now the $M_S = -2$ state and the Fukui function adopts a diffuse toroidal shape perpendicular to the field, signaling the loss of regioselectivity (green-shaded isosurface, third row, Figure~\ref{fig:no2m-fukui-bz}).

        We conclude the discussion for \ce{(NO2)-} with an important cautionary remark.
        The highly diffuse Fukui functions observed for strongly negative $M_S$ states should be interpreted carefully.
        Since \ce{(NO2)-} is an anion, its low-lying excited states are expected to be substantially more diffuse than those of neutral systems, reflecting the weaker binding of the excess electron.
        In neutral molecules, by contrast, the lowest unoccupied MOs (LUMOs) generally remain relatively localized, with significant diffuseness emerging only for higher unoccupied MOs.
        Consequently, similarly diffuse Fukui functions are likely to appear in neutral systems only for sufficiently high-energy states associated with increasingly negative $M_S$ values.
        In all such cases, careful analysis of frontier MOs is essential for extracting chemically meaningful interpretations from diffuse Fukui functions.

      \paragraph{Zero-field ambident character of \ce{SCN-}.}
        Figure~\ref{fig:scnm-fukui} shows the variation with magnetic-field strength and orientation of the Fukui function $f^-(\mathbf{r})$ for the three $M_S$ states in \ce{SCN-}.
        As before, $f^-(\mathbf{r})$ is expected to be totally symmetric with respect to the full magnetic group $\mathcal{M}$ of the system.
        However, the non-Abelianity of $\mathcal{M} = \mathcal{C}'_{\infty v}$ in the zero-field case requires special care when computing electron densities associated with degenerate wavefunctions, as outlined in Section~\ref{sec:degeneracy}.
        When this procedure was followed, the finite-difference $f^-(\mathbf{r})$ with $\Delta N_{\mathrm{e}} = -1$ (Equation~\ref{eq:fm-fd}) correctly retained cylindrical symmetry for all three $M_S$ states, even when the Kohn--Sham determinants of the $N_{\mathrm{e}}$- or $(N_{\mathrm{e}} - 1)$-systems possessed doubly degenerate $D[\Pi]$ symmetry.
        As a result, the pure-state-density-symmetry breaking previously observed in the finite-difference treatment with $\Delta N_{\mathrm{e}} = -1$\cite{article:Langenaeker1996} was entirely avoided.

        The zero-field Fukui function for the $M_S = 0$ state (red-shaded isosurface, top row, left-most column, Figure~\ref{fig:scnm-fukui}) clearly demonstrates the ambident nature of \ce{SCN-}, while also revealing a much more pronounced contrast in softness between the terminal \ce{S} and \ce{N} atoms compared to that observed in \ce{(NO2)-} between \ce{O} and \ce{N}.
        As expected from the theoretical and experimental data discussed in Section~\ref{sec:intro}, the \ce{S} terminus is much softer than the \ce{N} terminus.
        This qualitative pattern is also observed in the $M_S = -1$ and $M_S = -2$ states at zero field.

        \begin{figure*}[p]
          \centering
          \begin{subfigure}{\textwidth}
            \centering
            \includegraphics[width=\textwidth]{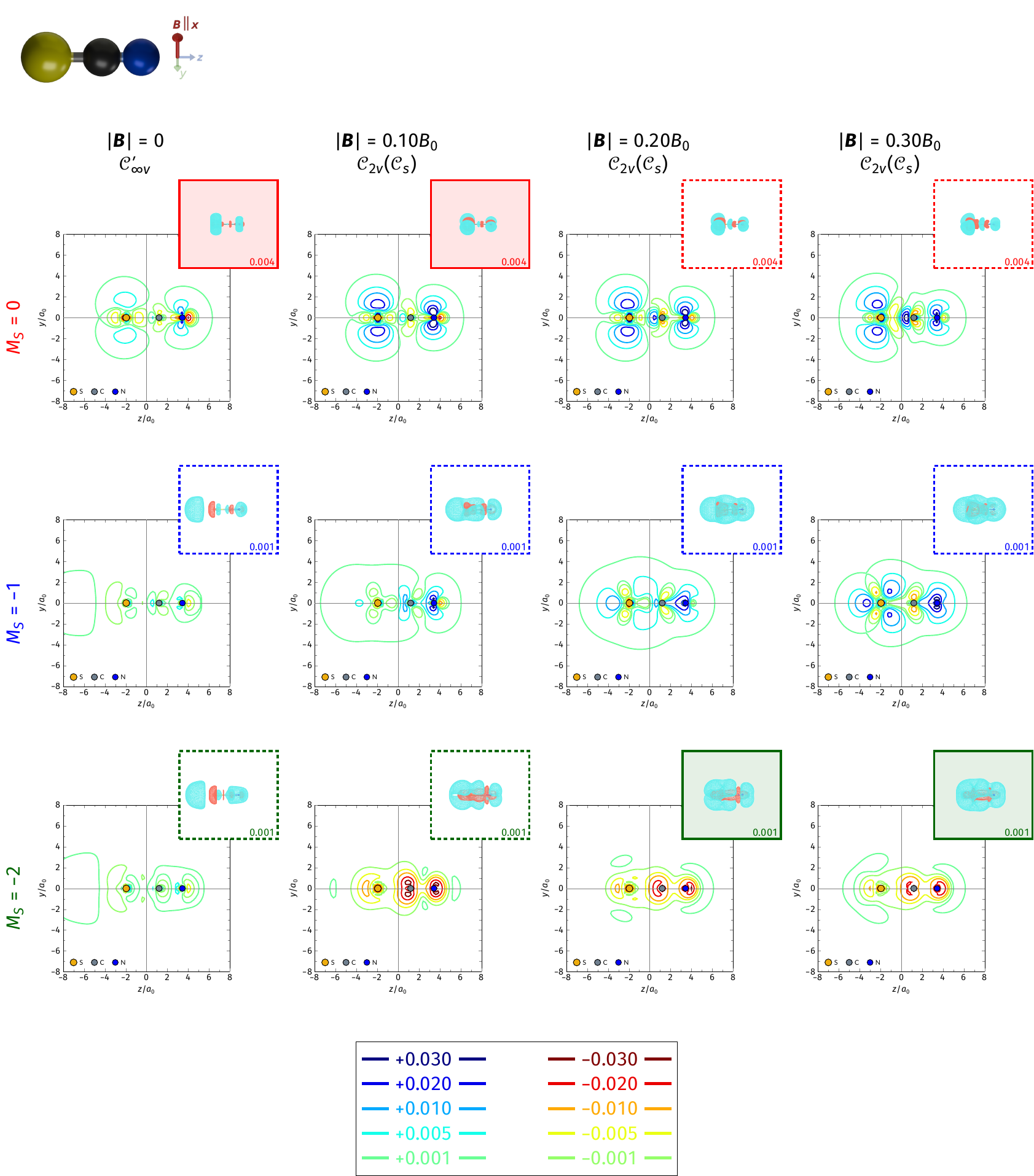}
            \caption{\ce{SCN-}, $\mathbf{B} \parallel \hat{\mathbf{x}}$.}
            \label{fig:scnm-fukui-bx}
          \end{subfigure}
        \end{figure*}
        \begin{figure*}[p]
          \ContinuedFloat
          \begin{subfigure}{\textwidth}
            \centering
            \includegraphics[width=\textwidth]{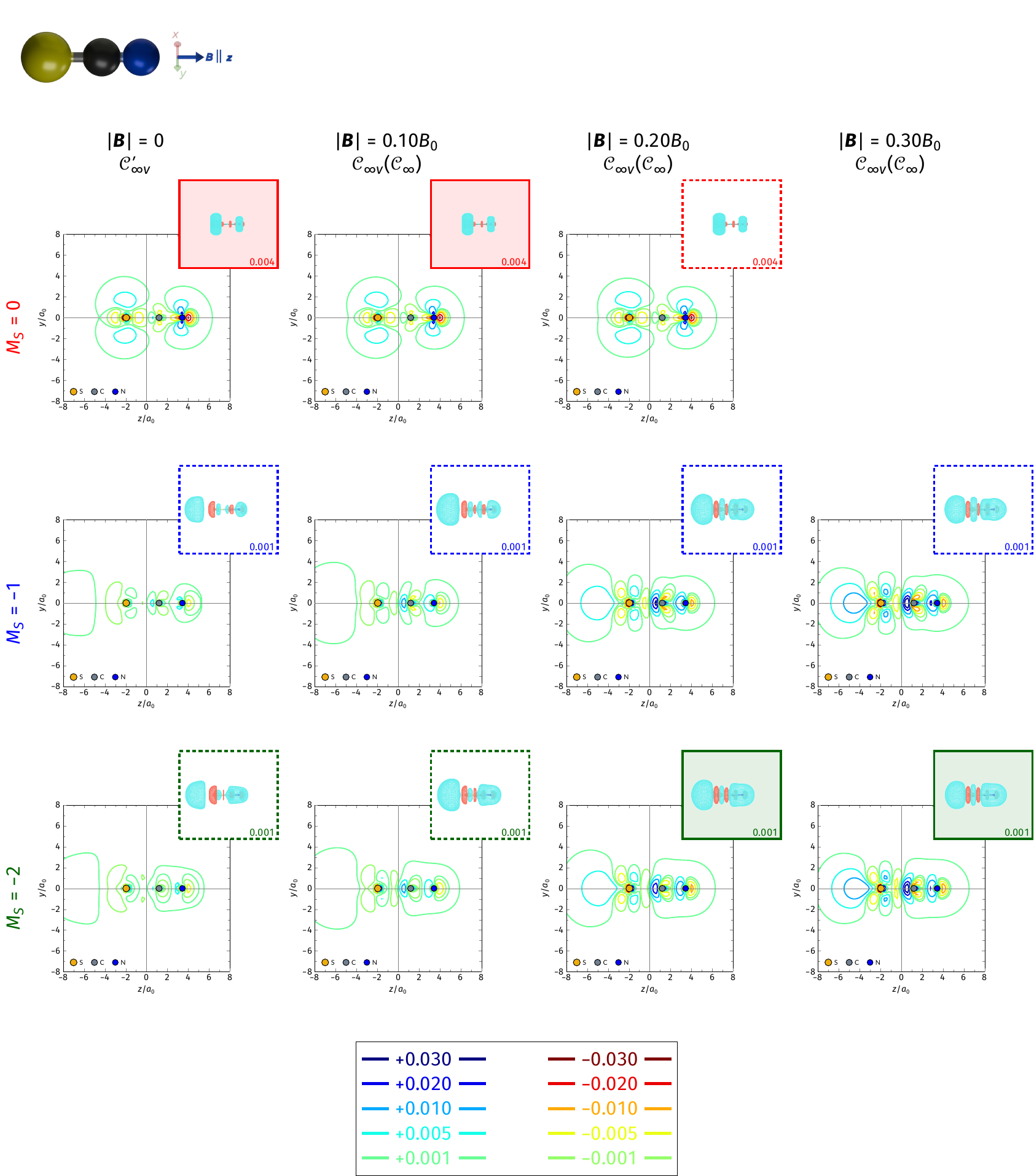}
            \caption{\ce{SCN-}, $\mathbf{B} \parallel \hat{\mathbf{z}}$.}
            \label{fig:scnm-fukui-bz}
          \end{subfigure}
          \caption{%
            Isosurfaces and contour plots of the Fukui function $f^-(\mathbf{r})$ for \ce{SCN-}.
            The isovalue $|f^-(\mathbf{r})|$ used for each isosurface is indicated in the bottom-right corner.
            At each field strength, the isosurface of the ground $M_S$ state is highlighted by a shaded box with solid borders.
          }
          \label{fig:scnm-fukui}
        \end{figure*}

      \paragraph{Field variations of ambident character of \ce{SCN-}.}
        Applying a magnetic field in any direction lifts all spatial degeneracies\cite{article:Pausch2021}.
        Consequently, the relevant unitary halving subgroups, namely $\mathcal{G} = \mathcal{C}_s$ for $\mathbf{B} \parallel \hat{\mathbf{x}}$ and $\mathcal{C}_{\infty}$ for $\mathbf{B} \parallel \hat{\mathbf{z}}$, are Abelian.
        As in the case of \ce{(NO2)-}, the corresponding magnetic groups $\mathcal{M}$ contain only irreducible corepresentations of the first kind, which are therefore non-degenerate.
        As a result, symmetry-conserved Kohn--Sham determinants for both $N_{\mathrm{e}}$- and $(N_{\mathrm{e}} - 1)$-systems must remain non-degenerate, implying that the associated pure-state electron densities, and hence the finite-difference Fukui functions, must be totally symmetric in $\mathcal{M}$ without requiring the degeneracy treatment in Section~\ref{sec:degeneracy}.
        Figure~\ref{fig:scnm-fukui} illustrates this behavior, while Figure~S2 in the Supplementary Material further demonstrates the lifting of spatial degeneracies in $\mathcal{C}_{\infty v}(\mathcal{C}_s)$ even though this magnetic group is isomorphic to a non-Abelian group.

        A perpendicular field ($\mathbf{B} \parallel \hat{\mathbf{x}}$) removes the $C_{\infty}$ axis of \ce{SCN-}, causing the Fukui function $f^-(\mathbf{r})$ to lose its zero-field cylindrical symmetry and instead adopt $\mathcal{C}_{2v}(\mathcal{C}_s)$ symmetry.
        At weak fields, where the $M_S = 0$ state remains the ground state, the ambident character of \ce{SCN-} is qualitatively similar to that at zero field, with \ce{S} remaining the softer nucleophilic site.
        However, the symmetry lowering from $\mathcal{C}'_{\infty v}$ to $\mathcal{C}_{2v}(\mathcal{C}_s)$ confines the nucleophilic regions to the molecular plane perpendicular to the magnetic field (\textit{i.e.}, the $yz$-plane), rather than distributing them cylindrically around the molecular axis as in the field-free case (red-shaded isosurface, top row, second column, Figure~\ref{fig:scnm-fukui-bx}).
        As the field becomes stronger, the $M_S = -1$ and subsequently the $M_S = -2$ states become energetically relevant, leading eventually to a more diffuse $f^-(\mathbf{r})$ (green-shaded isosurfaces, third row, Figure~\ref{fig:scnm-fukui-bx}).
        However, unlike in \ce{(NO2)-}, regioselectivity remains meaningful because $f^-(\mathbf{r})$ stays relatively localized, once again showing a preference for \ce{S} as the softer nucleophilic site.

        A parallel field ($\mathbf{B} \parallel \hat{\mathbf{z}}$) preserves the $C_{\infty}$ axis of \ce{SCN-} and hence the cylindrical symmetry of $f^-(\mathbf{r})$.
        At low field strengths, the $M_S = 0$ state retains the ambident character of \ce{SCN-}, with a clear preference for nucleophilic attack at the softer \ce{S} terminus (red-shaded isosurfaces, top row, Figure~\ref{fig:scnm-fukui-bz}).
        From moderate to high field strengths, the $M_S = -1$ and eventually the $M_S = -2$ states become the ground state, resulting in more diffuse Fukui functions.
        Although the ambident character persists, the preference for attack at \ce{S} becomes less pronounced, indicating a substantial modulation, though not a reversal, of the hard--soft contrast between the \ce{N} and \ce{S} sites.

        While this trend should be interpreted cautiously, it is qualitatively consistent with the atomic-level prediction that the hardness difference between \ce{N} and \ce{S} decreases at high magnetic fields. As in \ce{(NO2)-}, the increased diffuseness of the Fukui function can again be traced to the corresponding increased diffuseness of the relevant HOMO (Figure~\ref{fig:scnm-mos-zerofield}), as indicated by its increased contribution to the electric dipole moment.

      \paragraph{Comparison between \ce{(NO2)-} and \ce{SCN-}.}
        Overall, the effects of an external magnetic field on the ambident nucleophilicity of \ce{SCN-} are easier to interpret and rationalize within the HSAB principle than those of \ce{(NO2)-}.
        While both molecules are ambident, the hard--soft contrast between \ce{N} and \ce{S} is much more pronounced than that between \ce{N} and \ce{O} at zero field and remains appreciable even at increased field strengths.
        The contrast nevertheless decreases more significantly in the parallel-field case.

        The distinction between perpendicular and parallel field effects for \ce{(NO2)-} and \ce{SCN-} is less clear-cut, due primarily to the involvement of multiple $M_S$ states.
        In our view, the most striking feature is the \ce{SCN-} dipole reversal observed in the parallel-field case following the state crossing (Section~\ref{sec:dipole-moment-variations}), whereas within the adiabatic picture, such a reversal appears only for perpendicular fields, as previously found for the hydrogen halides\cite{article:Irons2022} and rationalized using symmetry\cite{article:Huynh2024}.
        The dipole moment reversal in \ce{HX} for a perpendicular field in the adiabatic approach\cite{article:Irons2022} is thereby contrasted with its reversal in \ce{SCN-} for a parallel field due to state crossing.

    \subsection{Charge versus Orbital Control}
    \label{sec:charge-orbital-control}

\begin{figure*}
        \centering
        \includegraphics[width=\textwidth]{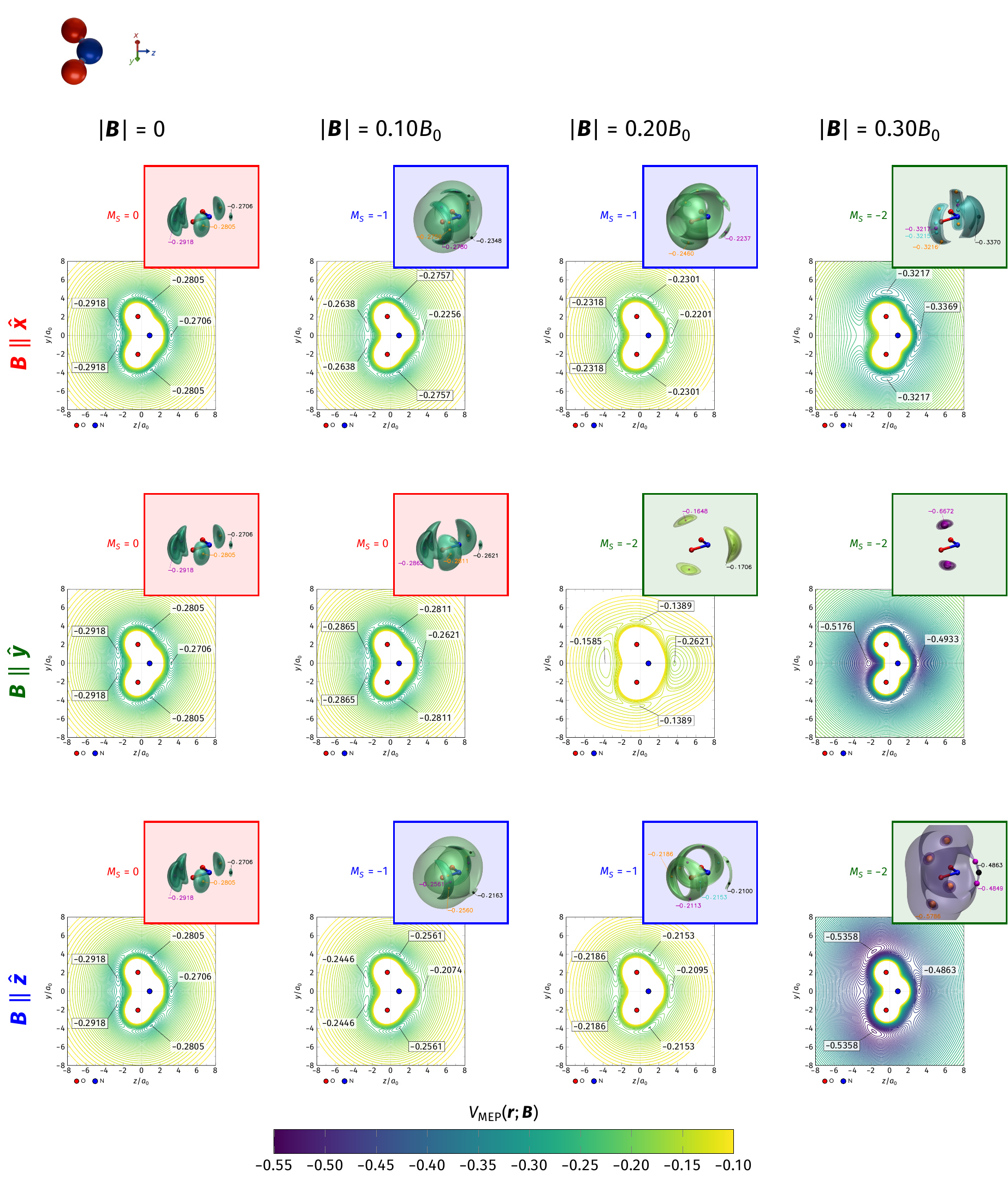}
        \caption{%
          Three-dimensional isosurfaces and in-plane contour plots of the molecular electrostatic potential, $V_{\mathrm{MEP}}(\mathbf{r}; \mathbf{B})$, for the ground spin state of \ce{(NO2)-} at different magnetic-field strengths and orientations.
          In the contour plots, in-plane local minima are labeled by their $V_{\mathrm{MEP}}(\mathbf{r}; \mathbf{B})$ values in $\si{\hartree}$, with the most negative ones shown in solid white boxes.
          In the isosurface plots, local minima are marked by colored spheres and annotated with their corresponding $V_{\mathrm{MEP}}(\mathbf{r}; \mathbf{B})$ values in $\si{\hartree}$; spheres of the same color denote symmetry-equivalent minima.
          All isosurfaces and contours are colored based on their values.
        }
        \label{fig:no2m-mep}
      \end{figure*}

      Thus far, the reactivity of \ce{(NO2)-} and \ce{SCN-} has been analyzed through the lens of Fukui functions, which primarily probe reactivity under \textit{orbital control}.
      In this final Section, we present a preliminary investigation of how an external magnetic field influences the electrostatic component of reactivity, which dominates under \textit{charge control}.
      To this end, we examine the variation of the molecular electrostatic potential $V_{\mathrm{MEP}}(\mathbf{r}; \mathbf{B})$ (Equation~\ref{eq:mep}) with magnetic field strength and orientation, for which the dipole moment results in Section~\ref{sec:dipole-moment-variations} provide a precursory rationale.
      We show this behavior for \ce{(NO2)-} in Figure~\ref{fig:no2m-mep}: given the well-known interplay between orbital and charge control in determining the ambident nucleophilicity of \ce{(NO2)-}, as emphasized by Klopman in Ref.~\citenum{book:Klopman1974}, we focus only on this anion, where the competition between these two reactivity paradigms is especially pronounced.

      At zero field, all local minima of $V_{\mathrm{MEP}}(\mathbf{r}; \mathbf{B})$ lie within the molecular plane (first column, any row, Figure~\ref{fig:no2m-mep}).
      The minima associated with the \ce{O} atoms are deeper than that adjacent to the \ce{N} atom by approximately $\SI{0.01}{\hartree} \approx \SI{6}{\kilo\cal\per\mole}$ to $\SI{0.02}{\hartree} \approx \SI{12}{\kilo\cal\per\mole}$, indicating a slight preference for the oxygen regions in charge-controlled electrophilic attacks.
      This is consistent with the orientation of the zero-field dipole moment discussed in Section~\ref{sec:dipole-moment-variations}.
      Together with the slight preference for a soft electrophilic attack on nitrogen revealed by the Fukui functions (Section~\ref{sec:ambidentcharacter}), these results provide a more nuanced picture of the ambident character of \ce{(NO2)-}: the \ce{N} and \ce{O} centers are almost equally reactive under both orbital and charge control, with only subtle preferences in accordance with Klopman's statement\cite{book:Klopman1974}.

      The planar arrangement of the local minima of $V_{\mathrm{MEP}}(\mathbf{r}; \mathbf{B})$ turns out to be a distinctive feature of the $M_S = 0$ state.
      As the ground state switches to the $M_S = -1$ and $M_S = -2$ states in all three field orientations, the local minima of $V_{\mathrm{MEP}}(\mathbf{r}; \mathbf{B})$ move above and below the molecular plane in a symmetrical manner governed by the underlying magnetic group, since $V_{\mathrm{MEP}}(\mathbf{r}; \mathbf{B})$, like the electron density, Fukui function, and electric dipole moment, is also a real-valued quantity.
      This might be traced back to the change in symmetry of the frontier MOs (Figure~\ref{fig:no2m-mos-zerofield}).
      Although the differences with the in-plane values remain relatively small in nearly all cases, the distances between the out-of-plane minima and the molecular plane can be substantial: for instance, the minima at $-0.2348$ in the $M_S = -1$ state of \ce{(NO2)-} for the $\mathbf{B} = 0.10\,B_0 \hat{\mathbf{x}}$ field are located at \SI{1.22}{\angstrom} above and below the molecular plane (top row, second column, Figure~\ref{fig:no2m-mep}).
      Consequently, sufficiently strong magnetic fields can redirect charge-controlled electrophilic attack away from the molecular plane, favoring reaction pathways that lead to non-planar products.
      A similar behavior is also observed for orbital-controlled reactivity: in the $M_S = -1$ and $M_S = -2$ states, the Fukui functions of \ce{(NO2)-} develop pronounced lobes above and below the molecular plane (Figure~\ref{fig:no2m-fukui}), indicating that strong magnetic fields can drive both charge- and orbital-controlled electrophilic attack into non-planar regions.

      At certain field strengths and orientations (\textit{e.g.}, $\mathbf{B} = 0.30\,B_0 \hat{\mathbf{x}}$ and $\mathbf{B} = 0.20\,B_0 \hat{\mathbf{y}}$), the deepest local minima are found in the vicinity of the \ce{N} site rather than the \ce{O} sites, indicating a reversal of the established charged-controlled regioselectivity.
      However, a systematic understanding of this behavior has yet to be elucidated.
      For instance, when the field is increased from $\mathbf{B} = 0.20\,B_0 \hat{\mathbf{y}}$ to $\mathbf{B} = 0.30\,B_0 \hat{\mathbf{y}}$, the ground state remains $M_S = -2$, yet the local minimum adjacent to \ce{N} disappears while the local minima in the \ce{O} regions deepen significantly.
      One possible reason for this behavior can be attributed to the diffuseness of the electron density in the $M_S = -2$ state and hence its susceptibility to current circulation induced by the external magnetic field.
      The resulting redistribution of electronic charge can then produce dramatic changes in the topology of the MEP, including the appearance, disappearance, and reordering of its local minima.

\section{Conclusion}
\label{sec:conclusion}

  The present study extends our previous investigations of magnetic-field effects on chemical reactivity descriptors, from atomic properties such as electronegativity and hardness to molecular systems ranging from small $\sigma$-bonded species, where magnetic-field-induced polarity changes play a central role, to $\pi$-bonded systems analyzed through the Fukui functions.
  By interpreting the results of current-DFT calculations through the lens of conceptual DFT, we exploit the ambident nucleophilicity and the associated regioselectivity of \ce{(NO2)-} and \ce{SCN-} as the ideal playground to study hardness-driven reactivity, particularly through the local softness towards electrophiles as reflected by the Fukui function $f^{-}(\mathbf{r})$.

  The main conclusion emerging from this work is that external magnetic fields can systematically modulate electronic structure and reactivity, both adiabatically and via spin-state crossing.
  The magnetic-field-induced adiabatic evolution of dipole moments and molecular polarity observed in earlier studies persists in these systems and is accompanied by substantial changes in the Fukui function.
  Extending the conventional adiabatic treatment to include higher spin states, as previously done for atoms, reveals significant changes in both global molecular polarity and local reactivity as reflected by the symmetry, shape, and diffuseness of the Fukui function.
  These changes can be traced to the structure of the frontier MOs of the systems and their evolution with magnetic field strength and direction.

  Despite these magnetic-field-induced modulations, the ambident character of both \ce{(NO2)-} and \ce{SCN-} is largely preserved, as is the preference for attack by soft electrophiles at the \ce{S} terminus of \ce{SCN-}.
  Particularly striking, however, are the pronounced changes in shape of the Fukui function already at field strengths as low as $0.1\,B_0$.
  These changes can be linked to alterations in the symmetry and character of the HOMO, which are a direct consequence of the change in the ground spin state facilitated by the magnetic field favoring unpaired electrons through the spin-Zeeman effect.
  In sufficiently high spin states where the HOMO becomes significantly diffuse, the electron density of this orbital becomes considerably susceptible to magnetically induced current circulation and therefore substantially sensitive to the applied field.
  As a result, key electronic-structure descriptors, including dipole moments and Fukui functions, exhibit dramatic variations with field strength.
  This behavior reflects a delicate balance between magnetic and electrostatic interactions in a regime where the external field becomes sufficiently strong to compete with the molecule's internal electronic forces.

  At higher field strengths, the Fukui function becomes increasingly diffuse and eventually expands preferentially in directions perpendicular to the applied field.
  This remarkable behavior suggests that the magnetic field has become the dominant influence on the electronic structure and reactivity, possibly accompanied by an increased importance of quadratic magnetic-field effects for highly diffuse orbitals.
  Under these conditions, the conventional notion of regioselectivity becomes progressively less meaningful, as the reactive regions cease to be localized at specific atomic sites and instead spread over large portions of the molecular framework.

  Ultimately, the high spin states stabilized by strong magnetic fields can be viewed as adiabatically linked to highly excited states at zero field, in which the valence electrons are much less tightly bound to the nuclei.
  It is therefore instructive to compare the high-field behaviors uncovered in the present work with the chemical reactivity of such highly excited states.
  Although this connection has not yet been addressed in detail within conceptual DFT, there is emerging evidence suggesting similarities between the two regimes\cite{manuscript:Wang2026}.

  This perspective is also consistent with discussions by Schmelcher\cite{article:Schmelcher2012} and Stopkowicz\cite{article:Stopkowicz2017} concerning the experimental realization of strong-field phenomena under laboratory conditions.
  They pointed out that Rydberg molecules constitute an ideal platform for probing the influence of strong magnetic fields on molecules in the laboratory, since the greatly weakened Coulomb interactions in highly excited Rydberg states\cite{book:Gallagher1994} allow the regime in which magnetic and Coulomb forces become comparable (as is the case in the present study) to occur at much lower field strengths.
  Consequently, as Schmelcher noted in his \textit{Science} Perspective\cite{article:Schmelcher2012}, ``standard laboratory magnetic fields of a few Teslas are sufficient to introduce novel magnetic properties into highly excited Rydberg states.''
  This observation suggests a promising route for connecting the strong-field phenomena discussed here with experimentally accessible molecular systems, which will form the basis for a future study.

  \section*{Acknowledgments}
    BCH thanks New College, University of Oxford for funding from the Oglander Fellowship.
    MW-T and AMW-T acknowledge financial support from the European Research Council under the European Union's H2020 research and innovation program/ERC Consolidator Grant topDFT [grant number 772259].
    FDP and PG are thankful to the Vrije Universiteit Brussel for a long term grant of a Strategic Research Program.
    The authors also thank Dr Bin Wang (Vrije Universiteit Brussel) for providing the preliminary MEP calculations and data for comparing the operational and theoretical expressions for the symmetrized electron density (Equations~\ref{eq:rhobar}, \ref{eq:rhobar_totsymproj} and \ref{eq:rhobar_degset}).

  \section*{Supplementary Material}
    The Supplementary Material contains plots showing the variation with magnetic fields strength of the energies of the frontier $\beta$-MOs in \ce{(NO2)-} and \ce{SCN-} for the field orientations and spin states considered in this work, and character tables for the magnetic groups involved in this work.

  \section*{Author Declarations}
    \paragraph{Conflict of interest.}
      The authors have no conflicts to disclose.

    \paragraph{Data availability.}
      The data that support the findings of this study are available from the corresponding author upon reasonable request.

  \appendix

\section{Electron Densities of Degenerate Wavefunctions}
\label{app:denprojection}

  \subsection{General Considerations}

    Let us consider an $N_{\mathrm{e}}$-electron wavefunction $\Psi^{(1)}(\mathbf{x}_1, \ldots, \mathbf{x}_{N_{\mathrm{e}}}) \equiv \Psi^{(1)}(\mathbf{x}^{N_{\mathrm{e}}})$ in a molecular system with (unitary or magnetic) symmetry group $\mathcal{H}$.
    Suppose that $\Psi^{(1)}$ is non-degenerate, which means that $\Psi^{(1)}$ transforms according to a one-dimensional irreducible representation $\Gamma^{(1)}$ of $\mathcal{H}$ if $\mathcal{H}$ is unitary, or a one-dimensional irreducible corepresentation $D[\Gamma^{(1)}]$ of $\mathcal{H}$ induced by a one-dimensional irreducible representation $\Gamma^{(1)}$ of its unitary halving subgroup if $\mathcal{H}$ is magnetic.
    For brevity, we shall denote both simply as $\Gamma^{(1)}$.
    In any case, the character function for $\Gamma^{(1)}$, which might be complex-valued, satisfies
    \begin{equation}
      \hat{h} \Psi^{(1)}(\mathbf{x}^{N_{\mathrm{e}}})
      = \chi^{\Gamma^{(1)}}(h) \Psi^{(1)}(\mathbf{x}^{N_{\mathrm{e}}}).
      \label{eq:1dchar-eigenvalue}
    \end{equation}
    and
    \begin{equation}
      | \chi^{\Gamma^{(1)}}(h) | = 1 \quad \textrm{for all $h \in \mathcal{H}$},
      \label{eq:1dchar}
    \end{equation}
    We note that Equation~\ref{eq:1dchar} holds even if $\mathcal{H}$ is a magnetic group and $\chi^{\Gamma^{(1)}}(a)$ is not invariant under unitary basis transformations for any antiunitary operation $a \in \mathcal{H}$, since such transformations only introduce a phase factor to $\chi^{\Gamma^{(1)}}(a)$\cite{article:Uhlmann2016}.
  
    The pure-state electron density corresponding to $\Psi^{(1)}$ has the form
    \begin{multline}
      \rho(\mathbf{r}) = N_{\mathrm{e}} \int
      \Psi^{(1)}(%
      \mathbf{r} s,
      \mathbf{x}_{2}
      \ldots,
      \mathbf{x}_{N_{\mathrm{e}}}
      )^*
      \ \Psi^{(1)}(%
      \mathbf{r} s,
      \mathbf{x}_{2}
      \ldots,
      \mathbf{x}_{N_{\mathrm{e}}}
      )
      \\ \D s
      \D \mathbf{x}_{2}
      \cdots
      \D \mathbf{x}_{N_{\mathrm{e}}}.
    \end{multline}
    Let us consider its transformation under the action of $\mathcal{H}$:
    \begin{align}
      \hat{h} \rho(\mathbf{r})
      &= \begin{multlined}[t]
        N_{\mathrm{e}} \int
          \hat{h} \Psi^{(1)}(%
          \mathbf{r} s,
          \mathbf{x}_{2}
          \ldots,
          \mathbf{x}_{N_{\mathrm{e}}}
          )^*
          \\ \hat{h} \Psi^{(1)}(%
          \mathbf{r} s,
          \mathbf{x}_{2}
          \ldots,
          \mathbf{x}_{N_{\mathrm{e}}}
          )
          \ \D s
          \D \mathbf{x}_{2}
          \cdots
          \D \mathbf{x}_{N_{\mathrm{e}}}
        \end{multlined}
      \nonumber \\
      &= \begin{multlined}[t]
        N_{\mathrm{e}} \int
          \chi^{\Gamma^{(1)}}(h)^* \Psi^{(1)}(%
          \mathbf{r} s,
          \mathbf{x}_{2}
          \ldots,
          \mathbf{x}_{N_{\mathrm{e}}}
          )^*
          \\ \chi^{\Gamma^{(1)}}(h) \Psi^{(1)}(%
          \mathbf{r} s,
          \mathbf{x}_{2}
          \ldots,
          \mathbf{x}_{N_{\mathrm{e}}}
          )
          \ \D s
          \D \mathbf{x}_{2}
          \cdots
          \D \mathbf{x}_{N_{\mathrm{e}}}
        \end{multlined}
      \nonumber \\
      &= \begin{multlined}[t]
        | \chi^{\Gamma^{(1)}}(h) |^2 N_{\mathrm{e}} \int
          \Psi^{(1)}(%
          \mathbf{r} s,
          \mathbf{x}_{2}
          \ldots,
          \mathbf{x}_{N_{\mathrm{e}}}
          )^*
          \\ \Psi^{(1)}(%
          \mathbf{r} s,
          \mathbf{x}_{2}
          \ldots,
          \mathbf{x}_{N_{\mathrm{e}}}
          )
          \ \D s
          \D \mathbf{x}_{2}
          \cdots
          \D \mathbf{x}_{N_{\mathrm{e}}}
        \end{multlined}
      \nonumber \\
      &= \rho(\mathbf{r}),
    \end{align}
    where the last equality follows from Equation~\ref{eq:1dchar}.
    The above relation holds for all $h \in \mathcal{H}$, establishing that $\rho(\mathbf{r})$ is indeed totally symmetric under the action of $\mathcal{H}$.

    Let us now consider a set of $d$-fold degenerate $N_{\mathrm{e}}$-electron wavefunctions
    \begin{equation}
      \Xi = \{
      \Psi^{(d)}_i(\mathbf{x}^{N_{\mathrm{e}}})
      \mid
      i = 1, \ldots, d
      \}, \quad d > 1,
      \label{eq:basis-d}
    \end{equation}
    that spans a $d$-dimensional \emph{irreducible} (co)representation $\Gamma^{(d)}$ of $\mathcal{H}$.
    The transformation of any wavefunction in this set under the action of $\mathcal{H}$ is given by
    \begin{equation}
      \hat{h} \Psi^{(d)}_i(\mathbf{x}^{N_{\mathrm{e}}})
      = \sum_{j=1}^{d}
      \Psi^{(d)}_j(\mathbf{x}^{N_{\mathrm{e}}})
      D_{ji}^{\Gamma^{(d)}}(h),
      \quad d > 1,
      \label{eq:deg-repmat}
    \end{equation}
    where $\mathbf{D}^{\Gamma^{(d)}}(h)$ is the (co)representation matrix of $h$ in the basis $\Xi$.
    The pure-state electron densities corresponding to the wavefunctions in this set are given by
    \begin{multline}
      \rho_i(\mathbf{r}) = N_{\mathrm{e}} \int
      \Psi_i^{(d)}(%
      \mathbf{r} s,
      \mathbf{x}_{2}
      \ldots,
      \mathbf{x}_{N_{\mathrm{e}}}
      )^*
      \\ \Psi_i^{(d)}(%
      \mathbf{r} s,
      \mathbf{x}_{2}
      \ldots,
      \mathbf{x}_{N_{\mathrm{e}}}
      )
      \ \D s
      \D \mathbf{x}_{2}
      \cdots
      \D \mathbf{x}_{N_{\mathrm{e}}}.
      \label{eq:rho_i}
    \end{multline}
    We shall also define the \emph{transition densities} between wavefunctions in this set as
    \begin{multline}
      \rho_{ij}(\mathbf{r}) = N_{\mathrm{e}} \int
      \Psi_i^{(d)}(%
      \mathbf{r} s,
      \mathbf{x}_{2}
      \ldots,
      \mathbf{x}_{N_{\mathrm{e}}}
      )^*
      \\ \Psi_j^{(d)}(%
      \mathbf{r} s,
      \mathbf{x}_{2}
      \ldots,
      \mathbf{x}_{N_{\mathrm{e}}}
      )
      \ \D s
      \D \mathbf{x}_{2}
      \cdots
      \D \mathbf{x}_{N_{\mathrm{e}}},
      \label{eq:transden}
    \end{multline}
    where $\rho_{ii}(\mathbf{r}) = \rho_{i}(\mathbf{r})$.
    The symmetry transformations of the pure-state densities are then given by
    \begin{align}
      \hat{h} \rho_i(\mathbf{r})
      &= \begin{multlined}[t]
        N_{\mathrm{e}} \int
          \hat{h} \Psi_i^{(d)}(%
          \mathbf{r} s,
          \mathbf{x}_{2}
          \ldots,
          \mathbf{x}_{N_{\mathrm{e}}}
          )^*
          \\ \hat{h} \Psi_i^{(d)}(%
          \mathbf{r} s,
          \mathbf{x}_{2}
          \ldots,
          \mathbf{x}_{N_{\mathrm{e}}}
          )
          \ \D s
          \D \mathbf{x}_{2}
          \cdots
          \D \mathbf{x}_{N_{\mathrm{e}}}
        \end{multlined}
      \nonumber \\
      &= \begin{multlined}[t]
        N_{\mathrm{e}}
          \sum_{jj'}^{d}
          D_{ji}^{\Gamma^{(d)}}(h)^*
          D_{j'i}^{\Gamma^{(d)}}(h)
          \\ \int
          \Psi_j^{(d)}(%
          \mathbf{r} s,
          \mathbf{x}_{2}
          \ldots,
          \mathbf{x}_{N_{\mathrm{e}}}
          )^*
          \ \Psi_{j'}^{(d)}(%
          \mathbf{r} s,
          \mathbf{x}_{2}
          \ldots,
          \mathbf{x}_{N_{\mathrm{e}}}
          )
          \\ \D s
          \D \mathbf{x}_{2}
          \cdots
          \D \mathbf{x}_{N_{\mathrm{e}}}
        \end{multlined}
      \nonumber \\
      &= \sum_{jj'}^{d}
      \rho_{jj'}(\mathbf{r})
      D_{ji}^{\Gamma^{(d)}}(h)^*
      D_{j'i}^{\Gamma^{(d)}}(h)
      \label{eq:g_rho_i}
      \\
      &\ne \rho_i(\mathbf{r}), \nonumber
    \end{align}
    clearly showing that the $\rho_i(\mathbf{r})$ are in general not totally symmetric under the action of $\mathcal{H}$, unless $\mathbf{D}^{\Gamma^{(d)}}(h)$ is the identity matrix for all $h \in \mathcal{H}$, but this would then mean that $\Gamma^{(d)}$ is a multiple of the totally symmetric irreducible representation of $\mathcal{H}$, which implies that the relation in Equation~\ref{eq:deg-repmat} would hold with $d = 1$, a contradiction.

  \subsection{Symmetry Projection of Electron Densities}
    We seek to construct a totally symmetric electron density based on the degenerate pure-state densities $\rho_i(\mathbf{r})$ given in Equation~\ref{eq:rho_i}.
    To this end, we shall make use of the projection operator for a representation $\Gamma$ with $d_{\Gamma}$ dimensions in the group $\mathcal{H}$:
    \begin{equation}
      \hat{\mathscr{P}}^{\Gamma} =
      \frac{d_{\Gamma}}{|\mathcal{H}|}
      \sum_{h \in \mathcal{H}} \chi^{\Gamma}(h)^* \hat{h}.
    \end{equation}
    We note that, since $\hat{\mathscr{P}}^{\Gamma}$ acts on the real linear space occupied by real-valued electron densities, we are at liberty to consider unitary representations of $\mathcal{H}$ even if $\mathcal{H}$ is a magnetic group.
    For the totally symmetric irreducible representation $\Gamma_{\mathrm{tot.sym.}}$, $d_{\Gamma_{\mathrm{tot.sym.}}} = 1$ and $\chi^{\Gamma_{\mathrm{tot.sym.}}}(h) = 1$ for all $h \in \mathcal{H}$, so the projection operator is reduced to
    \begin{equation}
      \hat{\mathscr{P}}^{\Gamma_{\mathrm{tot.sym.}}} =
      \frac{1}{|\mathcal{H}|}
      \sum_{h \in \mathcal{H}} \hat{h},
    \end{equation}
    and the totally symmetric electron density constructed from any of the degenerate pure-state densities $\rho_i(\mathbf{r})$ thus takes the form
    \begin{equation}
      \bar{\rho}(\mathbf{r})
      = \frac{1}{|\mathcal{H}|}
      \sum_{h \in \mathcal{H}} \hat{h}\rho_i(\mathbf{r}).
      \label{eq:rhobar_totsymproj}
    \end{equation}
    This is clearly an ensemble density constructed from $\rho_i(\mathbf{r})$ and all of its symmetry-equivalent partners in the group $\mathcal{H}$.
    
    The form of the totally symmetric electron density in Equation~\ref{eq:rhobar_totsymproj} is the \textit{operational} one as it shows how $\bar{\rho}(\mathbf{r})$ can be constructed from any of the degenerate pure-state densities $\rho_i(\mathbf{r})$ and its symmetry-equivalent partners.
    However, one question remains: is $\bar{\rho}(\mathbf{r})$ independent of the actual $\rho_i(\mathbf{r})$ chosen to be used in Equation~\ref{eq:rhobar_totsymproj}, because there is no \textit{a priori} requirement for the degenerate pure-state densities $\rho_i(\mathbf{r})$ to be linked by symmetry at all?
    To answer this, let us recall the Great Orthogonality Theorem for irreducible representations $\Gamma$ and $\Gamma'$\cite{book:Aktins2011}:
    \begin{equation}
      \sum_{h \in \mathcal{H}}
      D_{ij}^{\Gamma}(h)
      D_{i'j'}^{\Gamma'}(h)
      = \frac{|\mathcal{H}|}{d_{\Gamma}}
      \delta_{\Gamma \Gamma'} \delta_{ii'} \delta_{jj'}.
    \end{equation}
    If we now insert Equation~\ref{eq:g_rho_i} into Equation~\ref{eq:rhobar_totsymproj} and make use of the Great Orthogonality Theorem, we get
    \begin{align}
      \bar{\rho}(\mathbf{r})
      &= \frac{1}{|\mathcal{H}|}
      \sum_{h \in \mathcal{H}} \sum_{jj'}^{d}
      \rho_{jj'}(\mathbf{r})
      D_{ji}^{\Gamma^{(d)}}(h)^*
      D_{j'i}^{\Gamma^{(d)}}(h)
      \nonumber \\
      &= \frac{1}{d}
      \sum_{j=1}^{d}
      \rho_{j}(\mathbf{r}),
      \label{eq:rhobar_degset}
    \end{align}
    independent of the $\rho_i(\mathbf{r})$ we start with.
    This form of the totally symmetric electron density in Equation~\ref{eq:rhobar_degset} is the \textit{theoretical} one commonly seen in discussions of ensemble electron densities of degenerate states.
    It is easily seen from either Equation~\ref{eq:rhobar_totsymproj} or Equation~\ref{eq:rhobar_degset} that $\bar{\rho}(\mathbf{r})$ integrates to the same number of electrons $N_{\mathrm{e}}$ as any of the $\rho_i(\mathbf{r})$.
  
    What happens if we project $\rho_i(\mathbf{r})$ onto any non-totally-symmetric irreducible representations?
    We have, for $\Gamma \ne \Gamma_{\mathrm{tot.sym.}}$:
    \begin{equation}
      \rho^{\Gamma}(\mathbf{r})
      = \frac{d_{\Gamma}}{|\mathcal{H}|}
      \sum_{h \in \mathcal{H}}
      \chi^{\Gamma}(h)^*
      \hat{h}\rho_i(\mathbf{r}).
    \end{equation}
    Integrating the above over all space:
    \begin{align}
      \int \rho^{\Gamma}(\mathbf{r})\ \D\mathbf{r}
      &= \frac{d_{\Gamma}}{|\mathcal{H}|}
      \sum_{h \in \mathcal{H}}
      \chi^{\Gamma}(h)^*
      \hat{h} \int \rho_i(\mathbf{r})\ \D\mathbf{r}
      \nonumber \\
      &= \frac{N_{\mathrm{e}} d_{\Gamma}}{|\mathcal{H}|}
      \sum_{h \in \mathcal{H}}
      \chi^{\Gamma}(h)^*,
    \end{align}
    where we have used the fact that $N_{\mathrm{e}}$ is a scalar and must therefore be invariant with respect to all symmetry transformations.
    If we now recognize that $\chi^{\Gamma_{\mathrm{tot.sym.}}}(h) = 1$ for all $h \in \mathcal{H}$, we can write the above as
    \begin{equation}
      \int \rho^{\Gamma}(\mathbf{r})\ \D\mathbf{r}
      = \frac{N_{\mathrm{e}} d_{\Gamma}}{|\mathcal{H}|}
      \sum_{h \in \mathcal{H}}
      \chi^{\Gamma}(h)^* \chi^{\Gamma_{\mathrm{tot.sym.}}}(h).
      \label{eq:int_rho_gamma_intermediate}
    \end{equation}
    By making use of the Little Orthogonality Theorem for irreducible representations $\Gamma$ and $\Gamma'$\cite{book:Aktins2011},
    \begin{equation}
      \sum_{h \in \mathcal{H}}
      \chi^{\Gamma}(h)^*
      \chi^{\Gamma'}(h)
      = |\mathcal{H}| \delta_{\Gamma \Gamma'},
    \end{equation}
    Equation~\ref{eq:int_rho_gamma_intermediate} becomes
    \begin{equation}
      \int \rho^{\Gamma}(\mathbf{r})\ \D\mathbf{r}
      = 0, \quad \Gamma \ne \Gamma_{\mathrm{tot.sym.}}.
    \end{equation}
    This means that all non-totally-symmetric components of the degenerate pure-state densities $\rho_i(\mathbf{r})$ carry no electrons (and are thus not valid $N_{\mathrm{e}}$-electron densities).
    It is therefore physically meaningful and justifiable to consider only the totally symmetric component of $\rho_i(\mathbf{r})$.

  \bibliography{bib/ambidentnucleophiles}

  \clearpage

  \begin{figure*}[p]
    \centering
    \includegraphics[width=.6\textwidth]{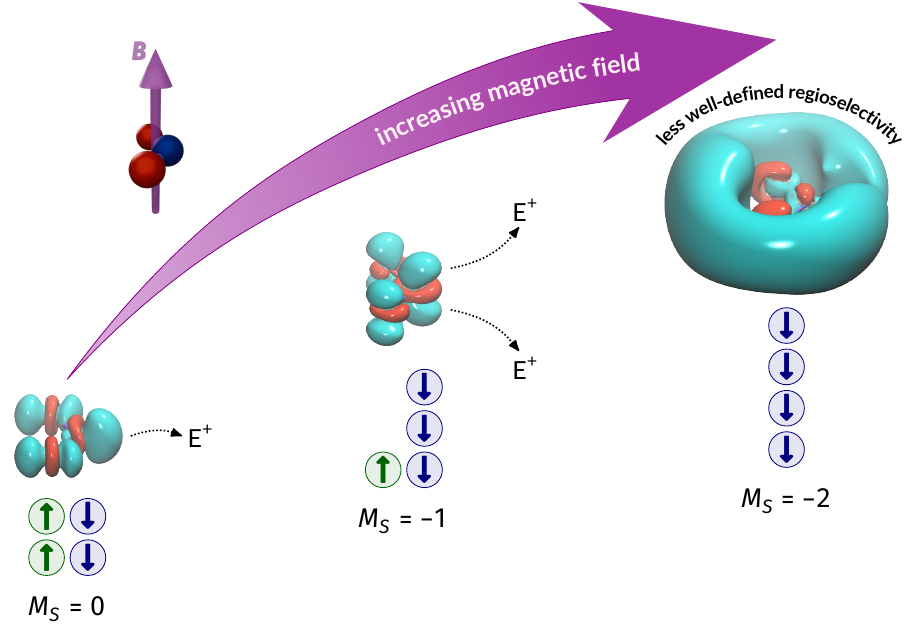}
    \caption*{Table of Contents Graphic.}
  \end{figure*}

\end{document}


\tableofcontents
  \clearpage

\section{Energy Variation of Frontier Molecular Orbitals with Magnetic Fields}
\label{suppinfo:mo-energies}

\begin{figure}
    \centering
    \includegraphics[width=\textwidth]{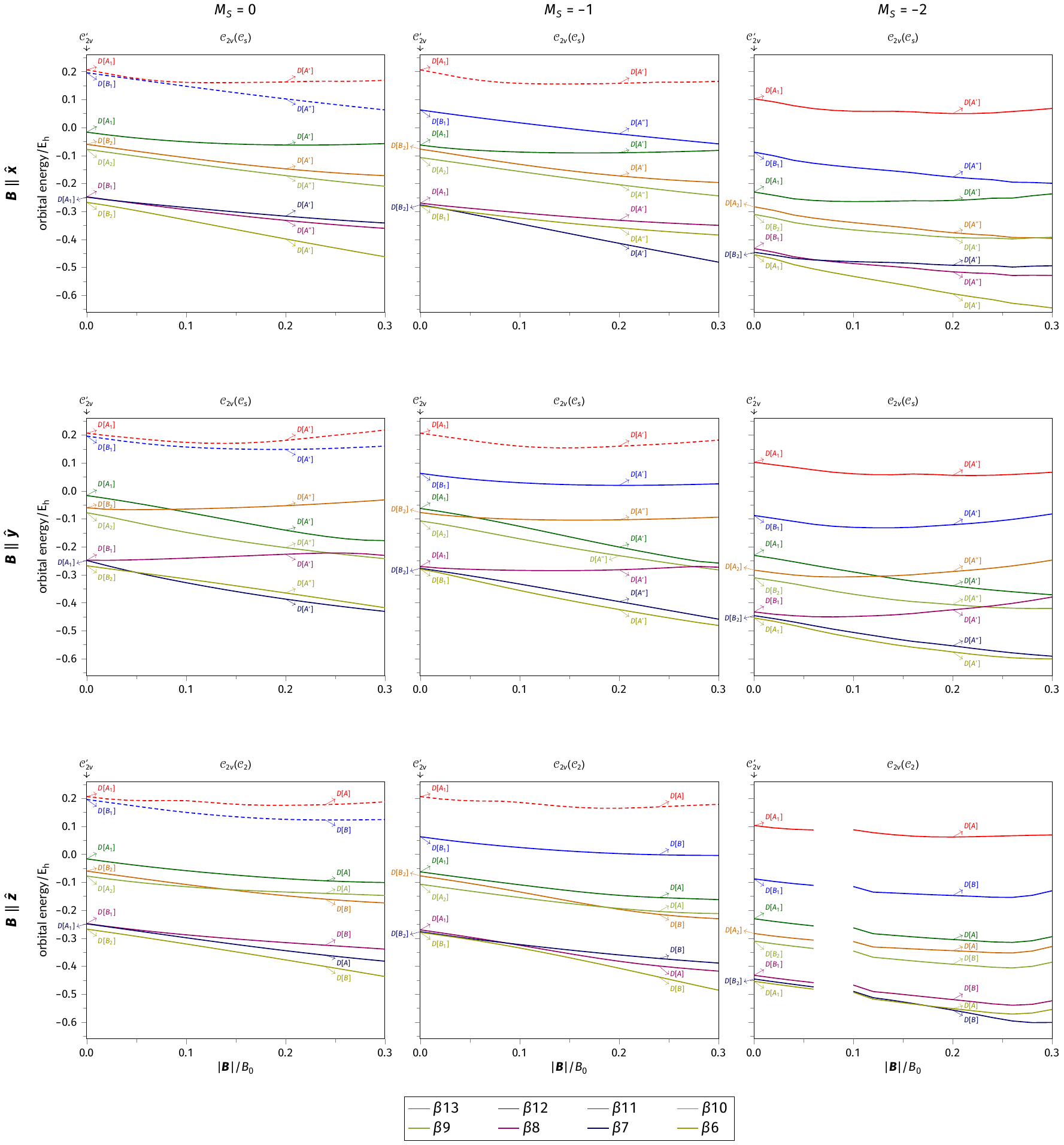}
    \caption{%
      Energy variation with respect to magnetic fields and corepresentation symmetry of frontier $\beta$-MOs in \ce{(NO2)-}.
      Occupied MOs are shown with solid curves, and unoccupied MOs are shown with dashed curves.\\[6pt]
      {\small Alt. text: Variation of frontier $\beta$-molecular orbital energies in \ce{(NO2)-} as a function of magnetic field strength for the three field orientations (perpendicular, in-plane, and parallel) and the three spin states ($M_S = 0$, $-1$, and $-2$).
      Orbital energies are labeled with their corepresentation symmetries; occupied orbitals are shown as solid curves and unoccupied orbitals as dashed curves.}%
    }
    \label{fig:no2m-mo_energies}
  \end{figure}

\begin{figure}
    \centering
    \includegraphics[width=\textwidth]{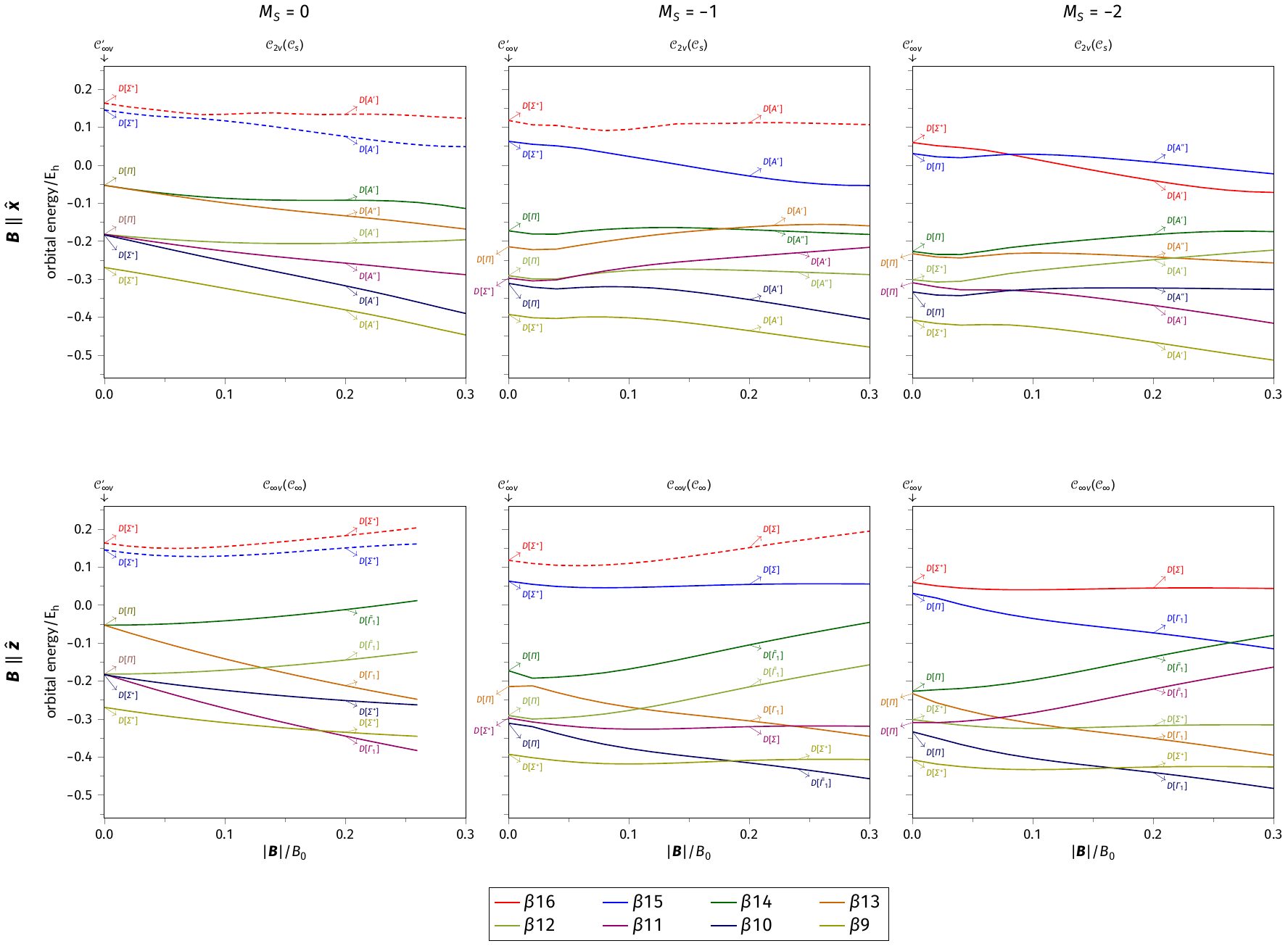}
    \caption{%
      Energy variation with respect to magnetic fields and corepresentation symmetry of frontier $\beta$-MOs in \ce{SCN-}.
      Occupied MOs are shown with solid curves, and unoccupied MOs are shown with dashed curves.\\[6pt]
      {\small Alt. text: Variation of frontier $\beta$-molecular orbital energies in \ce{SCN-} as a function of magnetic field strength for perpendicular and parallel magnetic fields and the three spin states ($M_S = 0$, $-1$, and $-2$).
      Orbital energies are labeled with their corepresentation symmetries; occupied orbitals are shown as solid curves and unoccupied orbitals as dashed curves.}%
    }
    \label{fig:scnm-mo_energies}
  \end{figure}

  Figures~\ref{fig:no2m-mo_energies} and \ref{fig:scnm-mo_energies} show the variation with magnetic fields strength of the energies of the frontier $\beta$-MOs in \ce{(NO2)-} and \ce{SCN-} for the field orientations and spin states considered in this work.

  \clearpage

\section{Character Tables for Selected Magnetic Groups}
\label{suppinfo:chartabs}

  Tables~\ref{tab:chartabs-c2v-prime}--\ref{tab:chartabs-cinfv-cinf} show the character tables for the magnetic groups involved in this work.

\begin{table}[h]
  \centering
  \caption{%
    Relevant character tables for the magnetic gray group $\mathcal{C}'_{2v}$.
  }
  \label{tab:chartabs-c2v-prime}
  \begin{subtable}[t]{.48\linewidth}
    \centering
    \subcaption{%
      Irreducible corepresentations.
      The irreducible corepresentation type gives the classification in Section~2.3.3 in the main text.
    }
    \label{tab:chartabs-c2v-prime-m}
    \scalebox{0.8}{
      \begin{tabular}[t]{l | R{0.6cm} R{0.6cm} R{0.6cm} R{0.6cm} | M{0.9cm}}
        \toprule
        $m\ \mathcal{C}'_{2v}$ & $\hat{E}$ & $\hat{C}_2$ & $\hat{\sigma}_v$ & $\hat{\sigma}'_v$ & Type \\ \midrule
        $D[A_1]$               & $+1$      & $+1$        & $+1$             & $+1$              & (i)  \\[3pt]
        $D[A_2]$               & $+1$      & $+1$        & $-1$             & $-1$              & (i)  \\[3pt]
        $D[B_1]$               & $+1$      & $-1$        & $+1$             & $-1$              & (i)  \\[3pt]
        $D[B_2]$               & $+1$      & $-1$        & $-1$             & $+1$              & (i)  \\ \bottomrule
      \end{tabular}
    }
  \end{subtable}
  \hfill
  \begin{subtable}[t]{.48\linewidth}
    \centering
    \subcaption{%
      Irreducible representations \textit{over a real linear space}.
      The $+/-$ presuperscripts give the parity of the irreducible representations under $\hat{\theta}$.
    }
    \label{tab:chartabs-c2v-prime-u}
    \scalebox{0.8}{
      \begin{tabular}[t]{l | R{0.6cm} R{0.6cm} R{0.6cm} R{0.6cm} R{0.6cm} R{0.6cm} R{0.6cm} R{0.6cm}}
        \toprule
        $u\ \mathcal{C}'_{2v}$ & $\hat{E}$ & $\hat{C}_2$ & $\hat{\sigma}_v$ & $\hat{\sigma}'_v$ & $\hat{\theta}$ & $\hat{\theta}\hat{C}_2$ & $\hat{\theta}\hat{\sigma}_v$ & $\hat{\theta}\hat{\sigma}'_v$ \\ \midrule
        $\prescript{+}{}{A}_1$ & $+1$      & $+1$        & $+1$             & $+1$              & $+1$           & $+1$                    & $+1$                         & $+1$                          \\[3pt]
        $\prescript{+}{}{A}_2$ & $+1$      & $+1$        & $-1$             & $-1$              & $+1$           & $+1$                    & $-1$                         & $-1$                          \\[3pt]
        $\prescript{+}{}{B}_1$ & $+1$      & $-1$        & $+1$             & $-1$              & $+1$           & $-1$                    & $+1$                         & $-1$                          \\[3pt]
        $\prescript{+}{}{B}_2$ & $+1$      & $-1$        & $-1$             & $+1$              & $+1$           & $-1$                    & $-1$                         & $+1$                          \\[3pt]
        $\prescript{-}{}{A}_1$ & $+1$      & $+1$        & $+1$             & $+1$              & $-1$           & $-1$                    & $-1$                         & $-1$                          \\[3pt]
        $\prescript{-}{}{A}_2$ & $+1$      & $+1$        & $-1$             & $-1$              & $-1$           & $-1$                    & $+1$                         & $+1$                          \\[3pt]
        $\prescript{-}{}{B}_1$ & $+1$      & $-1$        & $+1$             & $-1$              & $-1$           & $+1$                    & $-1$                         & $+1$                          \\[3pt]
        $\prescript{-}{}{B}_2$ & $+1$      & $-1$        & $-1$             & $+1$              & $-1$           & $+1$                    & $+1$                         & $-1$                          \\ \bottomrule
      \end{tabular}
    }
  \end{subtable}
  \\[6pt]
  {\small Alt. text: Character tables for the magnetic gray group $\mathcal{C}'_{2v}$, showing (a) irreducible corepresentations and their types and (b) irreducible representations over a real linear space, including parity under time reversal.}
\end{table}

\begin{table}[h]
  \centering
  \caption{Relevant character tables for the magnetic black-and-white group $\mathcal{C}_{2v}(\mathcal{C}_s)$.}
  \label{tab:chartabs-c2v-cs}
  \begin{subtable}[t]{.48\linewidth}
    \centering
    \subcaption{%
      Irreducible corepresentations.
      The irreducible corepresentation type gives the classification in Section~2.3.3 in the main text.
    }
    \label{tab:chartabs-c2v-cs-m}
    \scalebox{0.8}{
      \begin{tabular}[t]{l | R{0.6cm} R{0.6cm} | M{0.9cm}}
        \toprule
        $m\ \mathcal{C}_{2v}(\mathcal{C}_s)$ & $\hat{E}$ & $\hat{\sigma}_h$ & Type \\ \midrule
        $D[A']$                              & $+1$      & $+1$             & (i)  \\[3pt]
        $D[A'']$                             & $+1$      & $-1$             & (i)  \\ \bottomrule
      \end{tabular}
    }
  \end{subtable}
  \hfill
  \begin{subtable}[t]{.48\linewidth}
    \centering
    \subcaption{%
      Irreducible representations \textit{over a real linear space}.
      Since $\hat{\theta}\hat{C}_2$ is antiunitary, the principal rotation of this group has been chosen to be $\hat{E}$.
    }
    \label{tab:chartabs-c2v-cs-u}
    \scalebox{0.8}{
      \begin{tabular}[t]{l | R{0.6cm} R{0.6cm} R{0.6cm} R{0.6cm}}
        \toprule
        $u\ \mathcal{C}_{2v}(\mathcal{C}_s)$ & $\hat{E}$ & $\hat{\sigma}_h$ & $\hat{\theta}\hat{C}_2$ & $\hat{\theta}\hat{\sigma}_v$ \\ \midrule
        $A'_1$                               & $+1$      & $+1$             & $+1$                    & $+1$                         \\[3pt]
        $A'_2$                               & $+1$      & $+1$             & $-1$                    & $-1$                         \\[3pt]
        $A''_1$                              & $+1$      & $-1$             & $+1$                    & $-1$                         \\[3pt]
        $A''_2$                              & $+1$      & $-1$             & $-1$                    & $+1$                         \\ \bottomrule
      \end{tabular}
    }
  \end{subtable}
  \\[6pt]
  {\small Alt. text: Character tables for the magnetic black-and-white group $\mathcal{C}_{2v}(\mathcal{C}_s)$, showing (a) irreducible corepresentations and their types and (b) irreducible representations over a real linear space.}
\end{table}

\begin{table}[h]
  \centering
  \caption{%
    Relevant character tables for the magnetic black-and-white group $\mathcal{C}_{2v}(\mathcal{C}_2)$.
  }
  \label{tab:chartabs-c2v-c2}
  \begin{subtable}{.48\linewidth}
    \centering
    \subcaption{%
      Irreducible corepresentations.
      The irreducible corepresentation type gives the classification in Section~2.3.3 in the main text.
    }
    \label{tab:chartabs-c2v-c2-m}
    \scalebox{0.8}{
      \begin{tabular}{l | R{0.6cm} R{0.6cm} | M{0.9cm}}
        \toprule
        $m\ \mathcal{C}_{2v}(\mathcal{C}_2)$ & $\hat{E}$ & $\hat{C}_2$ & Type \\ \midrule
        $D[A]$               & $+1$      & $+1$  & (i)          \\[3pt]
        $D[B]$               & $+1$      & $-1$  & (i)          \\ \bottomrule
      \end{tabular}
    }
  \end{subtable}
  \hfill
  \begin{subtable}{.48\linewidth}
    \centering
    \subcaption{%
      Irreducible representations \textit{over a real linear space}.
      Since $\hat{C}_2$ is unitary, it is taken to be the principal rotation of this group.
    }
    \label{tab:chartabs-c2v-c2-u}
    \scalebox{0.8}{
      \begin{tabular}{l | R{0.6cm} R{0.6cm} R{0.6cm} R{0.6cm}}
        \toprule
        $u\ \mathcal{C}_{2v}(\mathcal{C}_2)$ & $\hat{E}$ & $\hat{C}_2$ & $\hat{\theta}\hat{\sigma}_v$ & $\hat{\theta}\hat{\sigma}'_v$ \\ \midrule
        $A_1$                                & $+1$      & $+1$        & $+1$                         & $+1$                          \\[3pt]
        $A_2$                                & $+1$      & $+1$        & $-1$                         & $-1$                          \\[3pt]
        $B_1$                                & $+1$      & $-1$        & $+1$                         & $-1$                          \\[3pt]
        $B_2$                                & $+1$      & $-1$        & $-1$                         & $+1$                          \\ \bottomrule
      \end{tabular}
    }
  \end{subtable}
  \\[6pt]
  {\small Alt. text: Character tables for the magnetic black-and-white group $\mathcal{C}_{2v}(\mathcal{C}_2)$, showing (a) irreducible corepresentations and their types and (b) irreducible representations over a real linear space.}
\end{table}

\begin{table}[h]
  \centering
  \caption{Relevant character tables for the magnetic grey group $\mathcal{C}'_{\infty v}$.}
  \label{tab:chartabs-cinfv-prime}
  \begin{subtable}[t]{\linewidth}
    \centering
    \subcaption{%
      Irreducible corepresentations.
      The irreducible corepresentation type gives the classification in Section~2.3.3 in the main text.
    }
    \label{tab:chartabs-cinfv-prime-m}
    \scalebox{0.8}{
      \begin{tabular}[t]{l l | R{0.6cm} L{1.5cm} R{0.6cm} R{0.8cm} | M{0.9cm}}
        \toprule
        \multicolumn{2}{c |}{$m\ \mathcal{C}'_{\infty v}$} & $\hat{E}$ & $2\hat{C}_z(\phi)$ & $\cdots$ & $\infty \hat{\sigma}_v$ & Type     \\ \midrule
        $D[\Sigma^+]$ & $D[A_1]$                           & $+1$      & $+1$               & $\cdots$ & $+1$                    & (i)      \\[3pt]
        $D[\Sigma^-]$ & $D[A_2]$                           & $+1$      & $+1$               & $\cdots$ & $-1$                    & (i)      \\[3pt]
        $D[\Pi]$      & $D[E_1]$                           & $+2$      & $2\cos \phi$       & $\cdots$ & $0$                     & (i)      \\[3pt]
        $D[\Delta]$   & $D[E_2]$                           & $+2$      & $2\cos 2\phi$      & $\cdots$ & $0$                     & (i)      \\[3pt]
        $D[\Phi]$     & $D[E_3]$                           & $+2$      & $2\cos 3\phi$      & $\cdots$ & $0$                     & (i)      \\[3pt]
                  \multicolumn{2}{c |}{$\cdots$}           & $\cdots$  & $\cdots$           & $\cdots$ & $\cdots$                & $\cdots$ \\ \bottomrule
      \end{tabular}
    }
  \end{subtable}

  \vspace{12pt}

  \begin{subtable}[t]{\linewidth}
    \centering
    \subcaption{%
      Irreducible representations \textit{over a real linear space}.
      The $+/-$ presuperscripts give the parity of the irreducible representations under $\hat{\theta}$.
    }
    \label{tab:chartabs-cinfv-prime-u}
    \scalebox{0.8}{
      \begin{tabular}[t]{l l | R{0.6cm} L{1.5cm} R{0.6cm} R{0.8cm} R{0.6cm} L{1.5cm} R{0.6cm} R{1.2cm}}
        \toprule
        \multicolumn{2}{c |}{$u\ \mathcal{C}'_{\infty v}$}
          & $\hat{E}$
          & $2\hat{C}_z(\phi)$
          & $\cdots$
          & $\infty \hat{\sigma}_v$
          & $\hat{\theta}$
          & $2\hat{\theta}\hat{C}_z(\phi)$
          & $\cdots$
          & $\infty \hat{\theta}\hat{\sigma}_v$ \\ \midrule
        $\prescript{+}{}{\Sigma}^+$
          & $\prescript{+}{}{A}_1$
          & $+1$
          & $+1$
          & $\cdots$
          & $+1$
          & $+1$
          & $+1$
          & $\cdots$
          & $+1$ \\[3pt]
        $\prescript{+}{}{\Sigma}^-$
          & $\prescript{+}{}{A}_2$
          & $+1$
          & $+1$
          & $\cdots$
          & $-1$
          & $+1$
          & $+1$
          & $\cdots$
          & $-1$ \\[3pt]
        $\prescript{+}{}{\Pi}$
          & $\prescript{+}{}{E}_1$
          & $+2$
          & $2\cos \phi$
          & $\cdots$
          & $0$
          & $+2$
          & $2\cos \phi$
          & $\cdots$
          & $0$ \\[3pt]
        $\prescript{+}{}{\Delta}$
          & $\prescript{+}{}{E}_2$
          & $+2$
          & $2\cos 2\phi$
          & $\cdots$
          & $0$
          & $+2$
          & $2\cos 2\phi$
          & $\cdots$
          & $0$ \\[3pt]
        $\prescript{+}{}{\Phi}$
          & $\prescript{+}{}{E}_3$
          & $+2$
          & $2\cos 3\phi$
          & $\cdots$
          & $0$
          & $+2$
          & $2\cos 3\phi$
          & $\cdots$
          & $0$  \\[3pt]
          \multicolumn{2}{c |}{$\cdots$}
            & $\cdots$
            & $\cdots$
            & $\cdots$
            & $\cdots$
            & $\cdots$
            & $\cdots$
            & $\cdots$
            & $\cdots$ \\[3pt]
          $\prescript{-}{}{\Sigma}^+$
          & $\prescript{-}{}{A}_1$
          & $+1$
          & $+1$
          & $\cdots$
          & $+1$
          & $-1$
          & $-1$
          & $\cdots$
          & $-1$ \\[3pt]
          $\prescript{-}{}{\Sigma}^-$
          & $\prescript{-}{}{A}_2$
          & $+1$
          & $+1$
          & $\cdots$
          & $-1$
          & $-1$
          & $-1$
          & $\cdots$
          & $+1$ \\[3pt]
          $\prescript{-}{}{\Pi}$
          & $\prescript{-}{}{E}_1$
          & $+2$
          & $2\cos \phi$
          & $\cdots$
          & $0$
          & $-2$
          & $-2\cos \phi$
          & $\cdots$
          & $0$ \\[3pt]
          $\prescript{-}{}{\Delta}$
          & $\prescript{-}{}{E}_2$
          & $+2$
          & $2\cos 2\phi$
          & $\cdots$
          & $0$
          & $-2$
          & $-2\cos 2\phi$
          & $\cdots$
          & $0$ \\[3pt]
          $\prescript{-}{}{\Phi}$
          & $\prescript{-}{}{E}_3$
          & $+2$
          & $2\cos 3\phi$
          & $\cdots$
          & $0$
          & $-2$
          & $-2\cos 3\phi$
          & $\cdots$
          & $0$  \\[3pt]
          \multicolumn{2}{c |}{$\cdots$}
          & $\cdots$
          & $\cdots$
          & $\cdots$
          & $\cdots$
          & $\cdots$
          & $\cdots$
          & $\cdots$
          & $\cdots$ \\ \bottomrule
      \end{tabular}
    }
  \end{subtable}
  \\[6pt]
  {\small Alt. text: Character tables for the magnetic gray group $\mathcal{C}'_{\infty v}$, showing (a) irreducible corepresentations for one- and two-dimensional symmetry species and (b) irreducible representations over a real linear space, including parity under time reversal.}
\end{table}

\begin{table}[h]
  \centering
  \caption{Relevant character tables for the magnetic black-and-white group $\mathcal{C}_{\infty v}(\mathcal{C}_{\infty})$.}
  \label{tab:chartabs-cinfv-cinf}
  \begin{subtable}[t]{.48\linewidth}
    \centering
    \subcaption{%
      Irreducible corepresentations.
      The irreducible corepresentation type gives the classification in Section~2.3.3 in the main text.
    }
    \label{tab:chartabs-cinfv-cinf-m}
    \scalebox{0.8}{
      \begin{tabular}[t]{l | R{0.6cm} L{1.8cm} R{0.6cm} | M{0.9cm}}
        \toprule
        $m\ \mathcal{C}_{\infty v}(\mathcal{C}_{\infty})$
          & $\hat{E}$
          & $2\hat{C}_z(\phi)$
          & $\cdots$
          & Type     \\ \midrule
        $D[\Sigma]$
          & $+1$
          & $+1$
          & $\cdots$
          & (i) \\[3pt]
        $D[\Gamma_1]$
          & $+1$
          & $\exp(+\mathrm{i} \phi)$
          & $\cdots$
          & (i) \\[3pt]
        $D[\bar{\Gamma}_1]$
          & $+1$
          & $\exp(-\mathrm{i} \phi)$
          & $\cdots$
          & (i) \\[3pt]
        $D[\Gamma_2]$
          & $+1$
          & $\exp(+2\mathrm{i} \phi)$
          & $\cdots$
          & (i) \\[3pt]
        $D[\bar{\Gamma}_2]$
          & $+1$
          & $\exp(-2\mathrm{i} \phi)$
          & $\cdots$
          & (i) \\[3pt]
        $D[\Gamma_3]$
          & $+1$
          & $\exp(+3\mathrm{i} \phi)$
          & $\cdots$
          & (i) \\[3pt]
        $D[\bar{\Gamma}_3]$
          & $+1$
          & $\exp(-3\mathrm{i} \phi)$
          & $\cdots$
          & (i) \\[3pt]
        $\cdots$
          & $\cdots$
          & $\cdots$
          & $\cdots$
          & $\cdots$ \\ \bottomrule
      \end{tabular}
    }
  \end{subtable}
  \hfill
  \begin{subtable}[t]{.48\linewidth}
    \centering
    \subcaption{%
      Irreducible representations \textit{over a real linear space}.
      Since $\hat{C}_z(\phi)$ is unitary, it is taken to be the principal rotation of this group.
    }
    \label{tab:chartabs-cinfv-cinf-u}
    \scalebox{0.8}{
      \begin{tabular}[t]{l l | R{0.6cm} L{1.8cm} R{0.6cm} R{1.2cm}}
        \toprule
        \multicolumn{2}{c |}{$u\ \mathcal{C}_{\infty v}(\mathcal{C}_{\infty})$}
        & $\hat{E}$
        & $2\hat{C}_z(\phi)$
        & $\cdots$
        & $\infty \hat{\theta}\hat{\sigma}_v$     \\ \midrule
        $\Sigma^+$
          & $A_1$
          & $+1$
          & $+1$
          & $\cdots$
          & $+1$ \\[3pt]
        $\Sigma^-$
          & $A_2$
          & $+1$
          & $+1$
          & $\cdots$
          & $-1$ \\[3pt]
        $\Pi$
          & $E_1$
          & $+2$
          & $2\cos \phi$
          & $\cdots$
          & $0$ \\[3pt]
        $\Delta$
          & $E_2$
          & $+2$
          & $2\cos 2\phi$
          & $\cdots$
          & $0$ \\[3pt]
        $\Phi$
          & $E_3$
          & $+2$
          & $2\cos 3\phi$
          & $\cdots$
          & $0$ \\[3pt]
        \multicolumn{2}{c |}{$\cdots$}
          & $\cdots$
          & $\cdots$
          & $\cdots$
          & $\cdots$ \\ \bottomrule
      \end{tabular}
    }
  \end{subtable}
  \\[6pt]
  {\small Alt. text: Character tables for the magnetic black-and-white group $\mathcal{C}_{\infty v}(\mathcal{C}_{\infty})$, showing (a) irreducible corepresentations and their types and (b) irreducible representations over a real linear space.}
\end{table}


  \clearpage